\documentclass[12pt]{JHEP3}
\usepackage{amsmath,amssymb,epsfig}

\newcommand{\be}{\begin{equation}}
\newcommand{\ee}{\end{equation}}
\newcommand{\beqa}{\begin{eqnarray}}
\newcommand{\eeqa}{\end{eqnarray}}

\def\GeV{{\rm GeV}}
\def\fm{{\rm fm}}
\def\MeV{{\rm MeV}}

\def\bra#1{\left\langle #1\right|}
\def\eeq{\end{equation}}

\def\ket#1{\left| #1\right\rangle}

\def\Tr{\mathop{\rm Tr}}

\relax

\title{Wilson loops in heavy ion collisions and their calculation in AdS/CFT}

\author{Hong Liu${}^{\,1}$, Krishna Rajagopal${}^{\,1}$ and
Urs Achim Wiedemann${}^{\,2}$\\
\vspace{0.1in}

${}^{\,1}$Center for Theoretical Physics, MIT,
Cambridge, MA 02139, USA
\vspace{0.1in}

${}^{\,2}$Department of Physics, CERN, Theory Division, CH-1211 Geneva 23
\vspace{0.1in}

E-mail addresses: {\tt hong\_liu@mit.edu, krishna@ctp.mit.edu, Urs.Wiedemann@cern.ch}
}

\abstract{Expectation values of Wilson loops define the nonperturbative
properties of the hot medium produced in heavy ion collisions that
arise in the analysis of both
radiative parton energy loss and quarkonium suppression.
We use the AdS/CFT correspondence to
calculate the expectation values of such Wilson loops
in the strongly coupled plasma of ${\cal N}=4$ super Yang-Mills (SYM) theory,
allowing for the possibility that the plasma may be moving with some collective
flow velocity as is the case in heavy ion collisions.
We obtain the ${\cal N}=4$ SYM values of
the jet quenching parameter $\hat q$, which describes the energy
loss of a hard parton in QCD, and of the
velocity-dependence of the quark-antiquark screening
length for a moving dipole as a function of the angle between
its velocity and its orientation.
We show that
if the quark-gluon plasma
is flowing with velocity $v_f$  at an angle $\theta$ with respect to the trajectory of
a hard parton, the jet quenching parameter $\hat q$ is modified  by a
factor $\gamma_f\left(1 - v_f\, \cos\theta \right)$, and show
that this result applies in QCD as in ${\cal N}=4$  SYM.
We discuss the relevance
of the lessons we are learning from all these calculations to heavy
ion collisions at RHIC and at the LHC.
Furthermore, we discuss the relation between our results and those
obtained in other theories with gravity duals, showing in particular that
the ratio between $\hat q$ in any two conformal theories with gravity duals
is the square root of the ratio of their central charges.
This leads us to conjecture that in nonconformal theories $\hat q$ defines
a quantity that
always decreases along renormalization group trajectories and
allows us to use
our calculation of $\hat q$ in ${\cal N}=4$ SYM to make a conjecture for
its value in QCD.
}

\preprint{MIT-CTP-3794\\ CERN-PH-TH/2006-257
\\ hep-ph/0612168}

\begin{document}
\def\vev#1{\langle#1\rangle}
\def\ov{\over}
\def\le{\left}
\def\ri{\right}
\def\ha{{1\over 2}}
\def\lam{{\lambda}}
\def\Lam{{\Lambda}}
\def\al{{\alpha}}
\def\ket#1{|#1\rangle}
\def\bra#1{\langle#1|}
\def\vev#1{\langle#1\rangle}
\def\det{{\rm det}}
\def\tr{{\rm tr}}
\def\Tr{{\rm Tr}}
\def\NN{{\cal N}}
\def\th{{\theta}}

\def\Om{{\Omega}}
\def \th{{\theta}}

\def \lam {\lambda}
\def \om {\omega}
\def \ra {\rightarrow}
\def \ga {\gamma}
\def\sig{{\sigma}}
\def\ep{{\epsilon}}
\def\apr{{\alpha'}}
\newcommand{\p}{\partial}
\def\LL{{\cal L}}
\def\TT{{\cal T}}
\def\CC{{\cal C}}

\def\tir{{\tilde r}}

\newcommand{\bea}{\begin{eqnarray}}
\newcommand{\eea}{\end{eqnarray}}
\newcommand{\nn}{\nonumber\\}
%%%%%%%%%%%%%%%%%%%%%%%%%%%%%%%%%%%%%%%%%%%%%%%%%%%%%%%%%%%%%%%%%%%%%%%%%%%%%%%%
%%%%%   Section 1. Introduction and Summary
%%%%%%%%%%%%%%%%%%%%%%%%%%%%%%%%%%%%%%%%%%%%%%%%%%%%%%%%%%%%%%%%%%%%%%%%%%%%%%%%

\section{Introduction}
Understanding the implications of
data from the Relativistic Heavy Ion Collider (RHIC) poses qualitatively new
challenges~\cite{RHIC}.   The characteristic features of the matter
produced at RHIC, namely its large and anisotropic collective flow and its
strong interaction with (in fact not so) penetrating hard probes,  indicate that
the hot matter produced in RHIC collisions must be described by QCD
in a regime of strong, and hence nonperturbative, interactions.
In this regime, lattice QCD has to date been the prime calculational tool based
solely on first principles.  On the other hand, analyzing the very same RHIC
data on collective flow, jet quenching and other hard probes requires
real-time dynamics: the hot fluid produced in heavy ion collisions is exploding rather
than static, and jet quenching by definition concerns probes of this fluid
which, at least initially,
are moving through it at close to the speed of light.
Information on real-time dynamics in a strongly interacting quark-gluon plasma
from lattice QCD is at present both scarce and indirect.
Complementary methods for real-time strong coupling calculations at finite temperature
are therefore desirable.

For a class of non-abelian thermal gauge field theories, the AdS/CFT conjecture provides
such an alternative~\cite{AdS/CFT}.
It gives analytic access to the strong coupling regime of finite temperature
gauge field theories in the limit of large number of colors ($N_c$)
by mapping nonperturbative problems at strong coupling
onto calculable problems  in the supergravity limit of a
dual string theory, with the background metric describing a curved
five-dimensional anti-deSitter
spacetime containing a black hole whose horizon is displaced
away from ``our'' 3+1 dimensional world in the fifth dimension.
Information about real-time dynamics within a thermal background can be
obtained in this set-up. The best-known example is the
calculation of the shear viscosity in several
supersymmetric gauge theories~\cite{Policastro:2001yc,Kovtun:2003wp,Buchel:2003tz,Buchel:2004hw,Buchel:2004di,Buchel:2004qq,Son:2006em,Benincasa:2006fu}.
It was found that the dimensionless ratio of the shear viscosity to the entropy density takes on the
``universal''~\cite{Kovtun:2003wp,Buchel:2003tz,Buchel:2004qq,Benincasa:2006fu} value $1/4\pi$ in the large number of colors ($N_c$) and
large 't Hooft coupling ($\lambda\equiv g_{\rm YM}^2 N_c$) limit of any gauge theory that admits a holographically dual supergravity description. Although
the AdS/CFT correspondence is not directly applicable to QCD, the universality of the result
for the shear viscosity and its numerical coincidence with
estimates of the same quantity in QCD made by comparing RHIC data with
hydrodynamical model analyses~\cite{Teaney:2003kp} have motivated further effort in
applying the AdS/CFT conjecture to calculate other quantities
which are of interest for the RHIC heavy ion program.
This has lead to the calculation of certain diffusion constants~\cite{Policastro:2002se} and thermal spectral functions~\cite{Teaney:2006nc}, as well as to first work~\cite{Sin:2004yx} towards a dual description of dynamics in heavy ion collisions themselves.
More recently, there has been much interest in the AdS/CFT calculation of
the jet quenching parameter which controls the description of
medium-induced energy loss for relativistic partons in QCD~\cite{Liu:2006ug,Buchel:2006bv,Vazquez-Poritz:2006ba,Caceres:2006as,Lin:2006au,Avramis:2006ip,Armesto:2006zv,Nakano:2006js} and the drag coefficient
which describes the energy loss for heavy quarks in ${\cal N}=4$ supersymmetric
Yang-Mills
theory~\cite{Herzog:2006gh,Gubser:2006bz,Casalderrey-Solana:2006rq,Drag1,Friess:2006aw}.
There have also been studies of the stability of heavy quark bound states
in a thermal environment~\cite{Rey:1998bq,Mateos:2006nu,Peeters:2006iu} with
collective motion~\cite{Peeters:2006iu,Liu:2006nn,Chernicoff:2006hi,Caceres:2006ta,Argyres:2006vs,Avramis:2006em,Friess:2006rk,Talavera:2006tj,Chernicoff:2006yp,SeeAlso}.

The expectation values of Wilson loops contain
gauge invariant information about
the nonperturbative physics of non-abelian gauge field theories.
When evaluated at temperatures above the crossover from hadronic
matter to the strongly interacting quark-gluon plasma, they can be related
to a number of different quantities which are in turn accessible in heavy
ion collision experiments.
In Section 2 of this paper, which should be seen as an extended introduction,
we review these connections. We review how the expectation value of
a particular time-like Wilson loop, proportional to $\exp(-i S)$ for some real $S$,
serves to define the potential between a static quark and antiquark in
a (perhaps moving) quark-gluon plasma.  However, in order to obtain
a sensible description of the photo-absorption cross-section in
deep inelastic scattering, the Cronin effect in proton-nucleus collisions,
and radiative parton energy loss and hence jet quenching in nucleus-nucleus
collisions, the expectation value of this Wilson loop must be proportional
to $\exp(-S)$ for some real and positive $S$ once the Wilson loop is taken to
lie along the lightcone.   In Section 3, we present the calculation of the
relevant Wilson loops in hot ${\cal N}=4$ supersymmetric Yang-Mills theory, using
the AdS/CFT correspondence, and show how its expectation value goes from
$\exp(-i S)$ to $\exp(-S)$ (as it must if this theory is viable as a model for
the quark-gluon plasma in QCD)
as the order of two non-commuting limits is exchanged.
The jet quenching parameter
$\hat q$, which describes the energy loss of a hard parton in QCD,
and the velocity-dependent
quark-antiquark potential for a dipole moving through the quark-gluon plasma
arise in different limits of the same Nambu-Goto action
which depends on the dipole rapidity $\eta$  and on $\Lambda$, the
location in $r$, the fifth dimension of the AdS space, of the boundary of the AdS
space where the dipole is located.
If we take $\eta\to\infty$ first, and only then take $\Lambda\to\infty$,
the Nambu-Goto action describes a space-like world sheet bounded
by a light-like Wilson loop at $r=\Lambda$, and defines the jet quenching parameter.
If instead we take $\Lambda\to\infty$ first, the action describes a time-like world sheet
bounded by a time-like Wilson loop, and defines the
$q\bar{q}$-potential for a dipole moving with rapidity $\eta$.
We review the calculation of both quantities.
In Section 4 we calculate the jet quenching parameter in a moving quark-gluon plasma,
and show that our result in this section is valid in QCD as in $\NN=4$ SYM.
In Section 5 we return to the velocity-dependent screening length, calculating
it for all values of the angle between the velocity and orientation of the
quark-antiquark dipole.

Section 6 consists of an extended discussion.
We summarize our results on the velocity-dependent screening length in Section 6.1.
In Section 6.2, we comment
on the differences between the calculation of the jet quenching parameter
and the drag force on a (heavy)
quark~\cite{Herzog:2006gh,Gubser:2006bz,Casalderrey-Solana:2006rq,Drag1,Friess:2006aw}.
In Section 6.3, we then compare our calculation of the jet quenching parameter to the
value of this quantity extracted in comparison with RHIC data.
The success of this comparison motivates us to, in Section 6.4,  enumerate
the differences between QCD and ${\cal N}=4$ supersymmetric Yang-Mills (SYM) theory, which
have qualitatively distinct vacuum properties, and the rapidly growing list of similarities
between the properties of the quark-gluon plasmas in these two theories.
A comparison between our result for the velocity scaling of
the quark-antiquark screening length and future data from RHIC and the LHC
on the suppression of high transverse momentum $J/\Psi$ and $\Upsilon$ mesons
could add one more entry to this list.
The single difference between ${\cal N}=4$ SYM and QCD which appears to us most likely
to affect the value of the jet quenching parameter is the difference in the number
of degrees of freedom in the two theories.  We therefore close in Section 6.5 by
reviewing the AdS/CFT calculations to date
of $\hat q$ in theories other than ${\cal N}=4$ SYM, and show that for any two conformal
field theories in which this calculation can be done, the ratio of $\hat q$ in one theory
to that in the other will be given by the square root of the ratio of the
central charges, and hence the number of degrees
of freedom.  This suggests that $\hat q$ in QCD is smaller than that in ${\cal N}=4$ SYM
by a factor of order $\sqrt{120/47.5}\sim 1.6$.  This conjecture can be tested by further
calculations in nonconformal theories.

A reader interested in our results and our perspective on our results should
focus on Sections 4 and 6. A reader interested in how we obtain our results
should focus on Section 3.

%%%%%%%%%%%%%%%%%%%%%%%%%%%%%%%%%%%%%%%%%%%%%%%
\section{Wilson loops in heavy ion collisions}
\label{sec2}

In this section, we consider Wilson lines
\begin{equation}
    W^r({\cal C}) ={\rm Tr}\, {\cal P} \exp \left[ i \int_{\cal C} dx_\mu\, A^\mu(x) \right]\, ,
    \label{2.1}
\end{equation}
where $\int_{\cal C}$ denotes a line integral along the closed path ${\cal C}$.
$W^r({\cal C})$ is the trace of an $SU(N)$-matrix
in the fundamental or adjoint representation, $r=F,A$, respectively.
The vector potential
$A^{\mu}(x) = A^{\mu}_a(x)\, T^a$ can be expressed in terms of the generators $T^a$ of the
corresponding representation, and ${\cal P}$ denotes path ordering. We discuss
several cases in which nonperturbative properties of interest in heavy ion physics
and high energy QCD can be expressed in terms of expectation values of (\ref{2.1}).

%
%%%%%%%%%%%%%%%%%%%%%%%%%%%%%%%%%%%%%%%%%%%%%%
%\begin{figure}[t]
\FIGURE[t]{
%\vskip-0.15in
% \hfill\hskip-0.6in
\includegraphics[scale=0.5,angle=0]{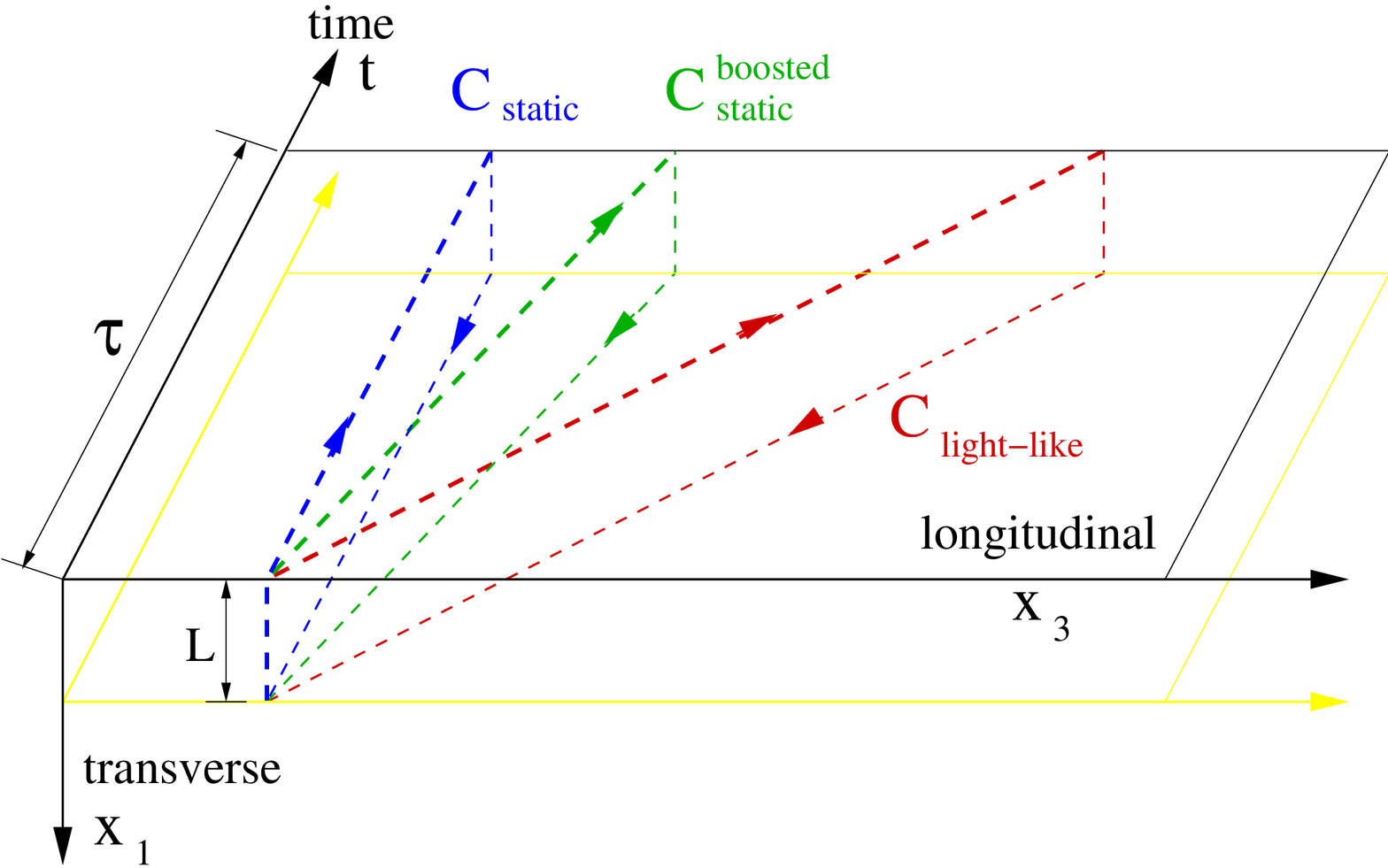}
%\hskip-0.1in
%\hfill
\caption{Schematic illustration of the shape of Wilson loops ${\cal C}$, corresponding to a $q\bar{q}$
dipole of size $L$, oriented along the $x_1$-direction, which is  (i) at rest with respect to the
medium (${\cal C}_{\rm static}$), (ii) moving with some finite velocity $v=\tanh\eta$ along the
longitudinal $x_3$-direction (${\cal C}_{\rm static}^{\rm boosted}$), or (iii) moving with
the velocity of light along the $x_3$-direction (${\cal C}_{\rm light-like}$).}
\label{fig0}
}
%\end{figure}
%%%%%%%%%%%%%%%%%%%%%%%%%%%%%%%%%%%%%%%%%%%%%%%%%
%

%%%%%%%%%%%%%%%%%%%%%%%%%%%%%%%%%%%%%%%%%%%%%%%
\subsection{The quark-antiquark static potential}

We shall use the Wilson loop
\begin{equation}
    \langle W^F({\cal C}_{\rm static}) \rangle = \exp \left[ - i\, {\cal T}\,
         \left(E(L) - E_{\rm ren}\right) \right]
    \label{2.2}
\end{equation}
to furnish a working
definition of
the $q\bar{q}$  static potential $E(L)$
for an infinitely heavy
quark-antiquark pair at rest with respect to the medium and
separated by a distance $L$.
Here, the closed contour ${\cal C}_{\rm static}$
has a short segment of length $L$ in the transverse direction,
and a very long extension
${\cal T}$ in the temporal direction, see Fig.~\ref{fig0}. The
potential $E(L)$ is defined
in the limit ${\cal T} \to \infty$.  The properties of the medium, including
for example its temperature $T$, enter into (\ref{2.2}) via
the expectation value $\langle \ldots \rangle$.
Here, $E_{\rm ren}$ is an $L$-independent renormalization, which
is typically infinite.
Eq. (\ref{2.2}) is written
for a Minkowski metric, as appropriate for our
consideration, below, of a quark-antiquark pair moving through the medium.
At zero temperature, the analytic continuation
$i{\cal T}\, \to {\cal T}$ of (\ref{2.2})
yields the standard relation between the static potential and
an Euclidean Wilson loop~\cite{Wilson:1974sk}.
In finite
temperature lattice
QCD~\cite{McLerran:1980pk,Philipsen:2002az,Kaczmarek:2002mc}, one
typically
defines a quark-antiquark static potential from the correlation
function of a pair of Polyakov loops
wrapped around the periodic Euclidean time direction. (For a
discussion of this procedure and alternatives to it,
see also Ref.~\cite{Laine:2006ns}.) In these Euclidean finite
temperature lattice calculations,
the corresponding quark-antiquark potential is renormalized such that
it matches  the zero temperature result at small
distances~\cite{Kaczmarek:2002mc}.
We shall use an analogous prescription.
We note that while (\ref{2.2})
is difficult to analyze in QCD,
its evaluation is straightforward for a class of strongly interacting
gauge theories
in the large number of colors limit
at both zero~\cite{Rey:1998ik} and nonzero temperature~\cite{Rey:1998bq},
as we shall see in Section 3.

The dissociation of charmonium and bottomonium bound states has been proposed as a signal for the formation of a hot and deconfined quark-gluon plasma~\cite{Matsui:1986dk}.
Recent analyses of this phenomenon are based on the study of
the quark-antiquark
static potential extracted from lattice QCD~\cite{Satz:2005hx}.
In these calculations of $E(L)$, the $q\bar{q}$-dipole
is taken to be at rest in the thermal medium,
and its temperature dependence is studied in detail.
In heavy ion collisions, however, quarkonium bound states are produced
moving with some velocity $v=\tanh\eta$ with respect to the medium. If the relative velocity of the quarkonium
exceeds a typical thermal velocity, one may expect that quarkonium suppression is
enhanced compared to thermal dissociation in a heat bath at
rest~\cite{Liu:2006nn}.
For a calculation of the velocity-dependent dissociation of such a
moving $q\bar{q}$-pair in a medium at rest in the $x_3$-direction, one has to evaluate
(\ref{2.2}) for the Wilson loop ${\cal C}_{\rm static}^{\rm boosted}$, depicted in Fig.~\ref{fig0}.
The orientation of the loop in the $(x_3,t)$-plane changes as a function of $\eta$.
This case is discussed in section~\ref{sec3c}. In section~\ref{sec4}, we discuss the
generalization to dipoles oriented in an arbitrary direction in the $(x_1,x_3)$-plane.

%%%%%%%%%%%%%%%%%%%%%%%%%%%%%%%%%%%%%%%%%%%%%%%
\subsection{Eikonal propagation \label{sec:eikon}}
We now recall cases of physical interest where, unlike in (\ref{2.2}),
the expectation value of a Wilson loop in Minkowski space is the exponent of a real quantity.
Such cases are important
in the high energy limit of various scattering problems.
Straight light-like Wilson lines
of the form
$W({\bf x}_i)={\cal P}\exp\{i\int dz^-T^aA^+_a({\bf x}_i, z^-)\}$
typically arise in such calculations when --- due to Lorentz contraction
--- the transverse position of a colored projectile does not change while propagating through the
target. The interaction of the projectile wave function with the target can then be described in
the eikonal approximation as a color rotation
$\alpha_i \to \beta_i$ of each projectile component $i$, resulting in an eikonal phase
$W_{\alpha_i \beta_i}({\bf x}_i)$.
A general discussion of this eikonal propagation approximation can be found in
Refs.~\cite{Kovner:2001vi,Kovner:2003zj}. Here, we describe two specific cases,
in which  expectation values of a fundamental and of an adjoint Wilson loop arise, respectively.

%%%%%%%%%%%%%%%%%%%%%%%%%%%%%%%%%%%%%%%%%
\subsubsection{Virtual photoabsorption cross section}

In deep inelastic scattering (DIS), a virtual photon $\gamma^*$
interacts with a hadronic target. At small Bjorken $x$, DIS can be formulated by
starting from the decomposition of the virtual photon into hadronic Fock states  and propagating
these Fock states in the eikonal approximation through the
target~\cite{Mueller:1989st,Nikolaev:1990ja,Piller:1999wx,Kovchegov:1999kx,Wiedemann:2000ez}.
However, in a DIS scattering experiment
the virtual photon
does not have time to branch into Fock states
containing many soft particles (equivalently, it does not have time to develop
a colored field) prior to interaction, as it would if it could propagate forever.
Instead, the dominant
component of its wave function which interacts with the target is its
$q\bar{q}$ Fock component:
\begin{equation}
  \vert\gamma^*\rangle =\int d^2({\bf x-y})\, dz\,
\psi({\bf x- y},z) \frac{1}{\sqrt{N}}
 \delta_{\alpha\, \bar{\alpha}}
                    |\alpha({\bf x})\, ,\bar{\alpha}({\bf y}),z\rangle\, .
\label{2.3}
\end{equation}
Here, $|\alpha({\bf x})\, ,\bar{\alpha}({\bf y}),z\rangle$ denotes a $q\bar{q}$-state, where
a quark of color $\alpha$ carries an energy fraction $z$ and propagates at transverse
position ${\bf x}$. The corresponding antiquark propagates at transverse position ${\bf y}$
and carries the remaining energy. The Kronecker $\delta_{\alpha\, \bar{\alpha}}$ ensures
that this state is in a color singlet.
$N$ is the number of colors; the probability that the photon splits into a quark antiquark
pair with any one particular color is proportional to $1/N$.
The wave function $\psi$ is written in the mixed
representation,  using configuration space in the transverse direction and momentum
space in the longitudinal direction.
It can be
calculated perturbatively from the $\gamma^* \to q \bar{q}$ splitting~\cite{Kovchegov:1999kx}.
Given an incoming state
$\vert \Psi_{\rm in} \rangle =  |\alpha({\bf x})\, ,\bar{\alpha}({\bf y})\rangle$, in the
eikonal approximation
the outgoing state
reads
$\vert \Psi_{\rm out} \rangle= W_{\alpha\gamma}^F({\bf x})\,
W_{\bar{\alpha} \bar{\gamma}}^{F\dagger}({\bf y})
|\gamma({\bf x})\, ,\bar{\gamma}({\bf y})\rangle$, and the total cross section is obtained by
squaring $\vert \Psi_{\rm tot} \rangle = \vert \Psi_{\rm out} \rangle - \vert \Psi_{\rm in} \rangle$.
{}From the virtual photon state (\ref{2.3}), one finds in this way the total virtual photoabsorption cross
section~\cite{Kovner:2001vi}
\begin{eqnarray}
\sigma^{DIS}&=&\int d^2{\bf x}\,d^2{\bf y}\,dz\,
\psi({\bf x- y},z)\,\psi^*({\bf x- y},z)\,
P^{q\bar q}_{\rm tot}({\bf x},{\bf y})\, ,
\label{2.4}\\
  P_{\rm tot}^{q\bar{q}} &=& \, \left\langle 2 -
  \frac{1}{N}
    {\rm Tr}\left[ W^F({\bf x})\, W^{F\dagger}({\bf y})\right]
- \frac{1}{N}
    {\rm Tr}\left[ W^F({\bf y})\, W^{F\dagger}({\bf x})
\right]
  \right\rangle\, .
  \label{2.5}
\end{eqnarray}
This DIS total cross section is written in terms of the expectation value of a
fundamental Wilson loop:
\begin{eqnarray}
\frac{1}{N}
\langle  {\rm Tr}\left[ W^{F\, \dagger}({\bf y})\, W^F({\bf x}) \right] \rangle
\longrightarrow
\left\langle{W^F({\cal C}_{\rm light-like})}\right\rangle
    = \exp\left[ - \frac{1}{8}  Q_s^2\, L^2 \right] + {\cal O}\left(\frac{1}{N^2} \right)\, .
\label{2.6}
 \end{eqnarray}
By the $\rightarrow$ we mean that in order to obtain a
gauge-invariant formulation, we have connected the two long light-like Wilson lines
separated by
the small transverse separation $L = \vert {\bf x} - {\bf y}\vert$
by two short transverse segments of length $L$, located a long distance $L^-\gg L$ apart.
This yields the closed rectangular loop ${\cal C}_{\rm light-like}$ illustrated in
Fig.~\ref{fig0}.
The expectation value
$\langle \dots \rangle $ denotes an average over the states of
the hadronic target; technically, this amounts to an average over the target color fields $A^\mu$ in the
Wilson line (\ref{2.1}).  If we could do deep inelastic scattering off a droplet of
quark-gluon plasma, the $\langle \dots \rangle $ would be a thermal expectation value.
We have parameterized $\langle{W^F({\cal C}_{\rm light-like})}\rangle$
in terms of the saturation scale $Q_s^2$. This is the standard parametrization
of virtual photoabsorption cross sections in the saturation physics approach to
DIS off hadrons and nuclei~\cite{Kovner:2005pe,Weigert:2005us,Iancu:2003xm}.
Although we do not know the form of the $1/N^2$ corrections to (\ref{2.6}), we
do know that they must be such that
$\langle W^F \rangle \rightarrow 0$  in the $L\rightarrow 0$ limit,
since in this limit $P_{\rm tot}^{q\bar{q}}$ must vanish.

We note that for small $L$, the $L^2$-dependence of the exponent in (\ref{2.6})
follows from general considerations. Since the transverse size of the $q\bar{q}$-dipole
is conjugate to the virtuality $Q$ of the photon,
$L^2 \sim 1/Q^2$, one finds
$P_{\rm tot}^{q\bar{q}} =  \frac{1}{4}  Q_s^2\, L^2 + O(L^4) \sim Q_s^2/Q^2$.
This is the
expected leading $Q^2$-dependence at high virtuality.

General considerations also indicate that the exponent in (\ref{2.6}) must have a
real part.
To see this, consider the limit of large $L$ and small virtuality, when the virtual photon is large in
transverse space, and its local interaction probability should go to unity. Since Eq. (\ref{2.5})
is the sum of the elastic and inelastic scattering probability, which are both normalized to one,
one requires $P_{\rm tot}^{q\bar{q}}  \to 2$ in this large-$L$ limit. This cannot be achieved with
an imaginary exponent in (\ref{2.6}).

The saturation momentum $Q_s$ is a characteristic property of any hadronic target. Qualitatively,
the gluon distribution inside the hadronic target is dense (saturated) as seen by virtual
photons
up to
a virtuality $Q_s$,
but it is dilute as seen at higher virtuality. As a consequence, a virtual photon has a
probability of order one for interacting with the target, if  --- in a configuration space
picture --- its transverse size is $\vert {\bf x} - {\bf y}\vert > 1/Q_s$, and it has a much smaller probability
of interaction for $\vert {\bf x} - {\bf y}\vert \ll 1/Q_s$. This is the physics behind (\ref{2.4}) and (\ref{2.5}).

%%%%%%%%%%%%%%%%%%%%%%%%%%%%%%%%%%%%%%%%
\subsubsection{The Cronin effect in proton-nucleus (p-A) collisions}
In comparing transverse momentum spectra from proton-nucleus and proton-proton
collisions, one finds that in an intermediate transverse momentum range of
$p_T \sim 1-5$ GeV, the hadronic yield in p-A collisions is enhanced~\cite{CRONIN}.
This so-called Cronin effect is typically understood in terms of the transverse momentum
broadening of the incoming partons in the proton projectile, prior to undergoing the hard interaction
in which the high-$p_T$ parton is produced. On the partonic level, this phenomenon
and its energy dependence have been studied by calculating
the gluon radiation induced by
a single quark in the
incoming proton projectile
scattering on a target of nuclear
size $A$ and corresponding saturation scale
$Q_s$~\cite{Kharzeev:2002pc,Baier:2003hr,Kharzeev:2003wz,Albacete:2003iq,Jalilian-Marian:2003mf,Blaizot:2004wu}.

One starts from the incoming wave function $\Psi_{\rm in}^{\alpha}$ of a bare
quark $\vert \alpha ({\bf 0})\rangle$, supplemented by the coherent state of quasi-real gluons
which build up its Weizs\"acker-Williams field $f({\bf x}) \propto g\frac{{\bf x}}{ {\bf x}^2}$.
Here, $g$ is the strong coupling constant and ${\bf x}$  and ${\bf 0}$ are the transverse positions
of the gluon and parent quark~\cite{Kovner:2001vi}.
Suppressing Lorentz and spin indices, one has
$\Psi_{\rm in}^{\alpha} = \vert \alpha({\bf 0}) \rangle + \int d{\bf x}\, d\xi\, f({\bf x})\, T_{\alpha\beta}^b
\vert \beta({\bf 0}); b({\bf x},\xi)\rangle$. The ket $\vert \beta({\bf 0}); b({\bf x},\xi)\rangle$
describes the two-parton state, consisting of a quark with color $\beta$ at transverse position
${\bf 0}$
and a gluon of color $b$ at transverse position ${\bf x}$. In the eikonal approximation,
the distribution of the radiated gluon is flat
in rapidity $\xi$. The outgoing wave function differs from $\Psi_{\rm in}^{\alpha}$
by color rotation with the phases $W^F_{\alpha\beta}$ for quarks and
$W^A_{bc}$ for the gluons:
\begin{eqnarray}
 \Psi_{\rm out}^\alpha &=&
  W^F_{\alpha\, \gamma}({\bf 0})\, \vert\gamma\rangle +
  \int   d{\bf x}\, f({\bf x})\, T_{\alpha\, \beta}^b
  W_{\beta\, \gamma}^F({\bf 0})\, W_{b\, c}^A({\bf x})\,
  \vert \gamma\, ;c({\bf x})\rangle\, .
  \label{2.7}
\end{eqnarray}
($\alpha$, $\beta$ and $\gamma$ are fundamental indices; $b$, $c$ and $d$ below
are adjoint indices.)
To calculate an observable related to an inelastic process, such as the number of gluons
$dN_{\rm prod}/d{\bf k}$ produced in the scattering, one first determines the component of the
outgoing wave function, which belongs to the subspace orthogonal to the incoming state
$|\delta\Psi\rangle = \left[1-|\Psi_{in}\rangle\langle\Psi_{in}|\right]| \Psi_{out}\rangle$. Next,
one counts the number of gluons in this state~\cite{Kovchegov:1998bi,Kovner:2003zj}
\begin{eqnarray}
  && \frac{dN_{\rm prod}}{d{\bf k}} = \frac{1}{N} \sum_{\alpha,d}
  \left\langle \delta \Psi_\alpha\vert a_d^\dagger({\bf k})\,
  a_d({\bf k})\vert\, \delta\Psi_\alpha \right\rangle
  \nonumber\\
  && \quad
  = \frac{\alpha_s\, C_F}{2\pi}\,
  \int d{\bf x}\, d{\bf y}\, e^{i{\bf k}\cdot({\bf x}-{\bf y})}
  \frac{ {\bf x}\cdot {\bf y}}{ {\bf x}^2\, {\bf y}^2}
  \frac{1}{N^2 - 1}\,
  \Bigg[ \left\langle
  {\rm Tr}\left[ W^{A\, \dagger}({\bf 0})\, W^A({\bf 0}) \right]
  \right\rangle
  - \left\langle
  {\rm Tr}\left[ W^{A\, \dagger}({\bf x})\, W^A({\bf 0}) \right]
  \right\rangle
  \nonumber\\
&& \qquad \qquad \qquad \qquad -
  \left\langle
  {\rm Tr}\left[ W^{A\, \dagger}({\bf y})\, W^A({\bf 0}) \right]
  \right\rangle
%   \nonumber \\
% && \qquad \qquad \qquad \qquad \qquad \qquad
  +
  \left\langle
  {\rm Tr}\left[ W^{A\, \dagger}({\bf y})\, W^A({\bf x}) \right]
  \right\rangle\Bigg]\, .
  \label{2.8}
\end{eqnarray}
Here, ${\bf x}$ and ${\bf y}$ denote the transverse positions of the gluon in the
amplitude and complex conjugate amplitude. The fundamental Wilson lines $W^F({\bf 0})$
at transverse position ${\bf 0}$, which appear in (\ref{2.7}), combine into an adjoint Wilson
line via the identity
$W^A_{ab}({\bf 0}) = 2 \,{\rm Tr}\left[ W^F({\bf 0})T^a W^{F\dagger}({\bf 0})T^b\right]$.
We now see that
the only information about the target which enters in (\ref{2.8})  is  that encoded in the transverse
size dependence of the expectation value of two light-like adjoint Wilson lines, which
we can again close to form a loop:
\begin{eqnarray}
\frac{1}{N^2 - 1}
\left\langle  {\rm Tr}\left[ W^{A\, \dagger}({\bf y})\, W^A({\bf x}) \right] \right\rangle
\longrightarrow
\left\langle{W^A({\cal C}_{\rm light-like})}\right\rangle
    = \exp\left[ - \frac{1}{4}  Q_s^2\, L^2 \right] + {\cal O}\left(\frac{1}{N^2}\right).
\label{2.9}
 \end{eqnarray}
Consistent with the identity ${\rm Tr} \,W^A = {\rm Tr}\, W^F\, {\rm Tr}\, W^F - 1$, the parameterization
of the expectation values of the adjoint and fundamental Wilson loops in (\ref{2.6}) and (\ref{2.9})
respectively differs in the large-$N$ limit only by a factor of 2 in the exponent.

Inserting (\ref{2.9}) into (\ref{2.8}), Fourier transforming the Weizs\"acker-Williams factors
and doing the integrals, one finds formally
\be
    \frac{dN_{\rm prod}}{d{\bf k}} = \frac{4\pi}{Q_s^2} \int d{\bf q}\,
        \exp\left[ - \frac{{\bf q}^2}{Q_s^2} \right]\,
        \frac{{\bf q}^2}{{\bf k}^2 \left({\bf q}-{\bf k}\right)^2}\, .
        \label{2.10}
\ee
To interpret this expression, we recall the high energy limit for gluon radiation in single
quark-quark scattering. For a transverse momentum transfer ${\bf q}$ between the
scattering partners, the spectrum in the gluon transverse momentum ${\bf k}$ is
proportional to the so-called Bertsch-Gunion factor
$\frac{{\bf q}^2}{{\bf k}^2 \left({\bf q}-{\bf k}\right)^2}$. Hence, Eq. (\ref{2.10}) indicates
that the saturation scale $Q_s$ characterizes the average squared transverse
momentum ${\bf q}^2$ transferred from the hadronic target to the highly energetic
partonic projectile. We caution the reader that the integrals in (\ref{2.8}) are divergent
and that the steps leading to (\ref{2.10}) remain formal since they were performed without
proper regularization of these integrals.  Furthermore, a more refined parametrization
of the saturation scale in QCD includes a logarithmic dependence of
$Q_s$ on the transverse separation $L$. Including this correction allows
for a proper regularization~\cite{Kovchegov:1998bi,Baier:2003hr}.
The analysis of (\ref{2.8}) is more complicated, but the lesson drawn from
(\ref{2.10}) remains unchanged: the saturation scale $Q_s^2$ determines the average
squared transverse momentum, transferred from the medium to the projectile.

The dependence of the saturation scale $Q^2_s$ on nuclear size $A$ is
$Q_s^2 \propto A^{1/3}$, i.e., $Q_s^2$ is linear in the in-medium path-length.
According to (\ref{2.10}), transverse momentum is accumulated in the hadronic target
due to Brownian motion, ${\bf q}^2 \propto A^{1/3}$.
In the discussion of high-energy scattering problems in heavy ion physics, where
the in-medium path length depends on the geometry and collective dynamics
of the collision region, it has proven advantageous to separate this path-length
dependence explicitly~\cite{Baier:2002tc}
\be
    Q_s^2 = \hat{q}\, \frac{L^-}{\sqrt{2}}\, ,
    \label{2.11}
\ee
in so doing defining a new parameter $\hat q$.
Here, we have expressed the longitudinal distance  $\Delta z = \frac{L^-}{\sqrt{2}}$ in terms
of the light-cone distance $L^-$.
The parameter $\hat q$ characterizes the average transverse momentum
squared transferred from the target to the projectile
per unit longitudinal distance travelled, i.e. per unit path length.
Note that $\hat q$ is well-defined for arbitrarily large $L^-$ in
an infinite medium, whereas $Q_s^2$ diverges
linearly with $L^-$ and so is appropriate only for a finite system.
We shall see in Section 2.3 that, when the expectation value in (\ref{2.9}) is evaluated
in a hot quark-gluon plasma rather than over the gluonic states of a cold nucleus as above,
the quantity $\hat q$ governs the energy loss of relativistic partons moving through
the quark-gluon plasma.  The simpler examples we have introduced here in Section 2.2
motivate the need for a nonperturbative evaluation of
the light-like Wilson loop $\left\langle{W({\cal C}_{\rm light-like})}\right\rangle$
in a background corresponding to a hadron or a cold nucleus, as in so doing one
could calculate the saturation scale and
describe DIS at small $x$ and the Cronin effect.  Unfortunately, although
hot ${\cal N}=4$ supersymmetric Yang-Mills theory describes a system with
many similarities to the quark-gluon plasma in QCD
as we shall discuss in Section 6,
it does not seem suited to modelling a cold nucleus.

%%%%%%%%%%%%%%%%%%%%%%%%%%%%%%%%%%%%%%%%%%%%%
\subsection{BDMPS radiative parton energy loss and the jet quenching parameter}
In the absence of a medium, a highly energetic parton produced in a hard process
decreases its virtuality by multiple parton splitting prior to hadronization. In a heavy ion collision,
this perturbative parton shower interferes with additional medium-induced radiation. The resulting interference pattern resolves longitudinal distances in the
target ~\cite{Baier:1996sk,Zakharov:1997uu,Wiedemann:2000za}. As a consequence,
its description goes beyond the eikonal approximation, in which the entire target acts totally
coherently as a single scattering center. As we shall explain now, this refined
kinematical description does not involve additional information about the
medium beyond that already encoded in the jet quenching parameter $\hat q$
that we have already introduced.

In the Baier-Dokshitzer-Mueller-Peigne-Schiff~\cite{Baier:1996sk} calculation of medium-induced
gluon radiation, the radiation amplitude for the medium-modified splitting processes $q \to q\, g$ or
$g \to g\, g$ is calculated for the kinematic region
\be
    E \gg \omega \gg \vert {\bf k}\vert, \vert {\bf q} \vert \equiv \vert \sum_i {\bf q}_i\vert \gg
                    T\, , \Lambda_{\rm QCD}.
    \label{2.12}
\ee
The energy $E$ of the initial hard parton is much larger than the energy $\omega$ of the
radiated gluon, which is much larger than the transverse momentum ${\bf k}$ of the
radiated gluon or the transverse momentum ${\bf q}$ accumulated due to many
scatterings of the projectile inside the target. This ordering is also at the basis of the
eikonal approximation. In the BDMPS formalism, however, terms which are subleading
in $O(1/E)$ are kept and this allows for a calculation of interference effects. To keep
$O(1/E)$-corrections to the phase of scattering amplitudes, one replaces eikonal
Wilson lines by the retarded Green's functions~\cite{Zakharov:1997uu,Kopeliovich:1998nw,Wiedemann:2000za}
\begin{equation}
    G(x^-_2,{\bf r}_2; x^-_1,{\bf r}_1\vert p) = \int_{{\bf r}(x^-_1)={\bf r}_1}^{{\bf r}(x^-_2)={\bf r}_2}
    {\cal D}{\bf r}(x^-)\, \exp\left[ \frac{ip}{2}\int_{x^-_1}^{x^-_2} dx^-\,
                                                   \left(\frac{d{\bf r}(x^-)}{dx^-}\right)^2
                                                   -i\, \int_{x^-_1}^{x^-_2} dx^-\, A^+(x^-,{\bf r}(x^-)) \right]\, .
     \label{2.13}
\end{equation}
Here, $p$ is the total momentum of the propagating parton, and the
color field $A^+ = A^+_a\, T^a$ is in the representation of the parton.
The integration goes over all possible paths ${\bf r}(x^-)$ in the light-like
direction between ${\bf r}_1 = {\bf r}(x_1^-)$ and ${\bf r}_2 = {\bf r}(x_2^-)$.
Green's functions of the form (\ref{2.13}) are solutions to the Dirac equation
in the  spatially extended target color field
$A^+$~\cite{Buchmuller:1995mr,Kopeliovich:1998nw,Wiedemann:2000ez}.
In the limit of ultra-relativistic momentum
$p \to \infty$, Eq. (\ref{2.13}) reduces to a Wilson line (\ref{2.1}) along an eikonal
light-like direction.
In the BDMPS formalism, the inclusive energy distribution of gluon
radiation from a high energy parton produced within a medium can be written in
terms of in-medium expectation values
of pairs of Green's functions of the
form (\ref{2.13}), one coming from the amplitude
and the other coming from the conjugate amplitude.
After a lengthy but purely technical calculation, it can be written in the
form~\cite{Wiedemann:2000za}
\begin{eqnarray}
  \omega\frac{dI}{d\omega\, d{\bf k}}
  &=& \frac{\alpha_s\,  C_R}{ (2\pi)^2\, \omega^2}\,
    2{\rm Re} \int_{\xi_0}^{\infty}\hspace{-0.3cm} dy_l
  \int_{y_l}^{\infty} \hspace{-0.3cm} d\bar{y}_l\,
   \int d{\bf u}\,
%   \int_0^{\chi \omega}\, d{\bf k}_\perp\,
  e^{-i{\bf k}\cdot{\bf u}}   \,
  \exp\left[ { -\frac{1}{4} \int_{\bar{y}_l}^{\infty} d\xi\, \hat{q}(\xi)\,
    {\bf u}^2 }\right] \,
  \nonumber \\
  && \times \frac{\partial }{ \partial {\bf y}}\cdot
  \frac{\partial }{ \partial {\bf u}}\,
  \int_{{\bf y}=0}^{{\bf u}={\bf r}(\bar{y}_l)}
  \hspace{-0.5cm} {\cal D}{\bf r}
   \exp\left[  \int_{y_l}^{\bar{y}_l} d\xi
        \left( \frac{i\, \omega}{2} \dot{\bf r}^2
          - \frac{1}{4} \hat{q}(\xi) {\bf r}^2  \right)
                      \right]\, .
    \label{2.14}
\end{eqnarray}
Here,  the Casimir operator $C_R$ is in the representation of the
parent parton. In the
configuration space representation used in (\ref{2.14}),
$\xi_0$ is the position at which the initial parton is produced in a hard process
and the internal integration variables $y_l$ and $\bar{y}_l$ denote the longitudinal
position at which this initial parton radiates the gluon in the
amplitude and complex conjugate amplitude,
respectively. (See Refs.~\cite{Wiedemann:2000za,Kovner:2003zj} for details.)
Since all partons propagate with the velocity of light, these longitudinal positions
correspond to emission times $y_l$, $\bar{y}_l$.

In deriving (\ref{2.14})~\cite{Wiedemann:2000za,Kovner:2003zj},
the initial formulation of the $q\to q\, g$ radiation amplitude of course involves
Green's functions (\ref{2.13}) in both
the fundamental and in the adjoint representation.
However, via essentially the same color algebraic identities which allowed us
to write the gluon spectrum (\ref{2.8})  in terms of expectation
values of adjoint Wilson loops only, the result
given in (\ref{2.14}) has been written in
terms of expectation values of adjoint light-like Green's functions of the form
(\ref{2.13}) only. These in turn
have been written in terms of the same jet quenching parameter
$\hat{q}$ defined as in (\ref{2.9}) and (\ref{2.11}),
namely via~\cite{Kovner:2003zj}
\begin{equation}
\left\langle{W^A({\cal C}_{\rm light-like})}\right\rangle
    = \exp\left[ - \frac{1}{4\sqrt{2}} \, {\hat q} L^- \, L^2 \right] +
{\cal O}\left(\frac{1}{N^2}\right),
\label{2.14a}
 \end{equation}
now with the expectation value of the light-like Wilson loop
evaluated in a thermal quark-gluon plasma
rather than in a cold nucleus.  The quantity $\hat q(\xi)$ which arises in (\ref{2.14})
is the value of $\hat q$ at the longitudinal position $\xi$, which
changes  with increasing $\xi$ as the plasma expands and dilutes.  In our analysis
of a static medium, $\hat q(\xi)=\hat q$ is
constant.

In QCD, radiative parton energy loss is the dominant energy loss mechanism
in the limit in which the initial parton has arbitrarily high
energy.
To see this, we proceed as follows.
Note first that in this high parton energy limit
the assumptions (\ref{2.12}) underpinning the BDMPS calculation
become controlled. And, given the ordering of energy scales in
(\ref{2.12}), the quark-gluon
radiation vertex should be evaluated with
coupling constant
$\alpha_s({\bf k}^2)$. The distribution of
the transverse momenta of the radiated gluon is
peaked around ${\bf k}^2\sim Q_s^2 = {\hat q} L^-/\sqrt{2}$~\cite{Salgado:2003rv}
which means that, in
the limit of large in-medium path length $L^-/\sqrt{2}$,
the coupling $\alpha_s$ is evaluated at a
scale ${\bf k}^2 \gg T^2$ at which it is weak,
justifying the
perturbative BDMPS formulation~\cite{Baier:1996sk}.  Next, we note that in the limit of large
in-medium path length the result (\ref{2.14})
yields~\cite{Baier:1996sk,Salgado:2003gb}
\be
    \omega \frac{dI}{d\omega} = \frac{\alpha_s C_R}{\pi} 2\, {\rm Re}\,
    \ln \left[ \cos\left( (1+i)\, \sqrt{\frac{\hat{q}\, {L^-}^2/2}{4\omega}}
                          \right) \right]\, .
                          \label{2.15}
\ee
Integrating this expression over $\omega$, one finds
that the average medium-induced
parton energy loss is given by
\be
\Delta E = \frac{1}{4} \alpha_s C_R \hat{q} \frac{{L^-}^2}{2}\ ,
\label{2.15a}
\ee
which
is independent of $E$ and {\it quadratic} in the path
length $L^-$.\footnote{For any finite
$L^-$, corrections to (\ref{2.14}) can
make the average energy loss $\Delta E$ grow logarithmically
with $E$ at large enough $E$~\cite{Zakharov:2000iz}.
% the argument of the $\cos$ in (\ref{2.15a}) is modified such that $\Delta E$
% in (\ref{2.15a}) grows logarithmically with $E$~\cite{Zakharov:2000iz}.
%the average energy loss $\Delta E$ can be shown to grow logarithmically
%with $E$
%~\cite{Zakharov:2000iz}, if eq. (\ref{2.14}) is modified to include
%contributions from single hard scattering contributions.}
}
This makes the energy lost by gluon radiation parametrically larger
in the high energy limit than that lost due to collisions alone, which
grows only linearly with path length, and makes radiative energy loss
dominant in the high parton energy limit.
Radiative parton energy loss has been
argued to be the dominant mechanism behind jet quenching at
RHIC~\cite{Zakharov:1997uu,Wiedemann:2000za,Gyulassy:2000er,jetquenchrev},
where
the high energy partons whose energy loss is observed in the data
have transverse momenta of at most
about 20 GeV~\cite{RHIC}.
At the LHC, the BDMPS calculation
will be under better control since the high energy partons
used to probe the quark-gluon plasma will then have transverse momenta
greater than 100 GeV~\cite{LHC}.

Although the BDMPS calculation itself is under control in the high
parton energy limit, a weak coupling calculation of
the jet quenching parameter $\hat q$ is not, as we now explain.
Recall that $\hat q$ is the transverse momentum squared
transferred from the medium to either the initial parton or
the radiated gluon, per distance travelled.
In a weakly coupled quark-gluon plasma, in which scatterings
are rare, $\hat q$ is given by
the momentum squared transferred in a single collision divided
by the mean free path between collisions.
Even though the total momentum transferred from the medium to the initial
parton and to the radiated gluon is perturbatively large since it
grows linearly with the path length,
the momentum transferred per individual scattering is only of order
$g(T)\, T$. So, a weak-coupling calculation of $\hat q$ is justified
only if $T$ is so large that physics at the scale $T$ is perturbative.
Up to a logarithm, such a weak-coupling calculation
yields~\cite{Baier:1996sk,Baier:2002tc,Baier:2006fr}
\begin{equation}
\hat q_{\rm weak-coupling} = \frac{8 \zeta(3)}{\pi} \alpha_s^2 N^2 T^3
\label{2.16}
\end{equation}
if $N$, the number of colors, is large.
However, given the evidence from RHIC data~\cite{RHIC} (the
magnitude of jet quenching itself;
azimuthal anisotropy comparable to that predicted by
zero-viscosity hydrodynamics) that the quark-gluon plasma is strongly
interacting at the temperatures accessed in RHIC collisions,
there is strong motivation to calculate $\hat q$ directly from
its definition via the light-like Wilson loop (\ref{2.14a}),
without assuming weak coupling.
If and when the quark-gluon plasma is strongly interacting,
the coupling constant involved in the multiple soft gluon exchanges
described by the weak-coupling calculation of $\hat q$ is in fact
nonperturbatively large, invalidating (\ref{2.16}).

To summarize, the BDMPS analysis of a parton losing
energy as it traverses a strongly interacting quark-gluon plasma
is under control in the high parton
energy limit, with gluon radiation the dominant energy loss mechanism
and the basic calculation correctly treated as perturbative.  In this
limit, application of strong coupling techniques to the entire radiation
process described by Eq. (\ref{2.14}) would be inappropriate, because
QCD is asymptotically free.  The physics of the strongly interacting
medium itself enters the calculation through the single jet quenching
parameter $\hat q$, the amount of transverse momentum squared picked
up per distance travelled by both the initial parton and the radiated
gluon.  A perturbative calculation of $\hat q$ is not under control,
making it worthwhile to investigate any strong coupling techniques
available for the evaluation of this one nonperturbative quantity.

%%%%%%%%%%%%%%%%%%%%%%%%%%%%%%%%%%%%%%%%%%%%%%%%%%%%%%%%%%%%%%%%%%%%%%%%%%%%%%%%
%%%%%   Section 3. Wilson loops from AdS/CFT
%%%%%%%%%%%%%%%%%%%%%%%%%%%%%%%%%%%%%%%%%%%%%%%%%%%%%%%%%%%%%%%%%%%%%%%%%%%%%%%%
\section{Wilson loops from AdS/CFT in ${\cal N}=4$ super
Yang-Mills theory \label{sec3}}

In section~\ref{sec2}, we have
recalled measurements of interest
in heavy ion collisions, whose description depends on thermal
expectation values of Wilson loops. For questions related to the
dissociation of quarkonium, the relevant Wilson loop is
time-like and $\langle W({\cal C}) \rangle$ is the exponent of an
imaginary quantity. Questions related to medium-induced energy
loss involve light-like Wilson loops and $\langle W({\cal C})
\rangle$ is the exponent of a real quantity.

In this section, we evaluate thermal expectation values of these
Wilson loops for thermal ${\cal N}=4$ super Yang-Mills (SYM)
theory with gauge group $SU(N)$ in the large $N$ and large 't~Hooft coupling limits, making use of the AdS/CFT
correspondence~\cite{AdS/CFT,Rey:1998ik}. In the present context,
this correspondence maps
the evaluation of
a Wilson loop in a hot strongly interacting gauge theory plasma
onto the much simpler problem
of finding the extremal area
of a classical string world sheet in a
black hole background~\cite{Rey:1998bq}.
We
shall find that the cases of real and imaginary exponents
correspond to space-like and time-like world sheets,
which both arise naturally as we shall describe.

%bounded by ${\cal C}$,
%which arise naturally in different ordered
%limits discussed below.

${\cal N}=4$ SYM is a supersymmetric gauge theory with one gauge
field $A_\mu$, six scalar fields $X^I, I=1,2,\cdots 6$ and four
Weyl fermionic fields $\chi_i$, all transforming in the adjoint
representation of the gauge group, which we take to be $SU(N)$.
The theory is conformally invariant and is specified by
two parameters: the rank
of the gauge group $N$ and the 't Hooft coupling $\lambda$,
 \begin{equation}
    \lambda = g_{YM}^2\, N\, .
    \label{3.1}
 \end{equation}
 (Note that the gauge coupling in the standard field theoretical
 convention $g_{YM}$, which we shall use throughout,
is related to that in the standard string theory convention $g_M$
by $g_{YM}^2=2\,g_M^2$.)

According to the AdS/CFT correspondence, Type IIB string theory in
an $AdS_5 \times S_5$ spacetime is equivalent to an $\NN=4$ SYM
living on the boundary of the AdS$_5$. The
string coupling $g_s$, the curvature radius $R$ of
the AdS metric
and the tension $\frac{1}{2\pi\alpha'}$ of the string are
related to the field theoretic quantities as
 \begin{equation}
 \frac{R^2 }{ \alpha'} = \sqrt{\lambda}\, ,
    \qquad
    4\pi\, g_s =  g_{YM}^2 = \frac{\lambda}{N}\, .
      \label{3.2}
 \end{equation}
Upon first taking the large $N$ limit at fixed $\lambda$ (which means
$g_s\rightarrow 0$) and then taking the large $\lambda$ limit (which
means large string tension)
$\NN=4$ SYM theory is
described by classical supergravity in $AdS_5 \times S_5$.
We shall describe the modification of this spacetime which corresponds
to introducing a nonzero temperature in the gauge theory below.

$\NN=4$ SYM  does not contain any fields in the fundamental
representation of the gauge group. To construct the Wilson loop
describing the phase associated with a particle in the fundamental
representation, we introduce a probe D3-brane at the boundary of
the AdS$_5$ and lying along $\vec n$ on $S^5$, where $\vec n$ is a
unit vector in ${\bf R}^6$~\cite{Rey:1998ik}.
The D3-brane (i.e. the boundary of the AdS$_5$)
is at some fixed, large value of $r$, where $r$
is the coordinate of the 5th dimension of AdS$_5$, meaning that
the space-time within the D3-brane is ordinary $3+1$-dimensional
Minkowski space.
The fundamental
``quarks'' are then given by the ground states of strings
originating on the boundary D3-brane and extending towards the
center of the AdS$_5$.\footnote{By the standard IR/UV connection~\cite{IRUV},
the boundary of the AdS$_5$ at some large value of $r$ corresponds
to an ultraviolet cutoff in the field theory.  The Wilson loop must
be located on a D3-brane at this boundary, not at some smaller $r$, in
order that it describes a test quark whose size is not resolvable.
Evaluating the expectation value of a Wilson loop then corresponds to
using pointlike test quarks to
probe physics in the field theory at length scales longer
than the ultraviolet cutoff.}
The
corresponding Wilson loop operator has the form
 \be \label{Wils}
 W(\CC) = {1 \ov N} \Tr P \exp \le[i \oint_{\CC} ds \, \le( A_\mu \dot x^\mu +
 \vec n \cdot \vec X \sqrt{\dot x^2} \ri) \ri]
 \ee
which, in comparison with (\ref{2.1}), also contains scalar fields
$\vec X = (X^1, \cdots X^6)$.    In the large $N$ and large $\lam$ limits, the
expectation value of a Wilson loop operator (\ref{Wils}) is given by the
classical action of a string in $AdS_5 \times S_5$,
with the boundary condition that the string world sheet
ends on the curve ${\cal C}$ in the probe brane.
The contour ${\cal C}$ lives within the $3+1$-dimensional
Minkowski space defined by the D3-brane, but the string
world sheet attached to it hangs ``down'' into the bulk of the
curved five-dimensional AdS$_5$ spacetime.
The classical string action is obtained by extremizing the
Nambu-Goto action. More explicitly, parameterizing the
two-dimensional world sheet by the coordinates $\sigma^{\alpha}=(\tau,\sigma)$,
the location of the string world sheet in the five-dimensional spacetime with coordinates
$x^\mu$ is
\begin{equation}
    x^\mu = x^\mu(\tau,\sigma)\, ,%\qquad \sigma^{\alpha} = (\tau,\sigma)\, ,
    \label{3.8}
\end{equation}
and the Nambu-Goto action for the string world sheet is given
by
 \begin{equation}
S =- \frac{1 }{ 2 \pi \alpha'} \int d\sigma d \tau \, \sqrt{ - \det
g_{\alpha \beta}}\, . \label{3.9}
 \end{equation}
Here,
\begin{equation}
  g_{\alpha \beta} = G_{\mu \nu} \partial_\alpha x^\mu \partial_\beta x^\nu\,
  \label{3.10}
\end{equation}
is the induced metric on the world sheet and $G_{\mu \nu}$ is the metric of the
$4+1$-dimensional AdS$_5$ spacetime.
The action (\ref{3.9}) is invariant under coordinate changes of
$\sigma^\alpha$. This will allow us to
pick world sheet coordinates $(\tau,\sigma)$ differently
for convenience in different calculations. Upon denoting
the action of the surface which
is bounded by ${\cal C}$ and
extremizes the Nambu-Goto action (\ref{3.9})
by $S(\CC)$, the
expectation value of the Wilson loop (\ref{Wils}) is given by~\cite{Rey:1998ik}
 \begin{equation}
\langle{W({\cal C})}\rangle = \exp\left[ i\, \lbrace S ({\cal C})  - S_0 \rbrace \right] \, ,\label{3.7}
 \end{equation}
where the subtraction
$S_0$ is the action of two disjoint strings, as we shall discuss in detail below.

To evaluate the expectation value of a Wilson loop at nonzero temperature
in the gauge theory, one replaces
AdS$_5$ by
an AdS Schwarzschild black hole~\cite{Witten:1998zw}. The metric
of the AdS black hole background is given by
 \begin{eqnarray}
ds^2 & = &  - f dt^2 + \frac{r^2 }{ R^2} (dx_1^2 + dx_2^2 + dx_3^2)+
\frac{1 }{ f} dr^2 = G_{\mu \nu} dx^\mu dx^\nu \, ,
\label{3.3}\\
f &\equiv& \frac{r^2 }{ R^2} \left(1 - \frac{r_0^4 }{ r^4} \right)\, .
\label{3.4}
 \end{eqnarray}
Here, $r$ is the  coordinate of the 5th dimension and the black hole
horizon is at $r = r_0$.
According to the AdS/CFT correspondence, the
temperature in the gauge theory is equal to the Hawking temperature
in the AdS black hole, namely
\begin{equation}
 T = \frac{r_0 }{ \pi R^2}\, .
   \label{3.6}
 \end{equation}
The probe D3-brane at the boundary
of the AdS$_5$ space
lies at a fixed $r$ which we denote
$r = \Lambda\, r_0$.  $\Lambda$ can be considered
a dimensionless ultraviolet cutoff in the boundary
conformal field theory.
We shall call the three spatial directions
in which the D3-brane is extended
$x_1$, $ x_2$, and
$x_3$.
The fundamental ``quarks'', which are open strings ending on the
probe brane, have a mass proportional to $\Lam$.
In order to correctly describe a Wilson loop in the continuum
gauge theory, we must remove the ultraviolet cutoff by taking
the
$\Lambda\rightarrow\infty$
%\begin{equation}
%    \Lambda \rightarrow \infty\,
%    \label{3.5}
%\end{equation}
limit.
%However, we shall see below that in the analysis of a light-like Wilson loop
%this
%limit must be taken with some care, and we therefore keep $\Lambda$ finite
%for the present in our treatment, and take the limit (\ref{3.5})
%below only when it
%is safe to do so.

Now consider the set of rectangular Wilson loops shown in
Fig.~\ref{fig0}, with a short side of length $L$ in the
$x_1$-direction and a long side along a time-like direction in the
$t-x_3$ plane, which describe a quark-antiquark pair moving along the
$x_3$ direction with some velocity $v$.  Here, $v=0$ corresponds to
the loop ${\cal C}_{\rm static}$ in Fig.~\ref{fig0} whereas $0<v <1$ corresponds to
${\cal C}_{\rm static}^{\rm boosted}$ in the figure. To analyze these
loops, it is convenient to boost
the system to the rest frame $(t',x_3')$ of the quark pair
 \begin{eqnarray}
    dt &=& dt'\,  \cosh\eta - dx_3'\, \sinh\eta\, ,
    \label{3.11}\\
    dx_3 &=& - dt'\,  \sinh\eta + dx_3'\, \cosh\eta\, ,
    \label{3.12}
\end{eqnarray}
where the rapidity $\eta$ is given by $\tanh \eta = v$.   The loop is now static, but
the quark-gluon plasma is moving with velocity $v$ in the negative $x_3'$-direction.
This Wilson loop can be used to describe the potential between two heavy
quarks moving through the quark-gluon plasma or, equivalently, two heavy
quarks at rest in a moving quark-gluon plasma ``wind''.
In the primed coordinates, the long
sides of the Wilson loop lie along $t'$ at fixed $x_3'$.
We denote
their lengths by $\TT$, which is the proper time of the
quark-pair.\footnote{In terms of the
 time $t_{lab}$ in the rest frame of the medium, we have the standard relation
$ {\cal T} = \sqrt{1-v^2} \; t_{lab} = \frac{t_{lab} }{ \cosh\eta}$.}
 We assume that $\TT \gg L$, so that the string world sheet
attached to the Wilson loop along the contour ${\cal C}$
can be approximated as
time-translation invariant. Plugging
(\ref{3.11}) and (\ref{3.12}) into (\ref{3.3}) and dropping the
primes, we find
\begin{equation}
ds^2 = - A\, dt^2 - 2B\, dt\, dx_3 + C\, dx_3^2 + \frac{r^2}{R^2}
\left(dx_1^2 + dx_2^2 \right)+ \frac{1 }{ f} dr^2 \
 \label{3.13}
\end{equation}
with
 \begin{equation}
  \label{3.14}
A = {r^2 \over R^2} \left(1 - {r_1^4 \over r^4} \right), \qquad  B = {r_1^2
r_2^2 \over r^2 R^2} , \qquad C = {r^2 \over R^2} \left(1 + {r_2^4 \over
r^4} \right)\, ,
 \end{equation}
where
 \begin{equation}
  \label{3.15}
r_1^4 = r_0^4 \cosh^2 \eta, \qquad r_2^4 = r_0^4 \sinh^2 \eta\, .
 \end{equation}
To obtain the light-like Wilson loop along the contour ${\cal C}_{\rm light-like}$
in Fig.~\ref{fig0}, we must take the $\eta\rightarrow \infty$ limit.
We shall see that the $\eta\rightarrow \infty$ limit and the $\Lambda\rightarrow\infty$
limit do not commute. And, we shall discover that in order to have a sensible
phenomenology, we must reach the light-like Wilson loop by first taking
the light-like limit ($\eta\rightarrow\infty$) and only then taking the Wilson
loop limit ($\Lambda\rightarrow\infty$).  For the present, we keep both $\eta$
and $\Lambda$ finite.

We parameterize the two-dimensional world sheet (\ref{3.8}), using the coordinates
 \begin{equation}
    \tau = t, \qquad \sigma = x_1 \in [-{L \ov 2}, {L \ov 2}] \ .
    \label{3.16}
 \end{equation}
 By symmetry, we will take $x^\mu$ to be functions of $\sigma$ only and we set
 \begin{equation}
 \label{3.17}
    x_2 (\sigma) = {\rm const}\, , \qquad x_3 (\sigma) = {\rm const}\, ,
    \qquad r = r(\sigma)\, .
 \end{equation}
The Nambu-Goto action (\ref{3.9}) now reads
 \begin{eqnarray}
 \label{3.18}
 S & = & {{\cal T} \over 2 \pi \alpha'} \int_{-{L \ov 2}}^{{L \ov 2 }} d \sigma \,
\sqrt{A \left({(\partial_\sigma r)^2  \over f } + {r^2 \over
R^2}\right)} \ ,
 \end{eqnarray}
with the boundary condition $r (\pm {L \ov 2}) = r_0 \Lam$.
This boundary condition ensures that when the string world sheet ends on
the D3-brane located at $r=r_0 \Lambda$, it does so on
the contour ${\cal C}$ which is located  at $x_1=\pm {L\over 2}$.
Our task is to find $r(\sigma)$, the shape of the string world sheet
hanging ``downward in $r$'' from its endpoints at $r=r_0 \Lambda$,
by extremizing (\ref{3.18}).
Introducing dimensionless variables
 \be
 r= r_0 y, \qquad \tilde\sigma = \sigma{r_0\over R^2}, \qquad
 l = {L r_0\over R^2} = \pi L T, \label{dimensionless}
\ee
 where $T$ is the temperature (\ref{3.6}),
we find that, upon dropping the tilde,
\be \label{peo}
 S({\cal C})
= { \sqrt{ \lambda }{\cal T}  T} \int_0^{l \over 2} d
\sigma\; {\cal L}\,
 \ee
with ($y'= \p_\sig y$)
\begin{equation}
 {\cal L} = \sqrt{\left(y^4 - {\cosh^2 \eta } \right) \left(1 + {y'^2 \over y^4
 - 1} \right)}
 \label{3.20}
 \end{equation}
and the boundary condition $y \le(\pm {l \ov 2}\ri) = \Lam$.
In writing
(\ref{peo}) we have used the fact that, by symmetry, $y (\sig)$ is an even
function. It is manifest from (\ref{peo}) that all
physical quantities only depend on $T$ and not on $R$ or $r_0$
separately.
We must now determine $y(\sigma)$ by extremizing (\ref{3.20}).  This can
be thought of as a classical mechanics problem, with $\sigma$ the analogue
of time.
Since $\LL$ does not depend on $\sig$ explicitly, the
corresponding Hamiltonian
\begin{equation}
    {\cal H} \equiv {\cal L} - y' \frac{\partial {\cal L}}{\partial y'}
    = {y^4-\cosh^2 \eta \over {\cal L}} = {\rm const}
    \label{3.21}
\end{equation}
is a constant of the motion in the classical mechanics problem.

It is worth pausing to recall how it is that the calculation of a Wilson loop
in a strongly interacting gauge theory has been simplified to a classical mechanics
problem.  The large-$N$ and large $\lambda$ limits are both crucial.  Taking
$N\to\infty$ at fixed $\lambda$ corresponds to taking the string coupling to zero,
meaning that we can ignore the possibility of loops of string breaking off from
the string world sheet.  Then, when we furthermore take $\lambda\to\infty$, we
are sending the string tension to infinity meaning that we can neglect fluctuations
of the string world sheet.  Thus, the string world sheet ``hanging down'' from the
contour ${\cal C}$ takes on its classical configuration, without fluctuating or splitting
off loops.  If the contour ${\cal C}$ is a rectangle with two long sides, meaning that
its ends are negligible compared to its middle, then
finding this classical configuration is a classical mechanics problem no
more difficult than finding the catenary curve describing a chain suspended
from two points hanging in a gravitational field, in this case the gravitational
field of the AdS Schwarzschild black hole.

Let us now consider keeping $\Lambda$ fixed and $\gg 1$, while increasing
$\eta$ from $0$ to $\infty$.  We see
that the quantity inside the square root in
(\ref{3.20}) changes sign when $y$ crosses $\sqrt{\cosh \eta}$.
The string world sheet is time-like for real $\LL$ (i.e. for $y >
\sqrt{\cosh \eta}$) and  is space-like for imaginary $\LL$ (i.e.
for $y < \sqrt{\cosh \eta}$). Since $y = \Lambda$ at the boundary
${\cal C}$, the signature of the world sheet depends on the
relative magnitude of $\sqrt{\cosh \eta}$ and $\Lam$: it is
time-like when $\sqrt{\cosh \eta} < \Lambda$ and becomes space-like
when $\sqrt{\cosh \eta} > \Lambda$.
If the
world sheet in (\ref{3.9}) is time-like (space-like), the
expectation value (\ref{3.7}) of the fundamental Wilson loop is the
exponent of an imaginary (real) quantity.
We shall give a physical interpretation of this behavior in Section 3.3.
Here, we explain that
this behavior is consistent with all the phenomenology described
in Section~2.
For $\eta = 0$, the Wilson loop defines the static
quark-antiquark potential, see (\ref{2.2}), and thus should
and does correspond to a time-like world sheet.
If the quark-pair is not at
rest with respect to the medium, but moves with a small velocity
$v = \tanh \eta$, one still expects that the quark-pair remains
bound and the world-sheet action remains time-like. We shall see, however,
that for  large enough
$\eta$ a bound quark-antiquark state cannot exist.
Once we reach $\eta\equiv \infty$, namely the light-like Wilson loop which
we saw in Section~2 originates from eikonal
propagation in high energy scattering and
is relevant to deep inelastic scattering, the Cronin effect,
and jet quenching, in order to have a sensible description of
these phenomena we see
from (\ref{2.14a}) or equivalently (\ref{2.6}) that the expectation
value of the Wilson loop must be the exponent of a real quantity.
This expectation is met by (\ref{3.20}) since the
string  world sheet is space-like as long as $\sqrt{\cosh \eta} > \Lambda$.
This demonstrates that in order to
sensibly describe any of the applications of Wilson loops
to high energy propagation, including in particular in our nonzero temperature
context the calculation of the jet quenching parameter $\hat q$, we must
take $\eta\to\infty$ first, before taking the $\Lambda\to\infty$ limit.

In subsection~3.1, we shall review the calculation of the quark-antiquark potential
and screening length as a function of the velocity $v$.  In subsection~3.2, we
calculate the jet quenching parameter. And, in subsection~3.3,  we return to
the distinction between the time-like string world sheet of subsection~3.1 and
the space-like string world sheet of  subsection~3.2, and give a physical
interpretation of this discontinuity.

\subsection{Velocity-dependent quark-antiquark potential and screening length}
\label{sec3c}
In this subsection we compute the expectation values of Wilson loops for
$\sqrt{\cosh \eta} < \Lambda $, from which we extract the
velocity-dependent quark-antiquark potential and screening length.
At the end of the calculation we take the heavy quark limit $\Lam \to \infty$.
In fact, because we are interested in the
case $\sqrt{\cosh \eta} < \Lambda$, in this subsection we could
safely take $\Lambda\to\infty$
from the beginning. The results reviewed in this subsection were obtained
in Refs.~\cite{Liu:2006nn,Chernicoff:2006hi,Avramis:2006em}.

We denote the constant of the motion identified in Eq, (\ref{3.21}) by $q$,
and rewrite this equation as
 \begin{equation}
 \label{3.34}
y' = {1 \over q} \sqrt{(y^4- 1) (y^4 - y_c^4)}
 \end{equation}
with
 \begin{equation}
  \label{3.35}
y_c^4 \equiv \cosh^2 \eta + q^2.
 \end{equation}
Note that $y_c^4 > \cosh^2 \eta\ge 1$.  The extremal string world sheet
begins at $\sigma=-\ell/2$ where $y=\Lambda$, and ``descends'' in $y$ until
it reaches a turning point, namely the largest value of $y$
at which $y'=0$.  It then ``ascends'' from the turning point to its
end point at $\sigma=+\ell/2$ where $y=\Lambda$.  By symmetry,
the turning point must occur at $\sigma=0$.  We see from (\ref{3.34})
that in this case, the turning point occurs at $y=y_c$ meaning that the
extremal surface stretches between $y_c$ and $\Lambda$.
The integration constant
$q$ can then be determined\footnote{For equation
(\ref{3.34}) to be well defined, we need $0 < q^4 < \Lambda^4 -
\cosh^2 \eta$.} from the equation ${l \over 2} = \int^{l \over
2}_0 d \sigma$ which, upon using (\ref{3.34}), becomes
 \begin{equation}
 \label{3.36}
{l } =%2 q^2 \int_{r_c}^\Lam dr \, {1 \ov \sqrt{(r^4- r_0^4) (r^4 -
%r_c^4)} } =
 {2 q} \int_{y_c}^{\Lambda} dy \, {1 \over
\sqrt{(y^4- y_c^4) (y^4 - 1)} }\ .
 \end{equation}
The action for the extremal surface can be found by substituting
(\ref{3.34}) into (\ref{peo}) and (\ref{3.20}),
 \begin{equation}
 \label{3.37}
S(l) ={ \sqrt{ \lambda }{\cal T}  T}   \int_{y_c}^\Lambda dy \; {
y^4 - \cosh^2\eta \over \sqrt{(y^4 -1)(y^4-y_c^4)}} \ .
 \end{equation}
Equation (\ref{3.37}) contains not only the potential between
the quark-antiquark pair but also the static mass of the quark and antiquark
considered separately
in the moving medium. (Recall that we have boosted to the rest frame of
the quark and antiquark, meaning that the quark-gluon plasma is
moving.)
Since we are only
interested in the quark-antiquark potential, we need to subtract from (\ref{3.37})
the action $S_0$ of two independent quarks, namely
 \begin{equation}
    E(L) \TT = S(l) - S_0\, ,
    \label{3.38}
\end{equation}
where $E (L)$ is the quark-antiquark potential in the dipole rest
frame. The string configuration corresponding to a single quark
at rest in a moving
hot medium in $\NN=4$ SYM was found in
Refs.~\cite{Herzog:2006gh,Gubser:2006bz}, from which one finds that
  \be
 S_0 = \sqrt{\lambda}\,  {\cal T}\, T\,  \int_{1}^{\Lam} dy \, .
 \label{3.39}
 \ee
To be self-contained, in appendix~\ref{appa} we review the
solution of~\cite{Herzog:2006gh,Gubser:2006bz},
along with a family of new drag solutions
describing string configurations corresponding to mesons made from
a heavy and a light quark.

 %%%%%%%%%%%%%%%%%%%%%%%%%%%%%%%%%%%%%%%%%%%%%%
%\begin{figure}[t]
\FIGURE[t]{
%\vskip-0.15in
% \hfill\hskip-0.6in
\includegraphics[scale=0.7,angle=0]{fig-eta-dep.eps}
%\hskip-0.1in
\hfill
\caption{Left panel: the quark-antiquark separation $l(q)$ as a function of the integration
constant $q$ for a quark-antiquark dipole oriented orthogonal to the wind propagating
at different
velocities $v=\tanh \eta$.  We discuss the case where the dipole is not orthogonal to
the wind in Section 5.
Right panel: The $q\bar{q}$ static potential for the same
quark-antiquark configurations as in the left panel.
Note that the
potential is normalized such that the small-distance behavior of the potential
is unaffected by velocity-dependent medium effects.
}\label{fig5}
}%\end{figure}
%%%%%%%%%%%%%%%%%%%%%%%%%%%%%%%%%%%%%%%%%%%%%%%%%
%

To extract the quark-antiquark potential, we use (\ref{3.36}) to
solve for $q$ in terms of $l$ and then plug the corresponding $q(l)$
into (\ref{3.37}) and (\ref{3.38}) to obtain $E(L)$. We can safely take
the $\Lam\rightarrow\infty$ limit, and do so in all results we present.
We show results at a selection of velocities in Fig.~\ref{fig5}.
In the remainder of this subsection, we
describe general features of these results.

First, Eq. (\ref{3.37}) has no solution when
$l > l_{\rm max} (\eta)$, where $l_{\rm max}$ is the maximum of
$l(q)$.  We see that $l_{\rm max}$ decreases with increasing velocity.

We see from the left panel in Fig.~\ref{fig5} that
for a given $l < l_{\rm max} (\eta)$, there are two branches of
solutions. The branch with the bigger value of $q$, and
therefore the larger  turning point $y_c$, has the
smaller $E(L)$ --- corresponding to the lower branches
of each of the curves in the right panel of the figure.
The upper branches of each curve correspond to the solutions
for a given $l<l_{\rm max}$ with smaller $q$ and $y_c$.
Because they have higher energy, it is natural to
expect that they describe unstable
solutions sitting at a saddle point in configuration space~\cite{Chernicoff:2006hi,Argyres:2006vs}.
This has been
confirmed explicitly in Ref.~\cite{Friess:2006rk}.

When $\eta$ is greater than some critical value $\eta_c$,
$E(L)$ is negative for the whole upper branch.
When $\eta < \eta_c$, there exists a value $l_c (\eta) < l_{max}$ such that
the upper branch has an $E(L)$ which is negative for $l<l_c$ and positive
for $l>l_c$.    $l_c$ goes to
zero as $\eta$ goes to zero.
If $\eta<\eta_c$ and $l>l_c$, then if the
unstable upper branch configuration is perturbed,
after some time it could settle down either to the lower
branch solution or to two isolated strings each described
by the drag solution of Ref.~\cite{Herzog:2006gh,Gubser:2006bz}
and Appendix A.  (Note that $E>0$ means
that a configuration has more energy than two isolated strings.)
On the other hand, if $E(L)$ is negative for the upper branch,
when this unstable configuration
is perturbed, the only static solution we know of
to which it can settle after some time is the lower
branch solution.

We see from Fig.~\ref{fig5} that using the action of the dragging string solution
of Refs.~\cite{Herzog:2006gh,Gubser:2006bz} as $S_0$ as we do
and as was considered as an option in Ref.~\cite{Avramis:2006em}, ensures that the
small-distance behavior of the potential is velocity-independent.
This seems to us a physically reasonable subtraction condition;
it is analogous to the
renormalization criterion used to
define the quark-antiquark potential in lattice calculations, namely that
at short distances it must be medium-independent~\cite{Kaczmarek:2002mc}.
Choosing the velocity-dependent subtraction (\ref{aa.12}) instead,
considered as an option in Ref.~\cite{Avramis:2006em}, makes the unstable upper branch
have $\lim_{L\to 0}E(L)=0$ for all velocities, but in so doing
makes the stable lower-branch have a velocity-dependent
$E(L)$ at all $L$, including small $L$.

One can obtain an analytical expression for $l_{\rm max}$ in
the limit of high velocity. Expanding (\ref{3.36}) in powers of
$1/y_c^4$ gives
\be
    l(q) = \frac{2\sqrt{\pi} q}{y_c^3}
        \left( \frac{\Gamma\left(\frac{3}{4}\right)}{\Gamma\left(\frac{1}{4}\right)}
        + \frac{\Gamma\left(\frac{7}{4}\right)}{8\Gamma\left(\frac{9}{4}\right)}
            \frac{1}{y_c^4}
         + \frac{3\Gamma\left(\frac{11}{4}\right)}{32\Gamma\left(\frac{13}{4}\right)}
            \frac{1}{y_c^8} + O\left(\frac{1}{y_c^{12}}\right)
        \right)\ .
        \label{3.40}
\ee
Truncating this expression after the second term, we find for the
maximum
\bea
    l_{\rm max} &=& \frac{\sqrt{2\pi}\, \Gamma\left(\frac{3}{4}\right)}{
                             3^{3/4}\Gamma\left(\frac{1}{4}\right)}
                             \left( \frac{2}{\cosh^{1/2}\eta} + \frac{1}{5\cosh^{5/2}\eta}
                             + \cdots \right)
                             \nonumber \\
                          &=& 0.74333 \left( \frac{1}{\cosh^{1/2}\eta} + \frac{1}{10\, \cosh^{5/2}\eta}
                          +  \cdots \right)\ .
                          \label{3.41}
\eea

\noindent  Note that $L_{max} = {l_{max} \ov \pi T}$ can be
interpreted as the screening length in the medium, beyond
which the only solution is the trivial solution corresponding to
two disjoint world sheets and thus $E (L) =0$. The first term of
this expression was given in~\cite{Liu:2006nn} (see
also~\cite{Chernicoff:2006hi}), the second term
in~\cite{Avramis:2006em}.  As we shall discuss further at the end of
Section 5, if we set $\eta=0$ in (\ref{3.41}), this expression which
was derived for $\eta\rightarrow\infty$ is not too far off the $\eta=0$ result,
which is $\ell_{\rm max}=0.869$.  Hence, as discovered in Ref.~\cite{Liu:2006nn},
the screening length decreases with increasing velocity to a good approximation
according to the scaling
\begin{equation}
L_{\rm max}(v) \simeq \frac{L_{\rm max}(0)}{\cosh^{1/2}\eta} =
\frac{L_{\rm max}(0)}{\sqrt{\gamma}} ,
\label{velocityscaling}
\end{equation}
with $\gamma=1/\sqrt{1-v^2}$.
This velocity dependence suggests that
$L_s$ should be thought of
as, to a good approximation,  proportional to
(energy density)$^{-1/4}$, since the energy density increases like $\gamma^2$
as the wind  velocity is boosted.

If the velocity-scaling of $L_s$ that we have discovered holds for
QCD, it will have qualitative consequences for
quarkonium suppression in heavy ion collisions~\cite{Liu:2006nn}.
For illustrative
purposes, consider the explanation of the  $J/\Psi$ suppression
seen at SPS and RHIC energies proposed in Refs.~\cite{Karsch:2005nk,Satz}: lattice
calculations
%~\cite{Kaczmarek:2004gv,Kaczmarek:2005ui}
of the $q\bar q$-potential
indicate that the $J/\Psi$(1S) state dissociates at a temperature
$\sim 2.1 T_c$ whereas the excited  $\chi_c$(2P) and
$\Psi'$(2S) states cannot survive above $\sim 1.2 T_c$; so, if
collisions at both
the SPS and RHIC reach temperatures above $1.2 T_c$ but not above
$2.1 T_c$, the experimental facts (comparable anomalous
suppression of
$J/\Psi$ production at the SPS and RHIC) can be understood
as the complete loss  of the ``secondary'' $J/\Psi$'s that
would have arisen from the decays of the excited states,
with no suppression
at all of  $J/\Psi$'s that originate as $J/\Psi$'s.
Taking Eq.~(\ref{velocityscaling}) at face value, the
temperature  $T_{\rm diss}$ needed to
dissociate the $J/\Psi$ decreases $\propto (1-v^2)^{1/4}$.
This indicates that $J/\Psi$ suppression at RHIC may increase
markedly (as the $J/\Psi$(1S) mesons themselves dissociate)
for $J/\Psi$'s with transverse momentum $p_T$ above some threshold
that is at most $\sim 9$~GeV
and would be $\sim 5$ GeV if the temperatures reached at RHIC are
$\sim 1.5 T_c$. The kinematical range in which this  novel quarkonium
suppression mechanism is operational lies within experimental reach of
future high-luminosity runs at RHIC and will be studied thoroughly at
the LHC in both the $J/\Psi$ and Upsilon channels.
If the temperature of the medium produced in LHC collisions
proves to be large enough that the $J/\Psi$(1S) mesons dissociate
already at low $p_T$, the $p_T$-dependent pattern that the
velocity scaling (\ref{velocityscaling}) predicts in the $J/\Psi$
channel at RHIC should be visible in the Upsilon channel at the LHC.

As a caveat, we add that in
modelling quarkonium
production and suppression versus $p_T$
in heavy ion collisions,
various other effects remain to
be quantified. For instance, secondary production mechanisms
such as recombination may contribute significantly to the
$J/\Psi$ yield at low $p_T$, although the understanding of
such contributions is currently model-dependent.
Also, at very high $p_T$, $J/\Psi$ mesons could form
outside the hot medium~\cite{Karsch:1987zw}. Parametric estimates
of this effect suggest that it is important only at much higher
$p_T$ than is of interest to us, and we are not aware of model
studies which have been done that would allow one to go
beyond parametric estimates.
The quantitative importance
of these and other effects may vary significantly, depending on
details of their
model implementation. In contrast, Eq. (\ref{velocityscaling}) was
obtained directly from a field-theoretic calculation
and its implementation
will not introduce
additional model-dependent uncertainties. For this reason, the velocity
scaling established here must be included in all future model
calculations.
We expect that its effect is most prominent at intermediate
transverse momentum,
where contributions from secondary production die out or can be
controlled,
while the formation time of the heavy bound states is still short
enough to ensure that they
would be produced within the medium if the screening by the medium
permits.

%%%%%%%%%%%%%%%%%%%%%%%%%%%%%%%%%%%%%%%%%%%
\subsection{Light-like Wilson loop and the jet quenching parameter}
\label{sec3b}

In order to calculate the jet quenching parameter
we need to take the $\eta\to\infty$ limit in which the Wilson loop
becomes light-like first, with the location of the
boundary D3-brane $\Lambda$ large
and fixed, and only later take $\Lambda\to\infty$.  As we approach
the light-like limit,
it is necessary that
$\sqrt{\cosh \eta} > \Lam$. In this regime, as we
discussed below equation (\ref{3.21}), the world sheet is
space-like, meaning that the expectation value of the Wilson loop is
the exponential of a real quantity.  As we reviewed in Section~2, this
must be the case in order to obtain sensible results for both
medium-induced gluon radiation of Eq. (\ref{2.8}) and the
virtual photo-absorption cross section in deep inleastic scattering
of Eq.~(\ref{2.4}).

When $\sqrt{\cosh \eta} > \Lam$, the first order equation of
motion, given by (\ref{3.21}), reads
 \be
y'^2 = {1 \ov q^2} (y^4- 1) (y_m^4 - y^4) \label{3.22}
 \ee
with
 \be
y_m^4 = \cosh^2 \eta - q^2\ . \label{3.23}
 \ee
The consistency of (\ref{3.22}) requires that $y_m  > \Lambda$,
which implies that the integration constant $q$ is constrained to
 $0 \leq q^2 \leq \cosh^2 \eta- \Lam^4$. Equation (\ref{3.22}) has a trivial
 solution
 \be \label{trisol}
  y (\sig) =\Lambda = {\rm const}, \qquad q^2 = \cosh^2 \eta-
  \Lam^4 \ .
 \ee
 However, one can check that (\ref{trisol}) does not solve the second order Euler-Lagrange
 equation of motion derived from (\ref{3.20}) and thus should be discarded.
Because $y_m>\Lambda$,
 the nontrivial solution of
(\ref{3.22}) which descends from $y=\Lambda$ at $\sigma=-l/2$
descends all the way to $y=1$,
where $y'=0$.
Thus, for any value of $l$ the string
starts at $y= \Lam$ and descends all the way to the horizon, where
it turns around  and then ascends  back up to $y=\Lam$.
The integration constant
$q$ can be
determined from the equation ${l \over 2} = \int^{l \over 2}_0 d \sigma$, i.e
  \be
{l } =
 {2 q } \int_{1}^{\Lam} dy \, {1 \ov
\sqrt{(y^4_m- y^4) (y^4 - 1)} }\,  \label{3.25}
 \ee
upon using (\ref{3.22}). The action (\ref{3.18}) takes the form
 \be
S(l) =i { \sqrt{ \lambda }{\cal T}  T}  \int_{1}^\Lam dy \; {
\cosh^2\eta - y^4 \ov \sqrt{(y^4 -1)(y_m^4-y^4)}}\, . \label{3.24}
 \ee
This action is imaginary and corresponds to a space-like world
sheet.

To extract $\hat q$ introduced in (\ref{2.6}),  (\ref{2.11})
and (\ref{2.14a}), we first take $\eta\to\infty$, making the contour ${\cal C}$
light-like,
and only then
take the $\Lambda\to\infty$ limit needed to ensure that we are evaluating
$W({\cal C})$, with the end of the string on the D3-brane at $y=\Lambda$
following the contour ${\cal C}$ precisely.
$\hat q$ can be obtained by studying the
small $l$-dependence of the action~(\ref{3.24}), which can be done
analytically. We start
 from the expansion of (\ref{3.25}),
 \begin{eqnarray}
    {l }
%   &=& {2 q^2 } \int_{1}^{\Lam} dy \, {1 \ov
%       \sqrt{(\cosh^2\eta- y^4) (y^4 - 1)} } + O(q^6)
%      \nonumber \\
    &=& \frac{2q}{\cosh\eta} \int_1^\Lambda dy \frac{1}{\sqrt{y^4-1}}
      + {\cal O}\left(q^3,\frac{\Lambda^4}{\cosh^2\eta}\right)\, .
      \label{3.26}
 \end{eqnarray}
Upon defining
 \begin{equation}
   \alpha\equiv \lim_{\Lambda \to \infty} \int_1^\Lambda dy \frac{1}{\sqrt{y^4-1}}
        = \sqrt{\pi} \frac{\Gamma\left(\frac{5}{4}\right)}{\Gamma\left(\frac{3}{4}\right)}
       \, ,
        \label{3.27}
 \end{equation}
we find that in the small $l$ (equivalently, small $q$) limit
\begin{equation}
l = \frac{2\alpha q}{\cosh\eta}\ .
\label{3.27a}
\end{equation}
In the same limit, the action (\ref{3.24}) takes  the form
\begin{equation}
    S(l) = S^{(0)} + q^2S^{(1)} + O(q^4)\, ,
    \label{3.28}
\end{equation}
where
\begin{eqnarray}
    S^{(0)} &=&  i \sqrt{ \lambda }{\cal T}  T \int_{1}^\Lam dy \, \sqrt{\cosh^2 \eta - y^4 \ov y^4 -1}\, ,
    \label{3.29}\\
    q^2 S^{(1)}(l) &=&     \frac{ i \sqrt{ \lambda }{\cal T}  T}{2} q^2 \int_{1}^\Lam dy \,
     \frac{1}{\sqrt{\left( \cosh^2 \eta - y^4\right) \left( y^4 -1\right)}}       \nonumber \\
%       &=& \frac{ i K}{4} q^2\, l
%       \nonumber \\
       &=& \frac{ i \sqrt{ \lambda }{\cal T}  T q^2 \alpha}{2 \cosh\eta}
        =   i \frac{\sqrt{\lambda} \pi^2\, T^3}{8\, \alpha} \left( {\cal T} \cosh \eta\right) L^2\, ,
        \label{3.30}
\end{eqnarray}
where we have used (\ref{3.27}), (\ref{3.27a}) and
$l=\pi\, L\, T$. Also, we have kept the dominant
large $\eta$-dependence only.
We identify $\left( {\cal T} \cosh \eta\right) = L^-/\sqrt{2}$, where $L^-$ is the extension of the
Wilson loop in the light-like direction, entering in (\ref{2.11})
and (\ref{2.14a}).

%The expression for $S(l)$ that we have arrived at is perhaps more easily derived
%by working from start to finish with a light-like Wilson loop described
%in the rest frame of the medium using
%light-cone coordinates, as in Ref.~\cite{Liu:2006ug}.  The present derivation
%has the virtue of showing how the result arises as one reaches the
%light-like Wilson loop by taking the $\eta\rightarrow\infty$ limit
%of a time-like Wilson loop.

As in Section 3.1, in order to determine the expectation of the Wilson line we need
to subtract the action of two independent single quarks, this time moving at the speed of light.
 In Appendix~\ref{appa}, we analyze the string configuration
 corresponding a single quark moving at the speed of light.
 There we find a class of solutions with space-like
 world sheets and also a class of solutions with time-like world
 sheet. Our criterion to determine which solution to subtract
 is motivated from the physical expectation discussed in Section  2, i.e.
 \be
 \lim_{l\to 0}  \left[ S (l) - S_0\right] = S^{(0)} - S_0 = 0\, .
 \label{3.32}
 \ee
Among the classes of solutions discussed in Appendix A,
the only one satisfying (\ref{3.32}) is the space-like world sheet
described by Eqs. (\ref{aa.17}) and (\ref{aa.18}) with $p=0$.
In this configuration, $S_0$ is the action of
two straight strings extending from $y = \Lam$ to
$y=1$ along the radial direction and is given by
 \be \label{3.31}
S_0
 = i \sqrt{ \lambda }{\cal T}  T \int_{1}^\Lam dy \, \sqrt{\cosh^2 \eta - y^4 \ov y^4  -1}\, .
 \ee

The $L^2$-term in the exponent of (\ref{2.14a}) can then be identified
with the $O(L^2)$-term (\ref{3.30}) of the action $S(l)$, and we thus
conclude
that the jet quenching
parameter in (\ref{2.14a}) is given by
\be
 \hat{q}_{SYM} = \frac{\pi^{3/2} \Gamma\left(\frac{3}{4}\right)}{ \Gamma\left(\frac{5}{4}\right)}
    \sqrt{\lambda}\, T^3\, .
% \hat{q}_{SYM} = \frac{\pi^{3/2} \Gamma\left(\frac{3}{4}\right)}{\sqrt{2} \Gamma\left(\frac{5}{4}\right)}
%   \sqrt{\lambda}\, T^3\, .
    \label{3.33}
 \ee
 We have used the fact that, as in (\ref{2.9}), in the large-$N$ limit the
 expectation value of the adjoint Wilson loop
 which defines $\hat q$ in (\ref{2.14a}) differs from that of the fundamental
Wilson loop which we have calculated by a factor of 2 in the exponent $S$.

 In Ref.~\cite{Liu:2006ug}, the result (\ref{3.33}) was obtained
starting directly from the
 loop
 ${\cal C}_{\rm light-like}$, described in the rest frame of
the medium using light-cone coordinates. Here, we
 showed that one can obtain the same result by taking the $v \to 1$
 limit of a time-like Wilson loop. It is also easy to check that
 the trivial solution (\ref{trisol}) goes over to the constant
 solution discussed in~\cite{Liu:2006ug}, which has a
 smaller action than (\ref{3.24}). In~\cite{Liu:2006ug} this trivial solution
 was discarded on physical grounds. Here, we see that if we
 treat the light-like Wilson line as the $\eta \to \infty$ limit of a
 time-like one, this trivial solution does not even arise.
We also note that in the light-like limit, the coefficient in front of
the scalar field term in (\ref{Wils}) goes to zero and
(\ref{Wils}) coincides with (\ref{2.1}).

In Section 4 we shall determine how $\hat q$ changes if the medium in which
the expectation value of the light-like Wilson loop is evaluated has some
flow velocity at an arbitrary
angle with respect to the direction of the Wilson loop.  In Section 6 we shall
discuss the comparison between our result for $\hat q$ and that extracted
by comparison with RHIC data, as well as discuss how $\hat q$ changes
with the number of degrees of freedom in the theory.

\subsection{Discussion: time-like versus space-like world sheets \label{sec3.3}}

We have seen that as we increase $\eta$ from $0$ to $\infty$
while keeping $\Lam$ fixed and large,
the behavior of the string world sheet has a discontinuity
at $\sqrt{\cosh \eta} = \Lam$, below (above) which the world sheet
is time-like (space-like). Here we give a physical interpretation
for this discontinuity.  Recall first from Section 3.1 that if $\cosh\eta\gg 1$
but $\sqrt{\cosh\eta} < \Lam$, the screening length
$L_{\rm max}$ is given by
 \be \label{sche}
 L_{\rm max} ={0.743 \ov \pi \sqrt{\cosh{\eta}} \, T} \ .
 \ee
Next, note that
the size $\delta$ of our external quark on the D3-brane at $y=\Lambda$, i.e.
at $r=r_0\Lambda =\pi R^2 T \Lambda$
can be estimated
using the standard IR/UV connection, namely~\cite{IRUV}
 \be \label{comS}
 \delta \sim {\sqrt{\lam} \ov M} \sim {1 \ov \Lam \, T}\ ,
 \ee
 where $M = \ha \sqrt{\lam} T \Lam$ is the mass
  of an external quark
 as can be read from (\ref{3.39}).
 (The apparent $T$-dependence
 of (\ref{comS}) is due to our definition of $\Lam$, with
the ultraviolet cutoff given by $\Lam r_0$,
and does not reflect genuine temperature-dependence.)
Thus the condition $\sqrt{\cosh \eta} =
 \Lam$ corresponds to~\cite{Chernicoff:2006hi}
 \be \label{ncon}
 L_{max} \sim \delta \ .
 \ee
When $\sqrt{\cosh\eta}\ll\Lam$, meaning that
$\delta \ll L_{\rm max}$, we expect that if instead of merely analyzing
Wilson loops we were to actually study mesons, we would in fact
find a bound state of a quark and anti-quark.
In this regime, it is reasonable to expect that
the expectation value of the Wilson loop should yield
information about the quark-antiquark potential,
meaning that it must be the exponential of an imaginary quantity
meaning that the string world sheet must be time-like, as indeed we find.
On the other hand, when $\sqrt{\cosh\eta}\gg\Lam$, meaning
that $\delta \gg L_{\rm max}$,
the size of one quark by itself is much greater than the putative screening length.
This means that the quark and antiquark cannot bind for any $L$, meaning
that the transition at $\sqrt{\cosh\eta} \sim \Lam \sim M/(\sqrt{\lambda}T)$
can be thought of as a ``deconfinement'' or ``dissociation'' transition
for quarkonium mesons made from quarks with mass $M$.
Furthermore, in a regime in which the size of one quark is greater than
the putative screening length, the concept of a quark-antiquark potential (and
a screening length) makes no sense.  Instead, in this regime it is appropriate
to think of the quark-antiquark pair as a component of the wave function
of a virtual photon in deep inelastic scattering, and hence to think of
the Wilson loop as arising in
the eikonal approximation to this high energy scattering process, as
discussed in Section 2. From our discussion there,
it is then natural to expect a space-like world sheet, which gives
the desired $\langle W \rangle \sim \exp[-S]$ behavior with $S$
real.

Our discussion explains the qualitative change in physics, but it
does not explain
the sharpness of the discontinuity
that we find at
$\sqrt{\cosh\eta}=\Lam$,
which likely has to do with the classical
string approximation~(which corresponds to large $N$ and large
$\lam$ limit) we are using.
When $\sqrt{\cosh\eta}<\Lam$ there is a discontinuity
between $L<L_{\rm max}$ and $L>L_{\rm max}$
where the quark-antiquark potential goes from being nonzero to zero.
This discontinuity is smoothed out by finite $\lambda$ corrections, with the
exponentially small quark-antiquark potential at large distances corresponding
to physics that is nonperturbative in $\alpha'$.    Presumably the discontinuity
at $\sqrt{\cosh\eta}=\Lam$ is also smoothed out at finite $\lambda$ and $N$.
Further insight into this question could perhaps be obtained without relaxing the large-$N$
and large-$\lambda$ limits by studying mesons rather than Wilson loops.

The operational consequences of the discontinuity at $\sqrt{\cosh\eta}=\Lam$ are
clear.
To compute the quark-antiquark potential and the
screening length in a moving medium, we take $\Lam$ to infinity at fixed $\eta$.
To compute $\hat q$, we must instead {\it first} take the
$\eta \rightarrow \infty$ limit at finite $\Lambda$, and only then
take $\Lambda\rightarrow \infty$. The two limits do not commute.

%%%%%%%%%%%%%%%%%%%%%%%%%%%%%%%%%%%%%%%%%%%

%%%%%%%%%%%%%%%%%%%%%%%%%%%%%%%%%%%%%%%%%%%%%%%%%%%%%%%%%%%%%%%%%%%%%%%%%%%%%%%%
\section{The jet quenching parameter in a flowing medium \label{sec:tran}}
In section~\ref{sec3b}, we have evaluated the expectation value of
a light-like Wilson loop specified by the trajectory of a dipole moving
in a light-like
direction, $k^\mu = (1,\hat{n})$ with $\hat n$ a unit vector and hence
$k^2 = 0$.
%In the ordered
%limit $\eta \to \infty$, $\Lambda \to \infty$, t
The world sheet
defined by this light-like Wilson loop is space-like and the
behavior of the Nambu-Goto action in the limit of small dipole size
determines the jet quenching
parameter $\hat{q}$.
% Since the two long lines of a Wilson loop entering quenching calculations arise from
% the $q\to q\, g$ amplitude and its complex conjugate, respectively, the orientation of the
% dipole with respect to its light-like direction of propagation is of no physical relevance.
% This is in difference to calculations of the $q\bar{q}$ static potential.
It has been argued previously that the motion of the medium orthogonal to the trajectory of the dipole
can affect the value of $\hat q$ in a nontrivial
fashion~\cite{Armesto:2004vz,Renk:2005ta}.
Furthermore, if the medium is flowing parallel to or antiparallel to
the trajectory of the dipole with velocity $v_f = \tanh \eta_f$,
there is a straightforward effect on $\hat q$: the calculation goes through unchanged,
with $L^-$
understood to be the light-cone distance in the rest frame of the medium, but
the relation between $L^-$ and the distance $\Delta z$ travelled in the lab frame
is modified: $\Delta z =  (L^-/\sqrt{2}) \exp(\eta_f)$, where the sign convention is
such that $\eta_f>0$ corresponds to the dipole velocity and
flow velocity parallel (i.e. the dipole feels a ``tail wind'')
while $\eta_f<0$ means that the dipole feels a head wind.
Correspondingly, $\hat q$ is multiplied by a factor of $\exp(-\eta_f)$, meaning
that it increases in a head wind and decreases in a tail wind.
In this Section, we
calculate how the jet quenching parameter $\hat q$ depends on
the speed and direction of the collective flow of the medium, allowing for
any angle between the jet direction and the flow direction.

The calculation of the effect on
jet quenching parameter $\hat q$ due to the collective motion of the
medium turns out to be  straightforward, once the geometry of the problem
is set up.  We shall specify the light-like four-momentum
$k^\mu = (1,\hat{n})$ (the direction of motion of the hard parton which
is losing energy; the direction of propagation of the dipole
moving at the speed of light which defines the Wilson loop)
by taking $\hat{n}$ to point along the negative
$x_3$-direction. According to the way the BDMPS energy loss
calculation is set up, the dipole is always perpendicular to the direction
of its motion, so we choose the dipole orientation to point in a direction $\hat m$
which must lie in the $(x_1,x_2)$-plane.
Now, we set the medium in motion.
The most general ``wind velocity'' has components parallel to and orthogonal
to the dipole direction $\hat n$.  We choose $\vec v = v \hat l$
to lie in the $(x_2,x_3)$-plane.  Because we fix the orthogonal component
to lie along the $x_2$-direction, we must leave the direction of the dipole orientation $\hat m$
in the $(x_1,x_2)$-plane unspecified.
Thus the most general
configuration is described by four parameters, the transverse
separation $L$ of the Wilson loop in the lab frame and
\be
    \cosh\eta_f = {1 \ov \sqrt{1-v^2}}\, ,\qquad
    \theta = \angle (\hat{l},\hat{n}) \, ,\qquad
    \phi = \angle (\hat{m},\vec{x}_1)\, .
    \label{5.1}
\ee
In the lab frame, the trajectory of the end points of the dipole
can be written as
 \be
 A^\mu_\pm= k^\mu t \pm {L \ov 2} m^\mu \, ,
% = \left(t,\pm\frac{L}{2}\cos\phi,\pm\frac{L}{2}\sin\phi,-t\right)\, ,
 \label{5.2}
 \ee
 where
 \be
  \qquad k^\mu = (1, \hat n), \qquad
 m^\mu = (0, \hat m) , \qquad k \cdot  m = 0, \qquad k^2
 =0, \qquad m^2 = 1\, .
 \label{5.3}
 \ee
Now, we boost $A^\mu$ with $ \vec v = v \hat l$,
boosting into a frame in which the medium is at rest. We
obtain
 \be
  A'^\mu_\pm =  k'^\mu t \pm {L \ov 2}  m'^\mu\, ,
  %= \tilde k^\mu (t-t_0) \pm {L \ov 2}  m''^\mu
  \label{5.4}
 \ee
 for some $k'^\mu$ and $m'^\mu$ which again satisfy
 \be
   k'^2
 =0, \qquad  m'^2 = 1, \qquad m' \cdot k'=0\, .
 \label{5.5}
 \ee
In general, $m'^\mu$ has a nonzero $0$-th component and thus the
two ends of the dipole do not have the same time. To fix this we
write
 \be
 A'^\mu_\pm  =  k'^\mu (t \pm t_0) \pm {L \ov 2}  m''^\mu\, ,
 \label{5.6}
 \ee
where we have defined
 \be
 m''^\mu = m'^\mu - t_0 k'^\mu\, ,
 \label{5.7}
 \ee
and choose $t_0$ such that the zeroth component of
$m''$ is zero, making $m"^\mu$ purely spatial.  It is easy to confirm
that, given (\ref{5.6}), we now have
 \be
 k'^2=0, \qquad m''^2 =1, \qquad m'' \cdot k' =0\, .
 \label{5.8}
 \ee
We now have almost exactly the same Wilson loop configuration
as we had in our original calculation of Section 3.2
when the medium was at rest from the beginning, with the
only difference being
that the two long sides of the Wilson loop do
not start and end at equal times, due to the shift $t_0$. This is
immaterial when $L_-$ is big: in our evaluation of the Wilson loop
we always assumed time translational invariance anyway, neglecting
the contribution of the ``ends of the loop'' relative to that of
the long, time translation invariant, mid-section of the loop. We thus find that
in the presence of a wind velocity
 \be
 \langle{W^A({\cal C})}\rangle = \exp\left[ -  S ({\cal C})\right]
 \label{5.9}
 \ee
 with
 \be
 S ({\cal C}) = - {1 \ov 4\sqrt{2}} \hat q_0 (L^-)' L^2\, ,
 \label{5.10}
 \ee
where $\hat q_0$ is the value with no wind and where
 \be
 (L^-)' = \sqrt{2} k'^0 t = k'^0 L^-
 \label{5.11}
 \ee
is the light-cone distance travelled in the rest frame of the medium
whereas $L^-$ is the corresponding quantity in the lab frame.
We thus conclude that the {\it only} effect of the collective flow of the medium
on $\hat q$ is what we called the straightforward effect above, namely that
due to the Lorentz transformation of $L^-$.
{}From the standard
Lorentz transformation rule,
 \be
  k'^0= \cosh \eta_f \,- \sinh\eta_f (\hat l \cdot \hat n) = \cosh \eta_f -
  \sinh \eta_f \cos \th\, .
  \label{5.12}
 \ee
We thus find
 \be
  \hat q = (\cosh \eta_f -
  \sinh \eta_f \cos \th) \, \hat q_0\, .
  \label{5.13}
  \ee
  This result is independent of $\phi$.

We have established the transformation rule (\ref{5.13}) by
boosting to the rest frame of the medium. This
reduced the problem to one with no wind but with a Lorentz
transformed longitudinal extension (\ref{5.11}). Alternatively,
the same result (\ref{5.13}) can be obtained by starting from the
metric corresponding to the medium having a velocity $\vec v$ and
doing the Wilson loop computation in this metric.  We have confirmed
by explicit calculation for several examples that
the same result (\ref{5.13}) is obtained.

The
derivation of the scaling (\ref{5.13}) relied only on properties
of Lorentz transformations; {\it nothing} in the calculation of the underlying
$\hat q_0$  (which depends on $T$ and $N$ and $\lambda$
in ${\cal N}=4$ SYM and varies from one theory to the next as
we shall discuss in Section 6) comes in.  We conclude that the
scaling (\ref{5.13}), which describes how
the jet quenching parameter $\hat q$ depends
on the collective flow velocity of the medium doing the quenching,
applies in QCD also. R. Baier {\it et al.} have reached the same
conclusion independently~\cite{BaierPrivate}.

To get a sense
of the order of magnitude of the effect, we note that transverse flow velocities
in excess of half the speed of light are generated by the time the
matter produced in a heavy ion collision freezes out.
A velocity $v=0.5$ corresponds to $\eta_f=0.549$, which yields $\hat q = 1.732\, \hat q_0$
for a head wind ($\theta=\pi$), $\hat q=1.155\, \hat q_0$ for $\theta=\pi/2$,
and $\hat q\, =0.577$ for a tail wind ($\theta=0$).
An  investigation of the quantitative consequences of (\ref{5.13})
requires modelling of the geometry and time-development of the
collective flow in a heavy ion collision,
along the lines of the analysis in
Refs.~\cite{Renk:2005ta,BaierPrivate,Renk:2006sx}.

%%%%%%%%%%%%%%%%%%%%%%%%%%%%%%%%%%%%%%%%%%%%%%%%%%%%%%%%%%%%%%%%%%%%%%%%%%%%%%%%
%%%%%   Section 4.
%%%%%%%%%%%%%%%%%%%%%%%%%%%%%%%%%%%%%%%%%%%%%%%%%%%%%%%%%%%%%%%%%%%%%%%%%%%%%%%%

\section{The static $q\bar{q}$ potential for all dipole orientations with respect to the wind}
\label{sec4}

%
%%%%%%%%%%%%%%%%%%%%%%%%%%%%%%%%%%%%%%%%%%%%%%
%\begin{figure}[t]
\FIGURE[t]{
%\vskip-0.15in
% \hfill\hskip-0.6in
\includegraphics[scale=0.5,angle=0]{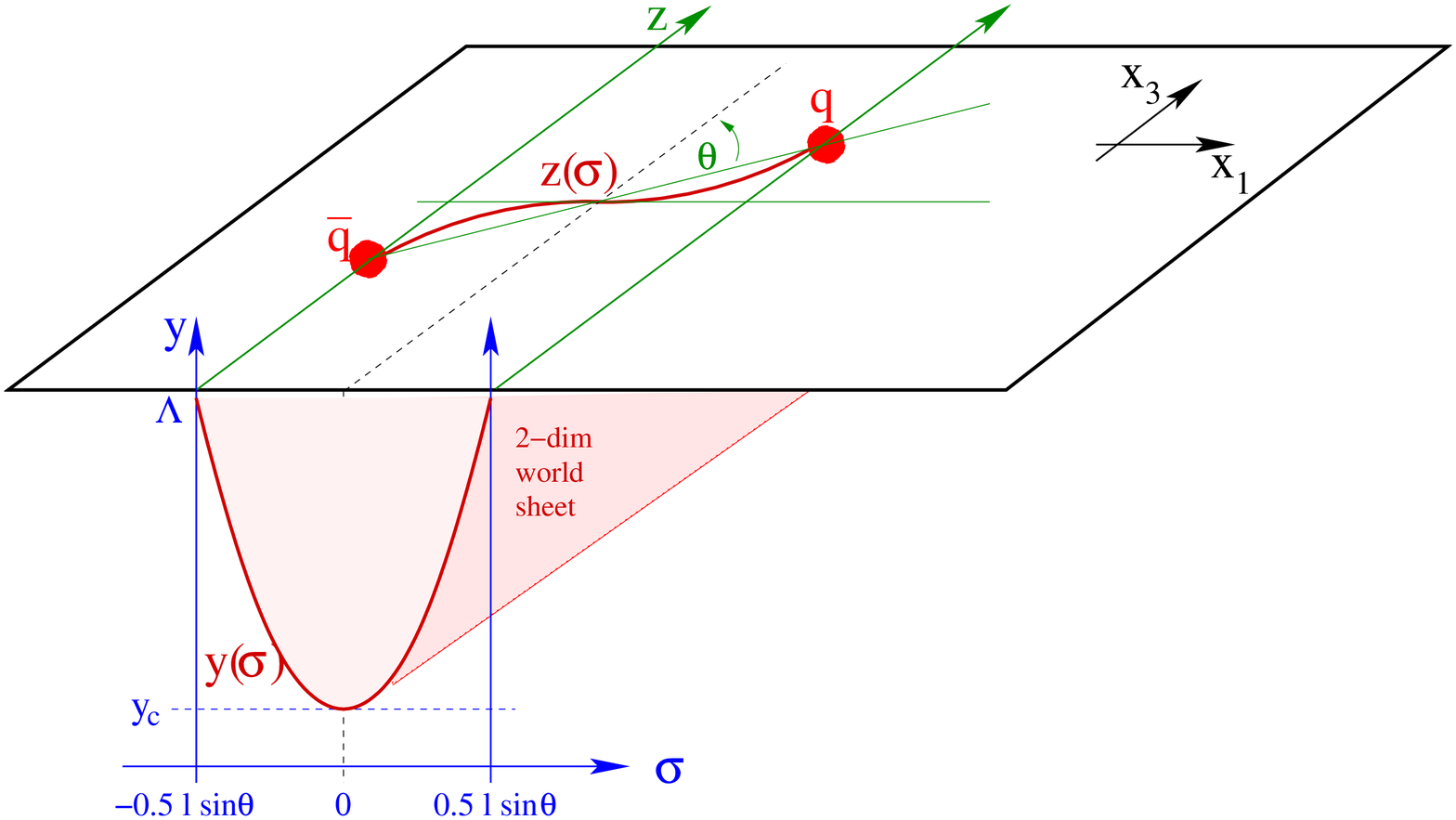}
%\hskip-0.1in
\hfill
\caption{Schematic picture of a $q\bar{q}$-dipole, moving along the $x_3$-direction. The
dipole is oriented along an arbitrary orientation $\theta$ in the $(x_1,x_3)$-plane. The
trajectories, along which the quark and antiquark propagate, specify the boundary ${\cal C}$
of a two-dimensional world sheet, which extends into the 5-th bulk dimension $y = r/r_0$.
The shape of this world sheet
is characterized by the functions $y(\sigma)$ and $z(\sigma)$,
where  $\sigma = x_1$.  By symmetry, $y$ has a turning point at $y(\sigma = 0)=y_c$.
For generic values of the angle $\theta$, the solution $z(\sigma)$ deviates from a straight line.
}\label{fig1}
}%\end{figure}
%%%%%%%%%%%%%%%%%%%%%%%%%%%%%%%%%%%%%%%%%%%%%%%%%

In section~\ref{sec3c}, we analyzed the quark-antiquark potential for a
$q\bar{q}$-dipole which was oriented in the $x_1$
direction and which propagated orthogonal to its orientation along the $x_3$ direction with velocity $v=\tanh\eta$.
%We took the regulator $\Lambda$ to infinity prior to increasing
%the rapidity $\eta$ to a finite, but perhaps arbitrarily large, value.
%In this way, we ensured that $\sqrt{\cosh{\eta}}<\Lambda$ and hence
%ensured that the expectation value of the Wilson loop was controlled by
%the action of a time-like string world sheet and hence was given by the
%exponential of an imaginary quantity.
%This
%allowed us to define the static $q\bar{q}$-potential in the presence of a
%thermal medium
%moving with velocity $v = \tanh \eta$ with respect to the dipole.
Here, we extend this analysis
to the case where the dipole is tilted by an arbitrary angle $\theta$ with respect to its direction
of motion, see Fig.~\ref{fig1}. For $\theta = \pi/2$, we recover the results obtained
in section~\ref{sec3c} above.

%

%
%
%%%%%%%%%%%%%%%%%%%%%%%%%%%%%%%%%%%%%%%%%%%%%%
%\begin{figure}[t]
\FIGURE[t]{
%\vskip-0.15in
% \hfill\hskip-0.6in
\includegraphics[scale=0.7,angle=0]{fig-wiggle.eps}
%\hskip-0.1in
\hfill
\caption{Solutions $y(\sigma)$, $z(\sigma)$ of the differential equations (\ref{4.6}) and (\ref{4.7})
with boundary conditions (\ref{4.2})
for different orientation angles $\theta$ and rapidity $\eta$. These solutions characterize the
two-dimensional world sheet bounded by the Wilson loop. For definitions, see Fig.~\ref{fig1} and
text.  Left panel: $y(\sigma)$ for $\eta=1$ for varying $\theta$. Middle panel:
$z(\sigma)-\sigma/\tan\theta$ for $\eta=1$ for varying $\theta$.
As Fig.~\ref{fig1} illustrates,
$z(\sigma)-\sigma/\tan\theta$ would be zero if the projection of the
string onto the D3-brane at $y=\Lambda$ were a straight line.
Third panel: $z(\sigma)-\sigma/\tan\theta$ for $\theta=\pi/4$ for varying $\eta$.
}\label{fig2}
}%\end{figure}
%%%%%%%%%%%%%%%%%%%%%%%%%%%%%%%%%%%%%%%%%%%%%%%%%
%

We work in the boosted metric (\ref{3.13}), in which the dipole is at rest. The dipole lies in the
$(x_1,x_3)$-plane and the parametrization of the two-dimensional world sheet of the corresponding Wilson loop is
\be
    \tau = t\, ,\qquad \sigma = x_1\, ,\qquad x_2 = {\rm const.}\, ,\qquad
        x_3 = x_3(\sigma)\, ,\qquad r=r(\sigma)\, .
        \label{4.1}
\ee
For a dipole with length $L$ whose orientation makes an angle $\theta$ with its direction
of propagation (the $x_3$-direction)  the projections of the dipole on
the $x_1$ and $x_3$
axis are of length $L \sin\theta$ and $L \cos\theta$, respectively.
We define dimensionless
coordinates
\be
y=\frac{r}{r_0}\, ,\qquad z=x_3\frac{r_0}{R^2}\, ,\qquad \tilde\sigma=\sigma \frac{r_0}{R^2}\, ,
\qquad l = L\frac{r_0}{R^2}\, ,
\ee
and drop the tilde.
The boundary conditions on $y(\sigma)$ and $z(\sigma)$ then become
\be
    y\left(\pm \frac{l}{2} \sin\theta \right) = \Lambda\, ,\qquad
    z\left(\pm \frac{l}{2} \sin\theta \right) =  \pm  \frac{l}{2} \cos\theta\, .
    \label{4.2}
\ee
Following the calculation of section~\ref{sec3c}, the Nambu-Goto action for (\ref{4.1})
can be written in the form (\ref{peo}), namely
$S({\cal C}) = \sqrt{\lambda} {\cal T} T \int_0^{l/2} d\sigma {\cal L}$,
with the Lagrangian reading
\begin{equation}
 {\cal L} = \sqrt{\left(y^4 - {\cosh^2 \eta } \right) \left(1 + {y'^2 \over y^4
 - 1} \right) + z'^2 \left(y^4 - 1\right)}\, ,
 \label{4.3}
 \end{equation}
 where $y'$ and $z'$ denote derivatives with respect to $\sigma$.
The Hamiltonian is
\begin{equation}
    {\cal H} = {\cal L} - y' \frac{\partial {\cal L}}{\partial y'}  - z' \frac{\partial {\cal L}}{\partial z'}
    = {y^4-\cosh^2 \eta \over {\cal L}} = q\, ,
    \label{4.4}
\end{equation}
a constant of the motion.  The momentum conjugate to $z$
\be
    \frac{\partial {\cal L}}{\partial z'} = \frac{y^4-1}{\cal L} z' = p\,
    \label{4.5}
\ee
is also a constant of the motion.
For a time-like world sheet, the constants of motion $q$ and $p$ must be real.
The equations of motion can be written in the form
\bea
    q^2 y'^2 &=& (y^4-\cosh^2\eta)(y^4-1-p^2) - q^2(y^4-1)\, ,
    \label{4.6}\\
    q^2 z'^2 &=& p^2\left(\frac{y^4-\cosh^2\eta}{y^4-1}\right)^2\, .
    \label{4.7}
\eea
Generic features of their solutions have been pointed out in Ref.~\cite{Liu:2006nn} already.
Fig.~\ref{fig2} shows numerical results. Since
$y'$ depends only on $y$ and since the boundary condition (\ref{4.2}) for $y$ is
symmetric under $\sigma \to - \sigma$, $y(\sigma)$ must be an even function of $\sigma$.
It descends ($y' < 0$) for $-l/2 \sin\theta < \sigma < 0$ and then ascends for
$0<\sigma< l/2 \sin\theta$. These features are clearly seen in Fig.~\ref{fig2}.
The turning point $y_c = y(0)$ satisfies the condition
\be
    \left(y_c^4-\cosh^2\eta\right) \left(y_c^4-1-p^2\right) - q^2\left(y_c^4-1\right)
    \equiv 0\, .
    \label{4.8}
\ee
Connecting the $q\bar{q}$-pair by a straight line in the $(x_1,x_3)$-plane would
correspond to $z(\sigma) = \sigma/\tan\theta$. To test for deviations of the string
world sheet away from this straight line, we plot $z(\sigma) - \sigma/\tan\theta$ in
Fig.~\ref{fig2}. We find a deviation of sinusoidal form for all angles except $\theta = 0, \pi/2$.
As an aside, we note that if one thinks of the two-dimensional world sheet as a flat piece
of paper, draws on it a straight line connecting $q$ and $\bar{q}$, and rolls it up as
depicted in Fig.~\ref{fig1},  then the projection of this straight line on the $(x_1,x_3)$-plane
would show a qualitatively similar sinusoidal wiggle.
However, the use of this analogy is limited, since we cannot specify in which sense or to
what extent the two-dimensional world sheet is flat.
Also, the observed deviation from the straight line behavior $z(\sigma) = \sigma/\tan\theta$
depends on rapidity. For $\eta = 0$, no deviation is possible since no direction in the
$(x_1,x_3)$-plane is singled out. For increasing values of $\eta$, the deviation increases
as seen in Fig.~\ref{fig2}.

%
%%%%%%%%%%%%%%%%%%%%%%%%%%%%%%%%%%%%%%%%%%%%%%
%\begin{figure}[t]
\FIGURE[t]{
%\vskip-0.15in
% \hfill\hskip-0.6in
\includegraphics[scale=0.5,angle=0]{p-versus-q.eps}
%\hskip-0.1in
\hfill
\caption{The integration constant $p$ as a function of the integration constant $q$
for $\eta=1$ and several fixed values of the orientation angle $\theta$.
This relation is defined by eqs. (\ref{4.9}) and (\ref{4.10}).  Along one of these curves,
$\ell$ changes with $q$ as shown in Fig.~\ref{fig4}.
}\label{fig3}
}%\end{figure}
%%%%%%%%%%%%%%%%%%%%%%%%%%%%%%%%%%%%%%%%%%%%%%%%%
%

The constants $q$ and $p$ must be related to the values of $l$ and $\theta$.  The
relationships are obtained by integrating the equations of motion (\ref{4.6})  and
(\ref{4.7}), giving
\be
    \frac{l}{2} \sin\theta =  q \int_{y_c}^\infty
        \frac{dy}{\sqrt{\left(y^4-\cosh^2\eta\right)\left(y^4-1-p^2\right) - q^2\left(y^4-1\right)}}
        \, ,\label{4.9}
\ee
\be
    \frac{l}{2} \cos\theta =  % \int_{y_c}^\infty   \frac{dz}{d\sigma} \frac{d\sigma}{dy} =
                  p \int_{y_c}^\infty   \frac{y^4-\cosh^2\eta}{y^4-1}
                  \frac{dy}{\sqrt{\left(y^4-\cosh^2\eta\right)\left(y^4-1-p^2\right) - q^2\left(y^4-1\right)}}\, .
                  \label{4.10}
\ee
With $q$ and $p$ determined,
the $q\bar{q}$ static potential in a moving thermal background then reads
\bea
    E(l,\theta; \eta) &=& S({\cal C}) - S_0
        \nonumber \\
            &=& \sqrt{\lambda} {\cal T} T \int_{y_c}^\infty
            \left[\frac{(y^4-\cosh^2\eta)}{\sqrt{\left(y^4-\cosh^2\eta\right)\left(y^4-1-p^2\right) - q^2\left(y^4-1\right)}} -1\right]\, dy
            \nonumber \\
            && - \sqrt{\lambda} {\cal T} T \left( y_c-1 \right)\, .
            \label{4.11}
\eea
Here, the subtraction term $S_0$, given in (\ref{3.39})
is the action for two isolated strings described by
the dragging solution of Refs.~\cite{Herzog:2006gh,Gubser:2006bz}
and Appendix A.

We have evaluated the potential $E(l,\theta; \eta)$ as a function of
the size $l$ of the dipole, its orientation $\theta$ with respect to its direction of motion, and
its velocity $v = \tanh \eta$ with respect to the thermal heat bath. Since the potential
in (\ref{4.11}) is written in terms of the integration constants $q$ and $p$, it is useful to
determine first how $p$ depends on $q$ for fixed $\theta$ and $\eta$. To do this,
we write $\tan\theta$ as the ratio of  Eqs. (\ref{4.9}) and (\ref{4.10}).  We find $p(q)$
to be a monotonously increasing function, whose
slope decreases with increasing angle $\theta$, see Fig.~\ref{fig3}.
For the maximal angle $\theta = \pi/2$, $p$ vanishes
independent of the value of $q$. This is the case
of a dipole oriented orthogonal to the wind, where Eq. (\ref{4.9}) reduces to Eq. (\ref{3.36}),
and the present calculation becomes that of
Section~\ref{sec3c}. For the opposite limit of a dipole oriented parallel to the wind,
$\theta = 0$, the parametrization (\ref{4.1}), (\ref{4.2}) of the two-dimensional world sheet
does not apply. However, the Nambu-Goto action is reparametrization invariant and,
as described in Appendix~\ref{appb}, in a parametrization which is
suitable for $0\leq \theta <\pi/2$
we find that the measurable quantity $E(l,\theta,\eta)$
depends smoothly on $\theta$ for $\theta \to 0$.

%%%%%%%%%%%%%%%%%%%%%%%%%%%%%%%%%%%%%%%%%%%%%%
%\begin{figure}[t]
\FIGURE[t]{
%\vskip-0.15in
% \hfill\hskip-0.6in
\includegraphics[scale=0.7,angle=0]{fig-theta-dep.eps}
%\hskip-0.1in
\hfill
\caption{Left panel: The size $l$ of the dipole and its orientation angle $\theta$
define the integration constants $q$ and $p$, see eqs. (\ref{4.9}) and (\ref{4.10}).
The plot shows $l$ as a function of $q$ for several fixed angles $\theta$.
Along each of these curves $p$ varies as shown in Fig.~\ref{fig3}.
Right panel: The static $q\, \bar{q}$-potential (\ref{4.11}) for rapidity $\eta=1$ and
different orientation angles $\theta$ of the dipole with respect to the direction of motion.
}\label{fig4}
}%\end{figure}
%%%%%%%%%%%%%%%%%%%%%%%%%%%%%%%%%%%%%%%%%%%%%%%%%%

Knowing $p(q)$ for fixed $\eta$ and $\theta$,  the rescaled dipole size $l(q)=l(q,p(q))$
can be written as a function of $q$ only. It takes values in the range
$l \in \left[ 0, l_{\rm max}\right]$. Here, the maximal dipole size $l_{\rm max}$
is the screening length above which bound states do not exist.  The value of $q$ at
which the maximum of $l(q)$ occurs
depends strongly on the angle $\theta$, as shown in the left
panel of Fig.~\ref{fig4}.
This is a feature of our parametrization: for smaller angle $\theta$,
$p(q)$ is a more steeply rising function (see Fig.~\ref{fig3}), and most of the
$\theta$-dependence
of $l(q,p(q))$ comes from the $p(q)$-dependence.
The value of $l_{\rm max}$ decreases
slightly with increasing angle $\theta$. This is consistent with the expectation that the
dipole is easier to dissociate if it is oriented orthogonal
to the direction of the wind, but
the effect is slight.

%
%%%%%%%%%%%%%%%%%%%%%%%%%%%%%%%%%%%%%%%%%%%%%%
%\begin{figure}[t]
\FIGURE[t]{
%\vskip-0.15in
% \hfill\hskip-0.6in
\includegraphics[scale=0.5,angle=0]{fig-lmax.eps}
%\hskip-0.1in
\hfill
\caption{The screening length $l_{\rm max}$ times its leading large-$\eta$
dependence $\sqrt{\cosh(\eta)}$. The exact results are given for dipoles oriented
perpendicular to the wind ($\theta=\pi/2$) and parallel to the wind ($\theta=0$).
The $\theta=\pi/2$ curve is compared to the analytical large-$\eta$ approximation
(\protect\ref{3.41}). Keeping only the first term in this analytical expression corresponds
to a horizontal line on the figure; including the term proportional to $(\cosh \eta)^{-5/2}$
improves the agreement with the exact result.
}\label{fig6}
}%\end{figure}
%%%%%%%%%%%%%%%%%%%%%%%%%%%%%%%%%%%%%%%%%%%%%%%%%
%

With $p(q)$ determined,
the $q\bar{q}$-static potential $E(l(q,p(q)),\theta,\eta)$ also becomes a function of $q$
only. This defines curves $\lbrace l(q), E(l(q)) \rbrace$, parametrized by the integration
constant $q \in \left[ 0,\infty\right]$. The $q\bar{q}$ static potential (\ref{4.11}) is a double-valued
function of $l$ in the range $l \in \left[ 0, l_{\rm max}\right]$, see the right panel of Fig.~\ref{fig4}.
The configurations whose energy is given by the
upper branch of $E(l)$ are presumably unstable, as has been shown
explicitly for $\theta=\pi/2$ in Ref.~\cite{Friess:2006rk}.
The lower branch displays
the typical short-distance behavior of a $q\bar{q}$ binding potential. For fixed
rapidity $\eta$, this potential shows the expected $\theta$-dependence: the $q\bar{q}$ pair
is more strongly bound if the dipole is aligned with the direction of motion, and this binding
decreases as the dipole presents itself at a larger angle with respect to the wind, see Fig.~\ref{fig3}.

In Fig.~\ref{fig5} in Section 3, we have explored the $\eta$-dependence of the $q\bar{q}$ static
potential for a dipole oriented orthogonal to the wind. The screening length displays the
dominant Lorentz-$\gamma$ dependence
$l_{\rm max}
%\simeq 0.74333/\sqrt{\cosh\eta}
\propto 1/\sqrt{\cosh\eta}  = 1/\sqrt{\gamma}$, as given in (\ref{3.41}).
This velocity dependence is much stronger than the angular dependence displayed in
Fig.~\ref{fig4}. The velocity dependence of the potential $E(l)$, shown on the right hand
side of Fig.~\ref{fig5}, also shows clearly that the short distance behavior of the potential
is not affected by velocity-dependent medium effects. This is a consequence of choosing
the regularization prescription (\ref{3.39}).

Finally, we show in Fig.~\ref{fig6} the screening length $l_{\rm max}$ multiplied by
$\sqrt{\cosh\eta}$. We include curves for $\theta=0$ and $\theta=\pi/2$; those
for other angles lie in between these two.  Note that both curves have the
same value of $l_{\rm max}$ in the $\eta\rightarrow 0$ limit as they must.
The flat behavior of these curves at large $\eta$
illustrates that $l_{\rm max} \propto 1/\sqrt{\cosh\eta}$ is the leading large-$\eta$ dependence
for all dipole orientations.
This leading behavior provides a numerically very accurate
approximation ($< 1 \%$ deviation) for $\eta > 2$, and even for $\eta = 0$, it is accurate
to within 20\% (note the suppressed zero in Fig.~\ref{fig6}).
Including the $(\cosh\eta)^{-5/2}$ term in the
analytical expansion (\ref{3.41}) improves the description.

%%%%%%%%%%%%%%%%%%%%%%%%%%%%%%%%%%%%%%%%%%%
%%%%%%%%%%%%%%%%%%%%%%%%%
%\section{Comparison with other recent results}

\section{Discussions and Conclusions}

In Section 2, we reviewed
the  physical arguments why
the expectation value of the time-like Wilson loop which describes the
quark-antiquark potential must be the exponential of an imaginary quantity
whereas the expectation value of the light-like Wilson loop which arises
in the physics of deep inelastic scattering, proton-nucleus collisions, and
the calculation of the jet quenching parameter relevant to parton energy
loss in heavy ion collisions is instead the exponential of a real quantity.
In Section 3, we saw how these results emerge by direct calculation in $\NN=4$ SYM
theory at strong coupling, where via the AdS/CFT correspondence the calculation
of the expectation values  of these two types
of Wilson  loops reduces to the evaluation of
the action of an extremal string world sheet, time-like in the first case and space-like in
the second.  These aspects of our paper are discussed at length in Section 3 and
we shall not discuss them further here.

This section incorporates several different discussions, while along the way
summarizing many of our conclusions.
In Section 6.1, we summarize what we have learned from our
calculations of screening in a hot wind.
We compare
our calculation of the jet quenching parameter $\hat q$ to the very different
approach to energy loss in
Refs.~\cite{Herzog:2006gh,Gubser:2006bz,Casalderrey-Solana:2006rq,Drag1,Friess:2006aw}
in Section 6.2.
In Section 6.3, we compare our calculation
of $\hat q$ in $\NN=4$ SYM to that extracted from RHIC data.
Given that we find surprisingly good agreement
between $\hat q_{SYM}$ and that extracted from RHIC data, in Section 6.4 we enumerate
the differences and similarities between $\NN=4$ SYM and QCD.  Finally, in Section 6.5 we
collect what is known about how $\hat q$ changes from the quark-gluon plasma of
one gauge theory to that of another, including deriving a new result which allows the
determination of $\hat q$ in any conformal theory with a gravity dual.  We use this
result to estimate $\hat q_{QCD}/\hat q_{SYM}$.

\subsection{Velocity dependence of screening length}

We can summarize what we have learned from our
calculation in Sections \ref{sec3} and \ref{sec4}
of the quark-antiquark potential and  screening
length in a hot wind as follows.
We find that the
screening length $L_{\rm max}$ of an external quark in an $\NN=4$ SYM
plasma with velocity $v = \tanh \eta$ can be written as
 \be
 L_{\rm max} = {f(\eta,\theta) \ov \pi T \sqrt{\cosh\eta}} \ ,
 \label{lmax}
 \ee
where $\th$ is the angle between the orientation of the dipole
and the velocity of the moving thermal medium in the rest frame
of the dipole.  $f (\eta, \th)$ is only weakly dependent on both of
its arguments. That is, it is close to constant.
In fact, for any values of $\eta$ and $\theta$, $f(\eta,\theta)$ lies between 0.74 and 0.87.
The limiting cases are
$f (\eta=0) \simeq 0.87$ for all $\theta$, and
$f(\infty, {\pi \ov 2}) \simeq 0.74$.
For a given $\eta$, $f (\eta, \th)$ is a monotonically
decreasing function of $\th$ as $\th$ varies from $0$ to $\pi/2$.
For a given $\th$, $f(\eta, \theta)$ is a monotonically
decreasing function of $\eta$. As $\eta \to \infty$, we find
$f (\eta, \th) = f( \infty, \th) (1 + {\cal O}(1/\cosh^2\eta) )$.

For $\NN=4$ SYM theory, since the energy density $\varepsilon\propto
T^4$, in the large $\eta$ limit equation~(\ref{lmax}) can also be
thought of as
 \be
 L_{\rm max} \propto {1 \ov \varepsilon (\eta)^{1 \ov 4}}\ ,
 \label{alam}
 \ee
where $\varepsilon (\eta) = \cosh^2 \eta \, \varepsilon(0)$ is the energy density of
the boosted medium.

As discussed in Ref.~\cite{Liu:2006nn}, if the velocity scaling of
$L_{\rm max}$ that we have found, namely (\ref{lmax}) and (\ref{alam}), holds for QCD, it
will have qualitative consequences for quarkonium suppression in
heavy ion collisions at RHIC and LHC. Since our discussion in
earlier sections only involves the AdS$_5$ part of the
geometry, the scaling (\ref{alam}) applies to any conformal
field theory with a gravity dual at finite temperature. To the extent that
the QGP of QCD at RHIC temperature is close to being conformal,
one is tempted to view this as a support of the applicability of
(\ref{lmax}) and (\ref{alam}) to QCD.
The results of Ref.~\cite{Peeters:2006iu} further support this view.
These authors studied large-spin mesons in a hot wind in a
confining, nonsupersymmetric theory and found that they dissociate
beyond a maximum wind velocity.
The relation
between the size $L$ of these mesons and their dissociation
velocity $v$ is consistent with $L\propto (1-v^2)^{1/4}$.

For more general theories with a gravity dual, one can use the
generic metric~(\ref{mp1}) which we introduce below
to study the screening length. A nice
argument presented by
Caceres, Natsuume and Okamura in Ref.~\cite{Caceres:2006ta} indicates that in
the large $\eta$ limit, one would generically have
  \be
 L_{\rm max} \propto {1 \ov \varepsilon (\eta)^{\nu}}
 \label{aam}
 \ee
for some index $\nu$. In particular, for any gauge theory which is
dual to an asymptotically AdS$_5$ geometry, one would find $\nu =
{1 \ov 4}$ as in~(\ref{alam}). Examples include $\NN=4$ SYM
with nonzero $R$-charge chemical potentials,
studied in Refs.~\cite{Caceres:2006ta,Avramis:2006em}.
(Note that since
chemical potentials introduce additional mass scales,
the dependence of
$L_{\rm max}$ on temperature is rather complicated and it is (\ref{alam}) which
generalizes, not (\ref{lmax}).)

For non-conformal
theories the scaling index $\nu$ can deviate from $1/4$. One measure
of the deviation from conformality is the deviation of the sound
velocity from the conformal value of $1/\sqrt{3}$.  Following
a similar argument in Ref.~\cite{Buchel:2006bv}
concerning the value of
$\hat q$ in non-conformal theories, Caceres, Natsuume and Okamura suggested
that for theories which are close to being conformal, the
index $\nu$ may profitably
be written as
 \be
 \nu = {1 \ov 4} + c \delta + \cdots, \qquad \delta = {1 \ov 3} - v_s^2
 \label{poh}
 \ee
with $c$ some constant. For the
cascading gauge theories of Ref.~\cite{Klebanov:2000hb},
$c=-9/16$ meaning that if $v_s^2\simeq 0.27-0.31$ in these theories
as is the case in QCD at $T\sim 1.5 T_c$~\cite{Karsch:2006xs},
the index is $\nu\simeq 0.21-0.24$.  This suggests that for QCD
the scaling (\ref{alam}) with $\nu=1/4$ should be a very good guide.

As discussed at the end of Section 3.1, the velocity scaling (\ref{alam})
describes how screening lengths
and, correspondingly, dissociation temperatures drop for quarkonia
moving through the thermal medium with some relative
velocity, and so
should be included in the modelling of quarkonium suppression. The
$p_T$-dependent pattern of quarkonium suppression predicted by
(\ref{alam}) will be tested in future heavy ion experiments at RHIC
and the LHC.

\subsection{Comparison with energy loss via drag}

Before comparing our result for $\hat q$ with that extracted from RHIC data,
which we shall do in Section 6.3, we discuss the differences
between our approach and recent
calculations~\cite{Herzog:2006gh,Gubser:2006bz,Casalderrey-Solana:2006rq,Drag1,Friess:2006aw}
of how an external quark loses energy while being dragged
through an $\NN=4$ SYM plasma.
In Appendix A (see (\ref{aa.9})) we have reproduced the
``dragging string solution'' of
Refs.~\cite{Herzog:2006gh,Gubser:2006bz} for
a string attached to and trailing behind
an isolated moving test quark,
since we need it as the subtraction
term in our calculation of the quark-antiquark potential in
Section 3.1.

To describe the results of
Refs.~\cite{Herzog:2006gh,Gubser:2006bz,Casalderrey-Solana:2006rq}
in their proper context, let us
start with the relativistic generalization of the Langevin
equations for a quark moving
through some thermal medium (see for example
Ref.~\cite{Moore:2004tg,Casalderrey-Solana:2006rq})
 \bea
 \label{lang1}
 {d p_L  \ov dt} & = & - \mu (p_L) p_L + \xi_L (t)\ , \\
 \label{lang2}
 {d p_T  \ov dt} & = &  \xi_T (t)\ ,
 \eea
where $p_L$ and $p_T$ are the longitudinal and transverse momentum
of the quark,
respectively.  (We have simplified the notation by dropping the spatial indices
on transverse quantities.)
Henceforth we shall denote $p_L$ by $p$.
$\xi_L$ and $\xi_T$ are random fluctuating forces in
the longitudinal and transverse directions, which satisfy
 \bea
 \label{lang3}
 \vev{\xi_L (t) \xi_L (t')} & = & \kappa_L (p) \delta (t-t') \ ,\\
 \label{lang4}
 \vev{\xi_T (t) \xi_T (t')} & = & \kappa_T (p) \delta (t-t')\ .
 \eea
$\kappa_L (p)$ and two times $\kappa_T (p)$
describe how much longitudinal and
transverse momentum squared is transferred to the quark per unit time.
Note that  at zero velocity, $\kappa_L(0) = \kappa_T (0)$ whereas
for $p>0$ one expects that $\kappa_L(p)\neq \kappa_T(p)$.
Also, upon assuming that the momentum fluctuations of the particle
are in equilibrium with the thermal  medium, as appropriate at zero
velocity, a fluctuation-dissipation theorem relates $\mu(0)$ to $\kappa_L(0)$
via the Einstein relation
\be
\mu(0) = \frac{\kappa_L(0)}{2mT}\ ,
\label{Einstein}
\ee
where $m$ is the static mass of the quark.
The relation (\ref{Einstein}) is also not expected to hold for $p>0$.  (See Ref.~\cite{Moore:2004tg}
for examples.)

The central result of
Refs.~\cite{Herzog:2006gh,Gubser:2006bz} is
that an isolated quark moving through the $\NN=4$ SYM plasma
with its trailing, dragging, string feels a drag force proportional to $p$,
described  by a
momentum independent  drag coefficient
 \be \label{enrl}
  \mu (p) = {\pi \sqrt{\lam} \ov 2m } T^2\ .
 \ee
Independently, $\kappa_L (0)$ in the $\NN=4$ SYM plasma was
calculated directly
in Ref.~\cite{Casalderrey-Solana:2006rq}. These authors found that
$\kappa_L(0)$ and $\mu(0)$ in (\ref{enrl}) indeed satisfy (\ref{Einstein}),
which can be considered a new consistency check of the
AdS/CFT framework.

Eq.~(\ref{enrl}) demonstrates that  in the high energy limit
the energy loss mechanism in strongly coupled $\NN=4$ SYM theory
is very different from that in QCD.  The origin of the difference
is the fact that $\NN=4$ SYM is not asymptotically free.
In QCD, as discussed in Section~\ref{sec2}, the
average energy loss of a parton in the high energy limit is
independent of $p$ (or at most logarithmically dependent on $p$)
and it is proportional to the {\it square} of the distance
travelled through the medium. There, the dominant mechanism by
which a high energy parton loses its energy is through radiating
gluons which have a high enough transverse momentum $k_T$ (and an
even higher energy) that $\alpha_s$ evaluated at $k_T$ is weak,
meaning that the dominant energy loss processes can be described
perturbatively, with nonperturbative physics at scales of order
the temperature coming in only via the description of the repeated
soft interactions between the radiated gluon and the medium, and
between the original high energy parton and the medium. The effect
of these nonperturbative  soft interactions is encoded in the
jet quenching parameter $\hat q$, which can be defined
nonperturbatively via a light-like Wilson loop as described in
Section 2.

In Ref.~\cite{Liu:2006ug} and in the present paper, we seek
insights about $\hat q$ in QCD by calculating the analogous
quantity in $\NN=4$ SYM.   We do not attempt to describe the full
process of energy loss in $\NN=4$ SYM because the asymptotic
freedom of QCD is crucial to the description of the radiative
parton energy loss process which dominates at high energies,
making it impossible to model the physics of QCD parton energy
loss at high energies in a theory like $\NN=4$ SYM which is
strongly interacting at all scales.
%Indeed, because $\NN=4$ SYM is
%strongly coupled at all scales it is not even possible to give a
%well-defined answer to the question of whether the  energy loss
%mechanism in $\NN=4$ SYM is or isn't dominated by gluon
%radiation~\cite{Herzog:2006gh}.

Even though the drag coefficient (\ref{enrl}) describes energy loss
in a $\NN=4$ SYM plasma even for quarks moving
relativistically~\cite{Herzog:2006gh},
it cannot be used to extract
$\hat q$ or $\kappa_T (p \to \infty)$.
As we
remarked after equation (\ref{lang4}), except
in the low-velocity  limit one does not expect
$\kappa_T(p)$ to be equal to $\kappa_L(p)$
or $\kappa_L(p)$ to be related to $\mu(p)$ via
the Einstein relation. Indeed, the direct calculation of $\kappa_L$
in Ref.~\cite{Casalderrey-Solana:2006rq}
manifestly requires modification at nonzero
$p$~\cite{Casalderrey-Solana-Private}.
The $\kappa_L (0)$
found in Ref.~\cite{Casalderrey-Solana:2006rq}
and indirectly in Ref.~\cite{Herzog:2006gh}
has the
same parametric dependence on $\lambda$ and $T$ as $\hat q$ in
(\ref{qhat}), and is smaller than $\hat q$ by a purely numerical  factor of
$\sim 1.20$.
This is curious, since the quantity $\kappa_L(0)=\kappa_T(0)$ has
no evident relation to
jet quenching (since jets by definition are relativistic) or to the jet quenching
parameter $\hat q$.
%, which is the amount of transverse momentum squared
%acquired by an ultrarelativistic quark per distance travelled and which is
%defined via a light-like Wilson loop.

To look for connections
between $\NN=4$ SYM energy loss described by (\ref{enrl}) and data
from RHIC one should then seek
circumstances in which all aspects of the energy
loss process {\it are} strongly coupled. Perhaps the energy loss
of quarks which are slowly moving and yet energetic --- i.e.
quarks which are heavy --- is the best example, as stressed by
many of the recent
papers~\cite{Herzog:2006gh,Gubser:2006bz,Casalderrey-Solana:2006rq,Drag1,Friess:2006aw},
although extracting the contribution from energy loss
to the medium modification of charmed meson production in
the regime in which the parent $c$-quark is slowly moving
presents considerable challenges.
Furthermore, precisely because the entire energy loss process is
treated in this approach, further questions like how the $\NN=4$
SYM medium responds to the dragging quark (i.e. ``where does the
lost energy go?'') can be addressed~\cite{Friess:2006aw}.

\subsection{Comparison of $\hat q$ of $\NN=4$ SYM with experimental estimate}

We turn now to the comparison between $\hat q$ in
$\NN=4$ SYM with that extracted from current RHIC
data. In Eq.~(\ref{3.33}) we found that
 \be
 \hat{q}_{SYM} = \frac{\pi^{3/2} \Gamma\left(\frac{3}{4}\right)}{ \Gamma\left(\frac{5}{4}\right)}
    \sqrt{\lambda}\, T^3 \approx 26.69 \sqrt{\alpha_{\rm SYM}N_c} \,T^3\,.
    \label{qhat}
 \ee
Taking $N=3$ and $\alpha_{\rm SYM}=\frac{1}{2}$,
thinking $\alpha_{\rm QCD}=\frac{1}{2}$ reasonable for
temperatures not far above the QCD phase transition, we shall use
$\lambda=6\pi$ to make
estimates.\footnote{This qualitative comparison will suffice for our purposes.
The question of how to relate the coupling strengths of different
thermal quantum field theories to each other is not guaranteed
to have an unambiguous answer. Here we note that,
unlike $\alpha_{\rm SYM}$,
the value of $\alpha_{\rm QCD}(T)$ is not
well-defined in a strongly interacting
quark-gluon plasma.
It may thus be preferable to make this comparison using
a quantity calculated nonperturbatively in both theories,  and compare the QCD
value for the QGP at a few times the QCD critical temperature $T_c$, without
reference to any $\alpha_{\rm QCD}$,
to the $\lambda$-dependent SYM value, thus fixing $\lambda$.
Because we want to keep the question of how to fix $\lambda$ separate from
the question of how the difference in the number of degrees of freedom in QCD
and $\NN=4$ SYM affects $\hat q$, it is also important
to choose a quantity which is independent of the number of degrees of
freedom. (For our purposes this rules out using the Debye mass, as suggested
in Ref.~\cite{Caron-Huot:2006te} in
a different context.)
One recent proposal~\cite{Gubser:2006qh} is to compare the shapes of the
static quark-antiquark potential.
The screening length
$L_{\rm max}$ is independent of $\lambda$ in $\NN=4$ SYM, so that cannot
be used.  Instead, one has to compare the shapes of the potentials themselves, which
is not straightforward.  Another possibility
is to use the lattice QCD calculation of the ratio of the energy density
of the QCD quark-gluon plasma to that of a non-interacting Stefan-Boltzmann gas
of quarks and gluons.
According to lattice QCD calculations done with two or three flavors of quarks,
this ratio rises to about 0.78-0.82
for $T\sim (1.5-2) T_c$ and then flattens at
higher $T$~\cite{Karsch:2001vs}.  In SYM, it is $\frac{3}{4}$ in the $\lambda\rightarrow\infty$
limit~\cite{Gubser:1996de}  and, using
the leading correction to this result which is
$+\frac{45}{32}\zeta(3)\lambda^{-3/2}$~\cite{Gubser:1998nz}, the range 0.78-0.82 corresponds to
$9<\lambda<15$.
The pressure in two- and three-flavor QCD also approaches about
0.8 times its Stefan-Boltzmann value, but
only at the higher temperature
$T \sim (2.5-3) T_c$~\cite{Karsch:2001vs}, suggesting that a comparison
of this sort could be more appropriate at the LHC than at RHIC.
%Even then a comparison like this could only be semi-quantitative:
%because there is no
%actual analogue of $\lambda$ in the strongly interacting QCD
%quark-gluon plasma, using a thermodynamic
%quantity to fix the value of $\lambda$ in the calculation of a dynamic
%quantity like $\hat q$ cannot be justified quantitatively.
}
{}From (\ref{qhat}), we find
 \bea
 \hat q_{\rm SYM} & =   4.5, 10.6, 20.7  \; \GeV^2/\fm \qquad {\rm for} \qquad  & T= 300, 400, 500 \, \MeV\,.
       %  & = 10.6 \;  \GeV^2/\fm \qquad & T= 400 \, \MeV \\
       %& = 20.7 \; \GeV^2/\fm \qquad & T= 500 \, \MeV
 \label{numV}
 \eea

We have computed (\ref{qhat}) in the large $N$ and large 't Hooft
coupling $\lam$ limit and thus it is only the leading order term in a
double expansion in $1/\sqrt{\lam}$ and $1/N^2$. There are two
sources of contributions to $1/\sqrt{\lam}$ corrections: from the
fluctuation of the string world sheet~(\ref{3.9}) and from the
modification of the background geometry (\ref{3.3}) due to $\apr$
corrections. The authors of Ref.~\cite{Armesto:2006zv} find that
the leading contribution of the second type
of correction is given by
 \be
 \hat q_{SYM} (\lam) = \hat q_{SYM} \le(1 - 1.765 \lam^{-{3 \ov 2}} +
 \cdots \ri)\ . \label{oneC}
 \ee
If one uses $\al_{\rm SYM}=\frac{1}{2}$, then the correction
from~(\ref{oneC}) is about $2\%$. The corrections due to
fluctuations of the world sheet, the leading order term of which is
expected to be of order $1/\sqrt{\lam}$, are harder to compute and
are not known at the moment. $1/N^2$ corrections, which require
string loop calculations, are also beyond currently available
technology. It is worth noting, however, that the
ratio of energy density and pressure to their Stefan-Boltzmann values
are both quite insensitive to changing $N$ from 3 to 4 to 8~\cite{Bringoltz:2005rr}.

In a heavy ion collision, $\hat q$ decreases with time $\tau$ as
the hot fluid expands and cools. The time-averaged $\hat q$ which
has been determined in comparison with RHIC data is
$\overline{\hat{q}} \equiv \frac{4}{(L^{-})^2}
\int_{\tau_0}^{\tau_0+ L^-/\sqrt{2}} \tau\, \hat{q}(\tau)\ d\tau$,
%the time-averaged $\hat q$,
found to be around
5-15~GeV$^2/$fm~\cite{Eskola:2004cr,Dainese:2004te}. If we assume
a one-dimensional Bjorken expansion with $T(\tau) = T_0
\left(\frac{\tau_0}{\tau}\right)^{1/3}$, take $\tau_0=0.5$~fm, and
take $L^-/\sqrt{2}=2$~fm, the estimated mean distance travelled in
the medium by those hard partons which ``escape'' and are
detected~\cite{Dainese:2004te}, we find that to obtain
$\overline{\hat{q}}=5$~GeV$^2/$fm from (\ref{qhat}) we need $T_0$
such that $T(1~{\rm fm})\approx 310$~MeV, only slightly higher
than that expected from hydrodynamic modelling~\cite{Kolb:2003dz}.
With $L^-/\sqrt{2}=1.5$~fm, we find
$\overline{\hat{q}}=5$~GeV$^2/$fm for
$T(1~{\rm fm})\approx 280$~MeV, fully consistent with expectations.
There are currently too many uncertainties in the various components
of this comparison to make a strong statement, but it seems clear that
the $\hat q$ given by (\ref{qhat}) that we have calculated in the
quark-gluon plasma of $\NN=4$ SYM is in qualitative agreement
with $\overline{\hat{q}}=5$~GeV$^2/$fm, which is in turn consistent with RHIC data.

The value obtained
from~(\ref{qhat}) assumes that the medium is static. However, in a
relativistic heavy ion collision, the medium itself develops
strong collective flow, meaning that
the hard parton is traversing a moving medium --- it
feels a wind. Thus to compare~(\ref{qhat}) with the
experimental estimate we should include the effects of the
wind on $\hat q$ that we
discussed in Section~\ref{sec:tran}. We found in Eq.~\ref{5.13} that
 \be
  \hat q = \gamma_f(1 -
  v_f \cos \th) \, \hat q_0\ ,
  \label{wind}
  \ee
where $v_f = \tanh \eta_f$ is the velocity of the wind, $\gamma_f=1/\sqrt{1-v_f^2}$,
and $\theta$
is the angle between the direction of motion of the hard parton
and the direction of the wind.
$\hat q_0$ is the value of $\hat q$ in the absence of a wind.
The result (\ref{wind}) for the dependence of $\hat q$ on collective flow
is valid in QCD and in $\NN=4$ SYM and in
the quark-gluon plasma of any other
gauge theory, since its derivation (see Section 4)
relies only on properties of Lorentz transformations.
If we crudely guess that head winds
are as likely as tail winds, and that the typical transverse wind velocity
seen by a high energy parton is about half the speed of light, $\hat q$ is
increased relative to that in (\ref{qhat}) by a factor of 1.16.
A credible evaluation of the consequences of (\ref{wind}) for the time-averaged $\hat q$
extracted from data will, however, require careful modelling of the geometry of
the collision and the time-development of the collective flow velocity, as
in Refs.~\cite{Renk:2005ta,BaierPrivate,Renk:2006sx}.

The weak-coupling QCD estimate of $\hat q$ given in (\ref{2.16})
is, when evaluated with $\alpha_s=1/2$, smaller than that in (\ref{numV}) by about a factor of 5.
This means that in order to make the weak-coupling estimate consistent
with data from RHIC, we would have to choose $\alpha_s > 1$, certainly beyond
weak-coupling.
The strong coupling calculation of $\hat q$ in $\NN=4$ SYM
is certainly in better agreement with data from RHIC than the weak-coupling calculation in QCD.
The obvious question, then, is how much the strong coupling result will change
as one modifies the theory from $\NN=4$ SYM towards, and ultimately to, QCD.
We shall describe the current state of our ability to answer this question in Section 6.5.

The BDMPS description of radiative parton energy loss, with the
nonperturbative physics of the medium entering through the jet
quenching parameter $\hat q$, is appropriate in the high parton
energy limit.  It is only by comparison to data that we can learn
whether the jets being quenched at RHIC are sufficiently energetic
for their energy loss to be described well by this formalism.  To
date this comparison has been broadly successful, albeit with
$\hat q$ seen as a free parameter. If we understood the QCD
prediction for $\hat q$ as a function of temperature in a strongly
interacting quark-gluon plasma even at the factor of two level,
this would make this comparison more stringent. And, it would turn
$\hat q$ into a ``thermometer'' with a calibration error of only a
factor of $2^{1/3}$, which would be an exceptionally valuable
addition since one of the biggest current weaknesses in our
understanding of RHIC phenomena is that we do not have an
experimental measure of the temperature at, say, a time of 1 fm
after the collision.   In the next two subsections, we first frame
the questions that need to be thought through if it is to  be
possible to go from our calculation of $\hat q$ in $\NN=4$ SYM to
a factor of two understanding of $\hat q$ in QCD, and then review
and extend the calculations of $\hat q$ in various other gauge
theories, yielding a conjecture for how to estimate $\hat q$ in
QCD at a semi-quantitative level, still with many caveats.
Our conjecture is that ${\hat q}_{QCD}/{\hat q}_{SYM}$ is of
the order of $\sqrt{47.5/120}\simeq 0.63$, namely the square root of the ratio of the
numbers of degrees of freedom in the two theories.

\subsection{$\NN=4$ SYM versus QCD}

We found in Section 6.3  that $\hat q$ calculated in $\NN=4$ SYM
theory is close to the value extracted from  RHIC data. Given that
RHIC is probing the quark-gluon plasma of QCD, is this agreement
meaningful or accidental?  In what respects can the strongly interacting
plasma of $\NN=4$ SYM theory give a reasonable
description of the quark-gluon plasma in QCD? After all, at a
microscopic level $\NN=4$ SYM is very different from
QCD:
\begin{itemize}
 \item The theory is conformal, supersymmetric and contains additional global
symmetry. The coupling does not run and there is no confinement.
 \item No dynamic quarks, no chiral symmetry and no chiral symmetry breaking.
 \item Additional scalar and fermionic fields in the adjoint representation.
\end{itemize}
These features of course make the vacuum sectors of the two theories
very different. However, if the quark-gluon plasma in QCD is
strongly interacting, as indicated by data from RHIC,
then one may ask whether the macroscopic properties, both
thermodynamic and dynamic, of quark-gluon plasma
at sufficiently strong coupling may be insensitive to
differences between the theories which seem stark in vacuum.
It is often the case that macroscopic properties of a
sufficiently excited many-body system are not sensitive to the
detailed underlying dynamics, with systems within the same
universality class exhibiting similar phenomena.
We are used to the idea that all metals, or all liquids, or
all ferromagnets have common, defining, characteristics
even though they may differ very significantly at a microscopic
level.  What we are asking here is what are the defining
commonalities of quark-gluon plasmas in different theories, and
in what instances do these commonalities
allow qualitative or semi-quantitative lessons learned
about the quark-gluon plasma of one theory to be applied
to that of another.

Returning to the differences between
QCD and $\NN=4$ SYM, many are obviously irrelevant to a
comparison between strongly-coupled plasmas
in the two theories.
After all, supersymmetry is explicitly and badly broken at high
temperature and, above $T_c$ in QCD, there is no confinement
and no chiral condensate.  Furthermore, in a strongly interacting
liquid there are, by definition, no well-defined, long-lived
quasiparticles anyway, making it plausible that observables or
ratios of observables
can be found which are insensitive to the differences between
microscopic degrees of freedom and interactions.
Because $\NN=4$ SYM is a conformal theory whereas QCD is not,
$\NN=4$ SYM cannot be used to describe QCD at or below its phase transition
at $T\sim T_c$, and cannot be used to describe QCD at asymptotically
high temperatures.  However, there are a variety of indications
from lattice QCD calculations (enumerated
below) that QCD thermodynamics is reasonably
well approximated as conformal in a range of temperatures
from about $2T_c$ up to some higher temperature not currently
determined.
It is not currently known whether the quark-gluon plasma
of QCD, as explored at RHIC and in lattice QCD calculations,
and that of $\NN=4$ SYM, as explored using AdS/CFT calculations,
are in the same universality class, or even in what sense
this question could be made precise.   However, given the
rapidly increasing list of similarities between the two quark-gluon plasmas,
it does not seem too far-fetched to imagine.  Here is a list of some
of the similarities between the quark-gluon plasmas of the two theories,
notwithstanding the stark differences between their vacua, with thermodynamic
comparisons listed first followed by dynamic comparisons:

\begin{itemize}

\item
Above about $1.2 T_c$, the ratio of the energy density $\varepsilon$ in
2- and 3-flavor QCD to that in
the absence of interactions is close to $T$-independent and takes on the
value of about 0.8~\cite{Karsch:2000ps,Karsch:2001vs,Karsch:2006xs}.
In zero-flavor QCD, this ratio is closer to 0.9~\cite{Boyd:1996bx}.  In $\NN=4$ SYM,
this ratio is $3/4$ in the $\lambda\rightarrow\infty$ limit~\cite{Gubser:1996de}
and is 0.8 for $\lambda\sim 11$~\cite{Gubser:1998nz}.

\item
Above about $2.5 T_c$, the ratio of the pressure $P$ in 2- or 3-flavor QCD to
that in the absence of interactions is also about 0.8.  In   zero-flavor QCD,
this ratio is closer to 0.9~\cite{Boyd:1996bx}.
In $\NN=4$ SYM, this ratio must be the same as that defined via the energy
density, a condition that is satisfied well in QCD.   Note, however, that
for $T\lesssim 2.5 T_c$, the deviation from conformality
parametrized by $\varepsilon-3P$ is significant.
This suggests that the use of a conformal theory like
$\NN=4$ SYM as a model for the quark-gluon plasma may be more
quantitatively reliable for heavy ion collisions at the LHC than at RHIC,
since RHIC is likely exploring temperatures that are less than $2 T_c$ whereas
the LHC can be expected to reach temperatures that are higher by about
a factor of two.

\item
All the results in $\NN=4$ SYM have been obtained in the $N\rightarrow\infty$ limit.
Although corrections are expected to be of order $1/N^2$, they have not been computed.
It is therefore very useful to test how the quantities on the QCD side of these
comparisons change with $N$.  One example of such a test is the calculation
of Ref.~\cite{Bringoltz:2005rr}, which finds that the ratio of the pressure to
its noninteracting value changes very little as $N$ is changed from 3 to 4 to 8.

\item
The square of the speed of sound in the QCD quark gluon plasma is close to $1/3$,
the value for a conformal theory,
for $T\gtrsim 2 T_c$; at $T=1.5 T_c$, it is already
$\simeq 0.27-0.31$~\cite{Boyd:1996bx,AliKhan:2001ek,Aoki:2005vt,Ejiri:2005uv,Karsch:2006xs}.

\item
The screening length defined by the potential between a quark and antiquark
at rest is $0.869/\pi T \simeq 0.28/T$ in $\NN=4$ SYM
in the large $N$ and $\lambda$ limits, as calculated in Section 3 and
as calculated first in Refs.~\cite{Rey:1998bq}.  In QCD the screening
length is not sharply defined, since the potential does not change suddenly
to zero, but operational definitions exist in the literature.
For QCD with zero~\cite{Kaczmarek:2004gv}
and  two~\cite{Kaczmarek:2005ui} flavors,
it is $\sim  0.7/T$ and $\sim 0.5/T$, respectively. QCD and $\NN=4$ SYM are therefore
qualitatively comparable in this regard, with the
quantitative difference between them plausibly reflecting
the larger number of degrees of freedom in $\NN=4$ SYM.

\item
Turning now to dynamic quantities, the shear viscosity in units  of the
entropy density is $1/4\pi$ in $\NN=4$ SYM in the $\lambda\rightarrow\infty$
limit~\cite{Policastro:2001yc}, and is $\simeq 1.25/4\pi$ for $\lambda=6\pi$~\cite{Buchel:2004di}.
Given the degree to which data on the azimuthal anisotropy of RHIC collisions
are well-described by zero-viscosity hydrodynamics, the ratio of the shear viscosity
to the entropy density has been estimated to be comparably small in
the quark-gluon plasma at RHIC~\cite{Teaney:2003kp}.  A quantitative extraction
of $\eta$ from RHIC data requires viscous hydrodynamic calculations, which are currently being
pursued by various groups~\cite{Muronga:2001zk}.

 \item
 In $\NN=4$ SYM, the quark-antiquark screening length scales with velocity according
 to $L_s(v)\sim L_s(0)/\sqrt{\gamma}$, as discussed in
 Ref.~\cite{Liu:2006nn} and Sections 3.1 and 5 above.  Comparison of this
predicted
scaling to QCD awaits data from RHIC on the $p_T$-dependence
 of $J/\Psi$ suppression
 at RHIC at $p_T> 5$~GeV and on the pattern of $p_T$-dependence
 of $J/\Psi$ and Upsilon suppression at the LHC.

\item
As we have discussed in Section 6.3, the jet quenching parameter $\hat q$ in
$\NN=4$ SYM is close to the value extracted from RHIC data~\cite{Liu:2006ug}.

\end{itemize}

To understand whether the above similarities are meaningful, one
avenue is to study strongly interacting quark-gluon plasmas in other
non-Abelian gauge theories with dual gravity descriptions and see whether
a general picture emerges.  In the case of the ratio of
the shear viscosity to the entropy density, it was indeed found that this
ratio is the same in a broad class of gauge theories~\cite{Kovtun:2003wp,Buchel:2003tz}.
A necessary (but not sufficient) condition for this striking lack of dependence
on the nature of the microscopic theory  is
that the dependence on the number of degrees of freedom cancels
in the dimensionless ratio.  In thinking about how the value of $\hat q$, a dimensionful quantity,
may change in going from $\NN=4$ SYM to QCD, it seems to us that the
two most pressing questions are how $\hat q$ depends on the number of
degrees of freedom and on the fact that QCD includes  fundamentals whereas
all the degrees of freedom in $\NN=4$ SYM are adjoints.  These seem to
have greater potential to change $\hat q$ significantly than do $1/\sqrt{\lambda}$
corrections, $1/N^2$ corrections or corrections due to the deviation away from
conformality.  We cannot currently address the effect of fundamentals.
In Section 6.5, we consider how $\hat q$ depends on the number of degrees of
freedom.

\subsection{The jet quenching parameter and degrees of freedom}

One qualitative feature of Eq.~(\ref{qhat}) is that at strong coupling
$\hat q$ is proportional to $\sqrt{\lambda}$, not to the number of
degrees of freedom $\sim N^2$.  This means that at strong coupling,
$\hat q$ cannot be thought of as ``measuring'' either the entropy
density $s$ or what is sometimes described as a ``gluon number
density'' or $\varepsilon^{3/4}$ as had been
expected~\cite{jetquenchrev}, since both $s$ and the energy
density $\varepsilon$ are proportional to $N^2 \lambda^0$.
Whereas the ratio of the shear viscosity to $s$ turns out to
be universal in theories with gravity duals, the ratio $\hat q/s$ vanishes
in the large-$N$ limit.  Even though (\ref{qhat}) upends
prior intuition on this point, it nevertheless seems that $\hat q$ should have some
straightforward dependence on the number of degrees of freedom in
the theory.  We shall now try to make this intuition precise.

We first examine how $\hat q$ of different conformal field
theories with a type IIB supergravity dual compare to each other,
using that of $\NN=4$ SYM as a reference point. The ten-dimensonal
metric dual to a conformal field theory at zero temperature can
be written in the form
 \be \label{warp}
 ds^2_{10} = \Om^2 (y) R^2 \le(ds^2_{AdS_5} + ds^2_{M_5} (y) \ri)\ ,
 \ee
where $R$ is the curvature radius of AdS$_5$ and the metric for
AdS$_5$ inside the parenthesis is normalized to have curvature
radius unity. $ds^2_{M_5}$ is the metric of an internal
five-dimensional manifold. The warp factor $\Om^2 (y)$ depends
only on the coordinates $y$ of the internal manifold.\footnote{The full
supergravity solution also involves a self-dual five-form, which
we will normalize to have the flux of $N$ D3-branes, and possibly
three-forms (when $\Om$ is nontrivial).} To put the theory at
finite temperature, one replaces $ds^2_{AdS_5}$ by the metric of
(\ref{3.3}) of an AdS-Schwarzschild black
hole. The computation of $\hat q$ is identical to what
we have done before and we find that (cf.~(\ref{3.33}))
 \be
 \hat{q}
  = \frac{\pi^{3/2} \Gamma\left(\frac{3}{4}\right)}{ \Gamma\left(\frac{5}{4}\right)}
    {\Om^2 (y) R^2 \ov \apr} \, T^3 \, .
\label{newqh}
 \ee
We shall compare different theories at fixed values of $N$, $\lam=4 \pi g_s N$
and $T$.  Note that the $\NN=4$ SYM relation $\sqrt{\lambda}=R^2/\apr$ is
modified in this more general context, as we shall see below in (\ref{doeo}).

Let us first consider theories in which the warp factor is
trivial, i.e. $\Om =1$.\footnote{In this case  $ds_{M_5}^2$ is an
Einstein manifold with curvature $R_\alpha{}^\beta = 4
\delta_\alpha^\beta$. If one requires the boundary theory be
supersymmetric, then $M_5$ needs to be a Sasaki-Einstein
manifold.} In addition to $S_5$, which corresponds to $\NN=4$ SYM
theory, an infinite number of examples of such dual pairs are now
known~\cite{gauntlett,ami}, with the boundary conformal field
theories being quiver gauge theories with product gauge groups. In
the simplest example, the manifold $M_5$ is a manifold with the
topology of $S_5$ known as $T^{1,1}$ and the corresponding
boundary theory is the Klebanov-Witten CFT~\cite{klebanov}. In
general,
Type IIB supergravity
equations of motion fix the curvature radius $R$ in terms of the
number $N$ of D3-branes as
  \be \label{doeo}
   R^4 = 4 \pi g_s N \apr^2 {\om_{S_5} \ov \om_{M_5}}\, ,
   %\qquad  \om_{S_5}
% = \pi^3
 \ee
where $\om_{M_5}$ and $\om_{S_5}=\pi^3$ are the volume of $ds_{M_5}^2$
and $ds_{S_5}^2$ respectively. Plugging~(\ref{doeo}) into~(\ref{newqh})
we find that (with $\Om =1$)
 \be
 {\hat q_{CFT} \ov \hat q_{\NN=4}} =
  \sqrt{\om_{S_5} \ov
 \om_{M_5}}  \ . \label{qcom}
 \ee
Noting that the
central charge $a$ of a CFT can be written
as~\cite{skenderis,gubser}
 \be
{a_{CFT} \ov a_{\NN=4}} = {\om_{S_5} \ov
 \om_{M_5}}\ ,
 \ee
 we can rewrite (\ref{qcom}) as
 \be
 {\hat q_{CFT} \ov \hat q_{\NN=4}} = \sqrt{a_{CFT} \ov a_{\NN=4}} \ . \label{cqa}
 \ee
For example, in
the Klebanov-Witten theory~\cite{klebanov},
 \be
 {\hat q_{KW} \ov \hat q_{\NN=4}} = \sqrt{27 \ov 16} \ .
 \ee
As another example, note that if you start with $\NN=4$ SYM, described
by Type IIB string theory on AdS$_5\times S_5$, and orbifold the $S_5$ by
$Z_2$, the central charge of the CFT doubles and $\hat q$
increases by a factor of $\sqrt{2}$.
To understand the implications of (\ref{qcom}), recall that
the entropy density of a CFT is related  to its central charge
such that
\be
\frac{s_{CFT}}{s_{\NN=4}}={a_{CFT} \ov a_{\NN=4}}\ ,
\ee
making it clear that the central charge counts the number of
thermodynamic degrees of freedom in the theory. We conclude that
in any conformal theory with a gravity dual  (\ref{warp}) with $\Om=1$,
 \be
 {\hat q_{CFT} \ov \hat q_{\NN=4}} = \sqrt{s_{CFT} \ov s_{\NN=4}} \ . \label{qhatversusentropy}
 \ee
Note that even though $\hat q \propto \sqrt{\lambda} N^0$ and $s\propto N^2 \lambda^0$,
these factors cancel in the ratios on the left and right hand sides of (\ref{qhatversusentropy}).
This equation should be read as saying that, in the relevant class of theories,
$\hat q/\sqrt{\lambda}$ is proportional to $\sqrt{s/N^2}$.

When the warp factor $\Om (y)$ in~(\ref{warp}) is nontrivial, the
value of $\hat q$ depends on where in  the internal space $M_5$
we put the probe brane. In these theories, different types of
quarks, which correspond to putting branes at different locations
in the internal manifold, have different values of $\hat q$, as was first pointed out in
Ref.~\cite{Vazquez-Poritz:2006ba}.  An
example of~(\ref{warp}) with nontrivial $\Om$ is the Pilch-Warner
geometry~\cite{pilch} which is dual to the $\NN=1$ superconformal
CFT of Leigh and Strassler~\cite{leigh}. In this case the computation of the normalization
condition and central charge are more complicated (see,
e.g., Ref.~\cite{mgauntlett}), but the result is again simple to state:
 \be
 {\hat q_{LS}  (y) \ov \hat q_{\NN=4}} = \Om^2 (y) \sqrt{a_{LS} \ov a_{\NN=4}} \ , \label{cqa2}
 \ee
with the ratio of $\hat q$'s again proportional to the square root of the
ratio of central charges.
For the Pilch-Warner geometry,\footnote{Note that our
normalization~(\ref{warp}) is different from that in~\cite{pilch}
and~\cite{mgauntlett}.} ${a_{LS} \ov a_{\NN=4}} = {27 \ov 32}$ and
 \be
 \Om^2 (y) = \le({3-\cos 2 \th \ov 2}\ri)^\ha, \qquad 0 \leq \th \leq {\pi \ov 2}\,,
 \ee
where $\theta$ is an angle specifying a location
within the internal $M_5$.

Thus for {\it any} two CFTs with a holographic dual, the ratio
of their jet quenching parameters $\hat q$ is proportional to the square root of the ratio of
their central charges, and
hence to the square root of the ratio of their
number of degrees of freedom.  ``Proportional to''  becomes ``equal to'' if $\Omega=1$.
In particular, if two CFTs are
connected by a renormalization group flow then $\hat q$ for the UV
theory is always larger than that of the IR theory.

Since QCD is not a CFT, we cannot directly apply the result we have just
derived to QCD.  However, to the extent that the quark-gluon plasma of QCD
is approximately conformal, as we have discussed in Section 6.4, perhaps our
result for comparing CFTs can be used as a guide.  In doing so it seems fair
to set $\Omega=1$ since there is no indication that  if QCD with $N_f$ massless flavors
of quarks had a holographic dual, there would be quarks with differing physics
corresponding to branes at differing locations in an internal manifold.
So, we conjecture that
\be
\frac{ \hat q_{QCD}}{\hat q_{\NN=4}} \sim \sqrt{\frac{ s_{QCD}}{ s_{\NN=4 }}} = \sqrt{\frac{47.5}{120}}
\simeq 0.63
\label{conjecture}
\ee
is a good estimate of the effect of the difference between the number of degrees
of freedom in the two theories on $\hat q$.   We have used $N=3$ in both theories, and
have used $N_f=3$ in QCD.\footnote{In QCD, the gluons contribute $2 (N^2-1)$ and
the $N_f$ flavors of quarks contribute $\frac{7}{8} 4 N N_f$.  In $\NN=4$ SYM,
the gauge bosons, fermions, and scalars contribute $2(N^2-1)$, $\frac{7}{8}8 (N^2-1)$
and $6(N^2-1)$, respectively.}
And, we make the comparison with $\lambda$
chosen in the $\NN=4$ SYM theory
such that the ratio of $s$ to its value in a noninteracting theory is the
same as that in the QCD quark-gluon plasma.  It would be good to ask
how (\ref{conjecture}) is affected by the fact that some of the degrees of freedom
in QCD are fundamentals. Unfortunately, we do not currently have any examples
of calculations of $\hat q$ in theories with fundamentals among the degrees of
freedom of the strongly interacting plasma.

Next, we ask how $\hat q$ is affected by deviations from conformality.
In a nonconformal theory with a dual gravity description,
the bulk
metric in the string frame can generically be written in the form
 \be \label{mp1}
 ds^2 = g(r,y) \left[-(1-f(r,y)) dt^2 + d\vec x^2 )\right] + {dr^2 \ov h (r,y)}  +
 ds^2_{M_5} (y,r)\ ,
 \ee
where $y$ again denotes coordinates of the internal manifold. The
corresponding $\hat q$ can be written
as~\cite{Buchel:2006bv,Armesto:2006zv}
 \be \label{moreq}
 \hat q ={1 \ov \pi \apr} \le( \int_{r_0}^\infty {dr \ov \sqrt{f h
 g^3}} \ri)^{-1}\ .
 \ee
It appears hard to extract a general story from~(\ref{moreq})
directly, without studying more examples.  We
can write $\hat q$ in the form
 \be
 \hat q = \sqrt{a(\lam, \frac{\mu}{T})}\, \sqrt{\lambda}\, T^3
\label{adefn}
 \ee
with $\mu$ some mass parameter(s) of the nonconformal theory.
Motivated from our
discussion of conformal theories it is tempting to speculate that
$a(\lam, \frac{\mu}{T})$ can be considered as a measure of the number of
degrees of freedom of a theory at an energy scale $T$.
The jet quenching parameter
$\hat q$ for the nonconformal cascading gauge
theories of Ref.~\cite{Klebanov:2000hb} is known in the high temperature
limit and the result is consistent with the hypothesis that
$a(\lam,\frac{\mu}{T})$
decreases with renormalization group flow~\cite{Buchel:2006bv}. It would also be interesting to
compute $\hat q$ for the geometry discussed in~\cite{FGPW} to see
whether the function $a(\lam, \frac{\mu}{T})$ decreases monotonically
with renormalization group flow. In other words,
the function $a(\lam,\frac{\mu}{T})$ defined by the jet quenching parameter
is a candidate resolution of the long-standing challenge to find
a four dimensional analogue
of the $c-$function of two dimensional conformal field theory.

We can also ask seek to evaluate how much $\hat q$ is affected
if the theory is ``as nonconformal'' as the quark-gluon plasma of
QCD is at a few times its $T_c$.
There is no one prescription for quantifying nonconformality. However,
the analysis of the
cascading gauge theories of
Ref.~\cite{Klebanov:2000hb} provides a nice example, as
the effect of the nonconformality
on $\hat q$ can be written~\cite{Buchel:2006bv}
\be
\frac{{\hat q}_{\rm cascading}}{{\hat q}_{KW}}
=\left( 1 - 3.12\left(\frac{1}{3}-v_s^2\right)\right)\,,
\label{cascadingqhat}
\ee
where $v_s$ is the speed of sound.  For $v_s^2$ in the range 0.27-0.31, as in
QCD at $T=1.5 T_c$~\cite{Karsch:2006xs}, the effect of the deviation from
nonconformality on $\hat q$ ranges from 6\% to 18\%.

We have obtained one nontrivial check in a nonconformal theory of
our conjecture that $a$, defined from the jet quenching parameter
via (\ref{adefn}), is a measure of the number of degrees of
freedom, as required if our specific conjecture (\ref{conjecture})
is to hold. Consider $\hat q$ for $(p+1)$-dimensional
super-Yang-Mills theories (with 16 supercharges) living at the
boundary of the geometry describing a large number of non-extremal
black Dp-branes~\cite{Itzhaki:1998dd}. We will restrict to $p <
5$. The case $p=3$ is $\NN=4$ SYM;  the cases $p=2$ and $p=4$
correspond to nonconformal theories in $2+1$- and
$4+1$-dimensions. The metric dual to these theories can be written
as
 \be \label{omet}
ds^2 = \apr { (d_p \tilde\lam z^{3-p})^{1 \ov 5-p}  \ov z^2} \le(-{\tilde f} dt^2
+ ds_p^2
  + \le({2 \ov 5-p}\ri)^2 {dz^2 \ov {\tilde f}} + z^2 d \Om_{8-p}^2
  \ri)\,,
 \ee
where\footnote{Here we are following standard
string theory convention and normalizing the gauge coupling
constant as $g_{M}^2 = (2 \pi)^{p-2} g_s \apr^{3-p \ov 2}$. For $p=3$,
the gauge coupling in the standard field theoretical
convention which we have used
elsewhere is $g^2_{YM}=2 g^2_{M}$, meaning
that $\tilde\lambda=\frac{1}{2}\lambda$.  For $p=3$, the  relation between
these coordinates and those in (\ref{3.3}) is $z=R^2/r$ with
$R^2 = \alpha' \sqrt{\lambda}$, meaning that $z_0=R^2/r_0$ and
$\tilde  f=R^2 f/r^2$.}
  \be
\tilde\lam = g_{M}^2 N, \qquad {\tilde f} = 1- \le({z \ov z_0} \ri)^{14-2p \ov
5-p } , \qquad d_p = 2^{7-2p} \pi^{9-3p \ov 2} \Gamma
 \le( 7 - p \ov 2\ri) \ .
  \ee
  $ds_p^2$ is the metric for flat
$p$-dimensional Euclidean space and $d \Om^2_{d}$ is the metric
for a $d$-dimensional sphere. Note that $g_{M}^2$, and hence $\lambda$, have mass
dimension $3-p$. The horizon is at $z = z_0$ and the boundary is
at $z=0$. The temperature can be obtained as
 \be
T = {7-p \ov 5-p} {1 \ov 2 \pi z_0} \ .
 \ee
The energy density $\varepsilon$ and entropy density $s$ of the
systems can be written as
 \be
 \label{enerdens}
 \varepsilon =  N^2 \, { 9-p \ov 14-2p} b_p  \lam_{\rm eff}^{p-3 \ov 5-p} (T) \,  T^{p+1}\ ,
\ee
 \bea
 s = N^2 \, b_p \lam_{\rm eff}^{p-3 \ov 5-p} (T) \, T^p \ ,
  \label{en}
  \eea
where
 \be
\lam_{\rm eff} (T)= \tilde\lam T^{p-3}, \qquad b_p = \le({2^{16-3p}
\pi^{13-3p \ov 2} \Gamma({7-p \ov 2}) \ov
 (7-p)^{7-p}}\ri)^{2 \ov 5-p} \ .
 \ee
$\lam_{\rm eff} (T)$ is the effective dimensionless coupling at
temperature $T$. Note that equation (\ref{en}) indicates that the
quantity $b_p \lam_{\rm eff}^{p-3 \ov 5-p} (T)$
characterizes the number of degrees of freedom at temperature
$T$.
By following the procedure
of Ref.~\cite{Liu:2006ug} or Section~\ref{sec3}  above, or by
simply applying~(\ref{moreq}) to~(\ref{omet}), we find that
 \be \label{o}
 \hat q =  {8 \pi^\ha \Gamma ({6-p \ov 7-p}) \ov \Gamma
({5-p \ov 14-2p} )} b_p^\ha
  \lam_{\rm eff}^{\frac{1}{2}\frac{p-3}{5-p}} (T) \, \sqrt{\lam_{\rm eff}(T)}\, T^3
  \ .
 \ee
We see from (\ref{o}) that the quantity $a(\lambda_{\rm eff})$
defined as in (\ref{adefn}) has the same dependence on
$\lambda_{\rm eff}$ that the entropy density (\ref{en}) has.

The calculation of $\hat q$ in  the  nonconformal $p=2$ and $p=4$
Dp-brane theories supports our conjecture
that $a$, defined from $\hat q$ via (\ref{adefn}),
measures the number of degrees of freedom at
temperature $T$.\footnote{
A relation analogous to
that between $\hat q$ and the number of degrees of freedom
may also be valid for the longitudinal drag coefficient $\mu_L$ of
Refs.~\cite{Herzog:2006gh,Gubser:2006bz}, defined in
(\ref{lang1}) and given by (\ref{enrl}) in $\NN=4$ SYM.
%, for theories with a supergravity dual.
In particular, following (\ref{adefn}) we can introduce
 \be
 \mu_L  = {1 \ov m} \sqrt{\tilde a (\lam, {\mu \ov T})} \sqrt{\lam} T^2
 \ee
and $\tilde a (\lam, {\mu \ov T})$ and $a (\lam, {\mu \ov T})$
could coincide up to some numerical constant. This holds for all
the CFTs with a IIB supergravity dual, meaning that in
all these theories $\tilde a$ is proportional to the
number of degrees of freedom just as $a$ is. This is also the case for the
$Dp$-brane, where we find that
 \be
 \mu_L  =  {1 \ov m}\le({5-p \ov 2}\ri)^{7-p \ov 5-p} b_p^\ha
  \lam_{\rm eff}^{\frac{1}{2}\frac{p-3}{5-p}} (T) \, \sqrt{\lam_{\rm
eff}(T)}\,
  T^2
  \ .
\ee
}
This conjecture can be further
tested by computing $\hat q$
in other nonconformal theories, like for example the $\NN=2^*$
theory of Refs.~\cite{Buchel:2003ah,Buchel:2004hw}.
This would also allow us to test our conjecture that
the function $a(\lam,\frac{\mu}{T})$
decreases
under renormalization group flow, and it would
test the conclusion indicated by (\ref{cascadingqhat}) that
deviations from conformality  with a magnitude comparable
to those in the QCD quark-gluon plasma
do not change $\hat q$ much.
%By studying the effect on $\hat q$
%of introducing a mass for some of the degrees of freedom in $\NN=4$ SYM and
%then dialing that mass from below $T$ to above $T$, we could test
%(\ref{conjecture}) in a nonconformal context.

If our conjecture relating $\hat q$ to the number of degrees
of freedom, valid for any conformal theory,
survives being further tested via the calculation
of $\hat q$ in more examples of nonconformal theories with gravity duals
and, even better, in theories with fundamentals
we will then have a new example of a common feature of strongly
interacting quark-gluon plasmas.
Furthermore,
the conjecture (\ref{conjecture})
together with our
result (\ref{qhat}) for $\hat q_{SYM}$
will then provide a
theoretical prediction for the jet quenching
parameter $\hat q$ in the strongly interacting quark-gluon plasma of QCD.
In order to make quantitative contact with data,
we will further need to model the
effects of collective flow on $\hat q$, using the result (\ref{wind})
which is valid in QCD.
On the experimental front, we can look forward to
studies of jet quenching being extended to
higher and higher transverse momentum jets as RHIC
runs at higher luminosities and
as the LHC comes  on line.  Furthermore, particularly at
the LHC new observables
sensitive to parton energy loss will be developed.
Making these assumptions about
(near) future theoretical and experimental developments,
we can look forward
to a stringent comparison between
experimental and theoretical
determinations of the jet quenching parameter
in the strongly interacting quark-gluon
plasma of QCD.

%%%%%%%%%%%%%%%%%%%%%%%%%%%%%%%%%%%%%%%%%%%%%%%%%%%%%%%%%%%%%%%%%%%%%%%%%%%%%%%%
\medskip
\section*{Note added}
\medskip

Shortly after the completion of our work, two
papers~\cite{Gubser:2006nz,calsa} appeared which calculated
the mean squared momentum transfer
$\kappa_T (p)$ in (\ref{lang4}) for a heavy external quark of mass $M$
moving with a velocity $v = \tanh \eta$ through the $\NN = 4$ plasma. 
These calculations are valid for the kinematic
regime~\cite{Gubser:2006nz,calsa}
 \begin{equation} \label{valid}
 \sqrt{\cosh\eta} < {M \ov \sqrt{\lam} T} = \Lam \ ,
 \end{equation}
meaning that they can be extended to $\eta\rightarrow\infty$ only
if the $M\rightarrow \infty$ limit has been taken first.
In the regime (\ref{valid}), one finds
 \be \label{TCG}
 \kappa_T = \sqrt{\cosh \eta} \,\sqrt{\lam} \,\pi\, T^3 \ .
 \ee
 In contrast, as discussed in Section
\ref{sec3.3} our calculation of $\hat q$ {\it requires}
$\sqrt{\cosh\eta} > \Lam$. Since this kinematic region does not
overlap with (\ref{valid}), a direct comparison of $\hat{q}$ and
$\kappa_T$ is difficult. 
However, we agree with 
Refs. ~\cite{Gubser:2006nz,calsa} that it would be  
desirable to have a better understanding of
whether there is a connection between $\hat{q}$ and
$\kappa_T$, and what this connection could be. To illustrate
this open issue, let us make the following remarks.  On the one hand,
it is obvious that $\kappa_T$ cannot play the role of $\hat{q}$
within the BDMPS energy loss formalism (\ref{2.14}), on which 
the definition of $\hat{q}$ is based. 
To see this, note first that  
$\hat{q}$ in (\ref{2.14}) is defined for a quark moving strictly
along the light-cone, $\eta \to \infty$, for which $\kappa_T$ in
(\ref{TCG}) diverges. Then, note that replacing $\hat q$ in (\ref{2.14}) by
a divergent quantity leads to an ill-defined expression.  
On the other hand, the value of $\hat{q}$ in 
(\ref{2.14}) is known to set the scale of the transverse momentum
broadening of the medium-modified gluon 
radiation. (For example, see Fig.~1 of Ref.~\cite{Salgado:2003rv}.)
So, while the value of $\hat{q}$ is known to set the transverse
momentum scale of BDMPS energy loss, a direct connection  
between $\hat{q}$ and $\kappa_T$ may not be straightforward.

We also stress that in the BDMPS formalism, the
quantity $\hat q$ which parameterizes radiative
energy loss of heavy quarks moving at high velocity
is {\it precisely} the same as that for massless quarks,
and is defined in terms of the 
short distance behavior of a strictly light-like Wilson loop (\ref{2.14a}).
One way of
seeing this is to recall from our discussion in Section 2 that the
two adjoint light-like Wilson lines in the Wilson loop which
defines $\hat q$ can be thought of (loosely) as representing the
radiated gluon (in the amplitude and in the conjugate amplitude in
the calculation of gluon emission) which is of course on the
light-cone independent of the mass of the quark. (See Section 2
and references therein for a more complete description.) Although
$\hat q$ itself is independent of the mass of the quark which is
losing energy, the relation between $\hat q$ and $\Delta E$, the
average energy lost, does depend on the quark mass.  The relation
(\ref{2.15a}), obtained from (\ref{2.14}), is only valid for
massless quarks. (For the analogous expressions for massive
quarks, see Ref.~\cite{Dokshitzer:2001zm}.) The BDMPS formalism
which relates the energy lost by massless or massive quarks to
$\hat q$ is only valid to leading order in $1/E$.  At higher
order, i.e. at lower energies, the energy loss will depend on more
properties of the medium than just the single parameter $\hat q$.
The divergent quantity $\kappa_T$ computed in Refs.~\cite{Gubser:2006nz,calsa} 
cannot serve to define the jet
quenching parameter $\hat q$ in the BDMPS radiation spectrum
(\ref{2.14}) for either massless or massive quarks.

Also after our paper appeared, Argyres {\it et al.} presented a
study of  ``upward going'' space-like string
configurations~\cite{Argyres:2006yz}. 
In the notation of our paper, these are found as follows. First, extend the
AdS$_5$
spacetime above the (no longer appropriately named) boundary
D3-brane at $r=\Lambda r_0$, where the quark that defines  the
Wilson loop is located.  Then, solve (\ref{3.22}) with the sign of
$y'$ at $\sigma=\pm \ell/2$ chosen so that the solution
$y(\sigma)$ to (\ref{3.22}) begins with $y(-\ell/2)=\Lambda$, then
{\it ascends} to a turning point at $y=y_m$ (with $y_m$ defined by
(\ref{3.23}) taking on a value which is just below
$\sqrt{\cosh\eta}$ for small $\ell$ and hence small $q$), and then
descends back down to $y(\ell/2)=\Lambda$.  We note that in the limit
in which $\sqrt{\cosh\eta} \to \infty$ at fixed $\Lambda$,  
the turning point up to which this solution ascends is
infinitely far above the ``boundary'' at  $\Lambda$.

Hence, according to the standard
IR/UV connection~\cite{IRUV}, these strings are probing physics at
length scales infinitely shorter than the thickness of a Wilson
line, in other words infinitely far to the ultraviolet of what is
normally considered to be the ultraviolet cutoff in the field
theory. It remains to be seen what field theory interpretation can
be given to these upward going strings, but they are certainly not
relevant to the evaluation of thermal expectation values of Wilson
loops, which are located on D3-branes which {\it bound} the AdS$_5$
spacetime. 

The  results of Ref.~\cite{Argyres:2006yz} themselves confirm the
conclusion, reached above via the IR/UV connection, that the
upward-going strings are not related to the light-like Wilson loop
which arises in the physics of deep inelastic scattering and
radiative energy loss in QCD and which we have calculated in
$\NN=4$ SYM. The action of the upward-going strings of
Ref.~\cite{Argyres:2006yz}  is the same as that of the
configuration (\ref{3.24}), namely $S= i \sqrt{\lambda} T L
L^-/2\sqrt{2}$, and is linear in $L$, the tranverse extent of the
Wilson loop.  If the expectation value of the light-like Wilson
loop  were to have this behavior, it would yield a photoabsorption
probability (\ref{2.5}) (for the thought-experiment of deep
inelastic scattering off quark-gluon plasma) which fell at large
virtuality only like $1/Q$, rather than the standard $1/Q^2$. [See
our discussion of Eqs. (\ref{2.4}) and (\ref{2.5}).] Furthermore,
if we subtract $S_0$ from the action $S$, with the finite $S_0$
given by (\ref{aa.18}) for any choice of $p$, this yields a
negative real $L$-independent term in the exponent of $\langle
W({\cal C}_{\rm light-like})\rangle$, making the photoabsorption
probability nonzero in the $Q\rightarrow \infty$ limit, meaning
that a dipole of zero size would have a nonzero absorption
probability.

%%%%%%%%%%%%%%%%%%%%%%%%%%%%%%%%%%%%%%%%%%%%%%%%%%%%%%%%%%%%%%%%%%%%%%%%%%%%%%%%
\medskip
\section*{Acknowledgments}
\medskip

We are grateful to  P. Argyres, N. Armesto, R. Baier, A. Buchel, J. Casalderrey-Solana,
D. Freedman, J. Friess,
S. Gubser, A. Guijosa, A. Kovner, P. Kovtun, E. Laermann, M. Laine, J. Maldacena,
H. Meyer, G. Moore, G. Michalogiorgakis, T. Renk, K. Sfetsos and B.~G.~Zakharov
for discussions.
HL is supported in part by the A.~P.~Sloan Foundation and the U.S.
Department of Energy (DOE) OJI program.  This Research was also
supported in part by the DOE cooperative research agreement
\#DF-FC02-94ER40818.

%%%%%%%%%%%%%%%%%%%%%%%%%%%%%%%%%%%%%%%%%%%%%%%%%%%%%%%%%%%%%%%%%%%%%%%%%%%%%%%%

%%%%%%%%%%%%%%%%%%%%%%%%%%%%%%%%%%%%%%%%%%%%%%%%%%%%%%%%%%%%%%%%%%%%%%%%%%%%%%%%
%%%%%   Appendix
%%%%%%%%%%%%%%%%%%%%%%%%%%%%%%%%%%%%%%%%%%%%%%%%%%%%%%%%%%%%%%%%%%%%%%%%%%%%%%%%
\appendix
\section{Single-string drag solutions and heavy-light mesons}  %Single string drag solutions as subtraction terms }
\label{appa}

The calculation of the expectation values of time-like
and light-like Wilson loops in Sections 3 and 5 required
the subtraction of terms corresponding to the action of a quark and antiquark
which propagate {\it independently} along the long sides of the Wilson loop, i.e. without
seeing each other.  In this Appendix, we enumerate the different
extremal string world sheets
that are possible (for different values of an integration constant)
given a single quark at rest on the D3-brane at $r=r_0\Lambda$
in the presence of a thermal medium moving with rapidity $\eta$
along the $x_3$-direction.

If $\Lambda > \sqrt{\cosh\eta}$, in addition to the drag solution
of Refs.~\cite{Herzog:2006gh,Gubser:2006bz}
we find solutions in which the string which begins on the D3-brane at $r=r_0\Lambda$
ends on a D3-brane located at $r=r_0 \sqrt{\cosh\eta}$.  Such solutions model mesonic
bound states of a heavy quark and a light quark in which the light quark
drags behind the heavy quark.

We discuss a single quark moving along the $x_3$-direction. The string world sheet of
this quark is of the form
 \be\label{aa.1}
 \tau = t, \qquad \sig = r , \qquad x_3 = x_3 (\tau, \sig)\ .
 \ee
 Calculating the components $g_{\al \beta}$ of the induced metric (\ref{3.10}) within
 this ansatz,
 % \bea
 % g_{\tau \tau} & = & -A  + 2 B \dot x_3 + C \dot x_3^2 \\
 % g_{\tau \sig} & = & B x_3' + C \dot x_3 x_3' \\
 % \nonumber
 % g_{\sig \sig} & = & C x_3'^2  + {1 \ov f}
 % \eea
one finds for the world-sheet action (\ref{3.9}) of {\it two} independent strings
 \be
S_0 = {2 \ov 2 \pi \apr} \int_{r_0}^{r_0 \Lam} dr \, \sqrt{{A \ov
f } - {2 B \ov f} \dot x_3 - {C \ov f} \dot x_3^2 + (AC+B^2)
x_3'^2 }\, .
\label{aa.2}
 \ee
 Here, $\dot x_3  \equiv \partial x_3/\partial t$ and $x'_3  \equiv \partial x_3/\partial r$.
 The notational short hands $f$ and $A$, $B$ and $C$ are defined in (\ref{3.4}) and (\ref{3.14}),
 respectively. We seek static profiles $x_3 = x_3 (\sig)$ that
 satisfy the equations of motion from (\ref{aa.2}).
We rescale the variables
 \be
 r= r_0 y\, , \qquad x_3 = {R^2 \ov r_0} z\, ,
 \label{aa.3}
 \ee
 and we introduce the notational short hands
 \be
 H = {y^4 - \cosh^2\eta}\, ,\qquad D = {y^4 - 1}\, .
 \label{aa.4}
 \ee
 The world-sheet action $S_0 $ takes the form ($z' \equiv \p_y z$)
  \be %\label{defep}
    S_0 = K \int_{1}^{\Lam} dy \, \sqrt{{H \ov D } + D z'^2 }\, ,
    \label{aa.5}
 \ee
 with $K=\sqrt{\lambda}T{\cal T}$, where ${\cal T}$ is the extension of the Wilson loop
in the $t$-direction.
The Euler Lagrange equations of motion imply $\frac{\partial {\cal L}}{\partial z'} = q = {\rm const.}$,
which leads to
 \be
 z'^2
 % & = &  q^2 {1 \ov D^2} {H \ov D -
 % q^2} \\
  =  q^2 {1 \ov (y^4 -1)^2} {y^4 - \cosh^2 \eta
 \ov y^4 -1 -   q^2}\, .
 \label{aa.6}
 \ee
We now classify the solutions to these equations of motion which begin from
$y=\Lambda$.

%. Because of the boundary conditions
%(\ref{3.19}), which hold also for the two independent strings, the solutions of (\ref{aa.6}) all start
%from $y=\Lambda$.

%%%%%%%%%%%%%%%%%%%%%%%%%%%%%%%%%%%%%%%%
%\begin{enumerate}

%\item \underline{Non-relativistic regime $\Lam > \sqrt{\cosh \eta}$:}

\subsection{Solutions in the  $\Lam > \sqrt{\cosh\eta}$ regime}

\begin{enumerate}

\item To have a solution stretching between $y=\Lam$ and $y=1$,
the only allowed value for $q$ is
 \be
 q^2 = \sinh^2 \eta\, ,
  \label{aa.7}
 \ee
which leads to
 \be
 z' = {\sinh \eta } {1 \ov y^4 -1}\, .
 \label{aa.8}
 \ee
 Integration of this equation gives the drag solution
 \be
 z(y) = {\rm const.} - \sinh\eta\, \left[ \arctan(y) + {\rm arccoth}(y) \right]\, .
 \label{aa.9}
 \ee
 This is the solution in the rest frame of the quark and its string. In the rest frame of the
 medium, (\ref{aa.9}) is multiplied by the Lorentz contraction factor
 $\sqrt{1-v^2} = 1/\cosh\eta$, and agrees with the solution
of Ref.~\cite{Herzog:2006gh,Gubser:2006bz}.
 % Note that in the solution, $z \propto \log (y-1) \to - \infty$ as $y \to 1$.
For the solution (\ref{aa.8}), the action (\ref{aa.5}) takes the form
 \be
 S_0 = \sqrt{\lambda}\,  {\cal T}\, T\,  \int_{1}^{\Lam} dy \, .
 \label{aa.10}
 \ee
 Since ${\cal T}$ is the proper time in the rest frame of the quark, it is related
 to the ``laboratory'' time in the rest frame of the medium via
 ${\cal T} = t_{\rm lab}/ \cosh \eta$. Hence, (\ref{aa.10}) is the relativistic boost of the
 action of a static quark.  This makes it the natural choice for the
 subtraction term
 in our analysis of
 the quark-antiquark potential in
Sections \ref{sec3c} and 5.  We saw in Fig.~\ref{fig5} that, with this choice of
subtraction, the quark-antiquark potential at small $L$ is independent of the velocity
of the medium, as is desirable on physical grounds.

 \item Solutions of (\ref{aa.6}), which stop at $y_1 = \sqrt{\cosh \eta}$,
 exist for values of $q$ satisfying
 \be
 0 \leq q^2 < \sinh^2 \eta\, .
 \label{aa.11}
 \ee
The actions for such solutions are
 \be
 S_0 =   K \int_{y_1}^\Lam dy \, \sqrt{y^4 - \cosh^2 \eta \ov y^4 -1 -  q^2 }\, .
 \label{aa.12}
 \ee
These solutions describe ``mesons'' made from a heavy quark and a light quark
with the light quark dragging behind the heavy quark.
\end{enumerate}

\subsection{Solutions in the $\sqrt{\cosh \eta} > \Lambda$ regime}

All solutions in this regime stretch between $y=\Lambda$ and the
horizon $y=1$.
In this regime, $y^4-\cosh^2\eta < 0$, since $y \leq \Lambda$. Requiring $z'^2 \geq 0$, one
finds from (\ref{aa.6}) that
\be
    \frac{q^2}{y^4 - \left(1 + q^2\right)} < 0\, .
    \label{aa.13}
\ee
This condition can be realized in two different ways:
\begin{enumerate}
\item Solutions with time-like world sheet.\\
Eq. (\ref{aa.13}) can be satisfied for
\be
 q^2 > 0\, ,\qquad  y^4 - (1+q^2) < 0\, \quad \longrightarrow \quad 1+q^2 > \Lambda^4\, .
 \label{aa.14}
\ee
%
% This implies $1+q^2 > \Lambda^4$.
The action for these solutions of (\ref{aa.6}) is time-like
\be
    S_0 =   K \int_{1}^\Lam dy \, \sqrt{ \cosh^2 \eta -y^4 \ov 1 +   q^2 - y^4 }\, .
        \label{aa.15}
\ee
For the value $q^2 = \sinh^2\eta$, this action coincides with (\ref{aa.10}).

\item Solutions with space-like world sheet.\\
Eq. (\ref{aa.13}) is also satisfied for
\be
 q^2 < 0\, ,\qquad  y^4 - (1+q^2) > 0\, \quad \longrightarrow \quad 1+q^2 < \Lambda^4\, .
 \label{aa.16}
\ee
In this case, $q=i\, p$ is purely imaginary. The equation of motion (\ref{aa.6}) becomes
\be
z'  =   {p \ov y^4 -1} {\sqrt{\cosh^2 \eta -
 y^4}
 \ov \sqrt{y^4 +   p^2 -1 }}\, ,
 \label{aa.17}
\ee
which has well-defined solutions for real values $p \geq 0$.
The action is that of a space-like world sheet
and is imaginary
\be
S_0 =  i K \int_{1}^\Lam dy \, \sqrt{ \cosh^2 \eta -y^4 \ov y^4 +  p^2 - 1 }\, .
\label{aa.18}
\ee
\end{enumerate}
In the calculation in Section \ref{sec3b} of the expectation value of the light-like Wilson loop
that defines the jet quenching parameter, we used (\ref{aa.18})
with $p=0$ as the $L$-independent
subtraction term because it
satisfies (\ref{3.32}).

%%%%%%%%%%%%%%%%%%%%%%%%%%%%%%%%%%%%%%%%%%%%%%%%%%%%%%%%%%%%%%%%%%%%%%%%%%%%%%%%
\section{Time-like Wilson loop with dipole parallel to the wind}
\label{appb}

In Section~\ref{sec4}, we calculated the static $q\bar{q}$-potential for (almost) all dipole
orientations with respect to the wind. While the parameterization (\ref{4.1}) of the
two-dimensional world sheet used there is applicable for arbitrarily small angles $\theta$,
it is not applicable for $\theta = 0$, the case where the dipole is parallel to the wind.
In this appendix, we repeat the calculation of Section~\ref{sec4} with a new parametrization
which works for $\theta=0$ and in fact for $0\leq \theta < \pi/2$ but which does
not work for $\theta=\pi/2$.

We start again from the boosted metric (\ref{3.13}), but in contrast to (\ref{4.1}),
we parametrize the world sheet by
 \be
\tau = t, \qquad \sig = x_3, \qquad x_2={\rm const}, \qquad x_1=x_1(\sigma),\qquad r=r(\sigma)\, .
\label{bb.1}
 \ee
The role of $x_1$ and $x_3$ are interchanged in this parametrization relative to
that in Section \ref{sec4}.

We define dimensionless
coordinates
\be
y=\frac{r}{r_0}\, ,\qquad w=x_1\frac{r_0}{R^2}\, ,\qquad \tilde\sigma=\sigma \frac{r_0}{R^2}\, ,
\qquad l = L\frac{r_0}{R^2}\, ,
\ee
and drop the tilde.
The boundary conditions on $y(\sigma)$ and $w(\sigma)$ then become
\be
    y\left(\pm \frac{l}{2} \cos\theta \right) = \Lambda\, ,\qquad
    w\left(\pm \frac{l}{2} \cos\theta \right) =  \pm  \frac{l}{2} \sin\theta\, .
    \label{bb.2}
\ee
The Nambu-Goto action takes the form
$S({\cal C}) = \sqrt{\lambda} {\cal T} T \int_0^{l/2} d\sigma {\cal L}$,
with the Lagrangian now given by
\begin{equation}
 {\cal L} = \sqrt{(y^4-1)+(y^4-\cosh^2\eta)w'^2 + \frac{y^4-\cosh^2\eta}{y^4-1}y'^2}\ ,
 \label{bb.3}
 \end{equation}
where $y'$ and $w'$ denote derivatives with respect to $\sigma$.
The constants of the motion are
\begin{eqnarray}
    {\cal H} = {\cal L} - y' \frac{\partial {\cal L}}{\partial y'}  - w' \frac{\partial {\cal L}}{\partial w'}
    &=& {y^4-1 \over {\cal L}} \equiv q\,,\nonumber\\
    \frac{\partial {\cal L}}{\partial w'} = \frac{y^4-\cosh^2\eta}{\cal L} z' \equiv p\, .
    \label{bb.4}
\end{eqnarray}
Note that the constants of the motion  $p$ and $q$ here are not the same as in
Section~\ref{sec4}.
The equations of motion can be written in the form
\bea
    q^2 y'^2 &=&\left( \frac{y^4-1}{y^4-\cosh^2\eta}\right)^2 \left[ (y^4-\cosh^2\eta)
    (y^4-1-q^2)-p^2(y^4-1)\right] \, ,
    \label{bb.4a}\\
    q^2 w'^2 &=& p^2\left( \frac{y^4-1}{y^4-\cosh^2\eta}\right)^2 .
    \label{bb.4b}
\eea
Since $y'$ becomes singular at $y^2=\cosh\eta$, the turning point $y_c$
which satisfies
\be
 (y_c^4-\cosh^2\eta)
    (y_c^4-1-q^2)-p^2(y_c^4-1)=0
    \label{bb.5}
    \ee
must also fulfill the condition
\be
\sqrt{\cosh\eta}<y_c<\Lambda\, .
\label{bb.5a}
\ee
The constants $q$ and $p$ are related to the values of $l$ and $\theta$
via
\bea
    \frac{l}{2} \sin\theta &=& \int_{y_c}^\Lambda \frac{dw}{dy}dy\nonumber\\
    &=&  p \int_{y_c}^\infty
        \frac{dy}{\sqrt{\left(y^4-\cosh^2\eta\right)\left(y^4-1-q^2\right) - p^2\left(y^4-1\right)}}
        \, ,\label{bb.6}
\eea
\be
    \frac{l}{2} \cos\theta =  % \int_{y_c}^\infty   \frac{dz}{d\sigma} \frac{d\sigma}{dy} =
                  q \int_{y_c}^\infty   \frac{y^4-\cosh^2\eta}{y^4-1}
                  \frac{dy}{\sqrt{\left(y^4-\cosh^2\eta\right)\left(y^4-1-q^2\right) - p^2\left(y^4-1\right)}}\, .
                  \label{bb.7}
\ee
We now see that the expressions (\ref{bb.5}), (\ref{bb.6}) and (\ref{bb.8}) differ
from their analogues in Section 5 simply by exchanging $q$ and $p$.

Results with $0<\theta<\pi/2$ can be obtained with either the parametrization
of the string world sheet in this Appendix or that in Section 5.    Let us now specialize
to $\theta=0$, the case that cannot be handled with the parametrization of
Section 5.  We see from (\ref{bb.6}) that $\theta=0$ corresponds to $p=0$,
which means that the turning point which solves (\ref{bb.5}) is given simply
by
\be
y_c^4=1+q^2\ .
\label{bb.8}
\ee
Now, the condition (\ref{bb.5a}) becomes a
restriction on the allowed values of $q$:
\be
\sinh^2 \eta  < q^2 < \Lam^4 -1\, .
\label{bb.9}
\ee
In contrast to the situation for any nonzero value of $\theta$, when $\theta=0$
the   constant $q$ cannot be taken to zero: if we were to
choose $q<\sinh \eta$, then $y^4$ would hit $\cosh^2 \eta$,
at which point $y'\rightarrow \infty$ and below which $y'$ is imaginary.
Instead, when we choose
$q>\sinh\eta$ we find a solution in which $y$ reaches a turning point at $y_c$
and safely begins to ascend, never reaching these pathologies.

Notice that there is ``almost'' another choice of $y_c$:  $y_c=\sqrt{\cosh\eta}$ does
satisfy (\ref{bb.5}) with $p=0$, but it just barely fails to satisfy (\ref{bb.5a}).
For arbitrarily small but nonzero values of $p$, however, if $q < \sinh\eta$
there {\it is} a legitimate turning point at a  $y_c$ just above $\sqrt{\cosh\eta}$, and
a solution can be found.  So, it is only for $\theta\equiv 0$ and hence $p\equiv 0$
that there is an inaccessible range of small values of $q$.   If we sit at a $q$
which is less than $\sinh\eta$ and take $p\rightarrow 0$, what happens
to the solution is that the string world sheet develops a cusp at its turning point.
For $p$ small but nonzero, the shape of the function $y(\sigma)$ near its minimum
looks like a very slightly rounded ``V'', with the amount of rounding controlled by $p$.
So, for $p\equiv 0$ there is no solution in this regime of small $q$.   We shall see momentarily
that this regime of $q$ corresponds to a part of the unstable higher energy branch
of solutions.

%
 %%%%%%%%%%%%%%%%%%%%%%%%%%%%%%%%%%%%%%%%%%%%%%
%\begin{figure}[t]
\FIGURE[t]{
%\vskip-0.15in
% \hfill\hskip-0.6in
\includegraphics[scale=0.7,angle=0]{fig-zerodegree.eps}
%\hskip-0.1in
\hfill
\caption{Same as Fig.~\ref{fig5}, but now for dipole orientation $\theta = 0$.
Note that solutions with $\theta=0$ exist only for $q>\sinh\eta$.}
\label{figappb}
}%end{figure}
%%%%%%%%%%%%%%%%%%%%%%%%%%%%%%%%%%%%%%%%%%%%%%%%%
%

Let us return to the case with $\theta=p=0$ keeping $q>\sinh\eta$.
The solution evidently has $w'=0$ throughout.
Furthermore, although there is no reason of symmetry to expect it,
given the wind blowing in the $x_3$ direction, we can see that the solution
$y(\sigma)$ will be $\sigma \rightarrow -\sigma$ symmetric.  This follows
from the fact that $y'^2$ depends only on $y$, not explicitly on $\sigma$,
and from the symmetric boundary condition (\ref{bb.2}).  This means
that the descending half of the $y(\sigma)$ curve and the ascending
half must have the same shape, implying that the turning point at which
$y=y_c$ must be at $\sigma=0$, half way between the boundaries
at which the boundary condition (\ref{bb.2}) fixes $y$.
$q$ can be determined in terms of $l$
from (\ref{bb.7}),  which with $p=0$ is simply
the equation ${l \ov 2} =\int^{l \ov 2}_0 d \sig$
and becomes
 \be \label{peo2}
{l } =%2 q^2 \int_{r_c}^\Lam dr \, {1 \ov \sqrt{(r^4- r_0^4) (r^4 -
%r_c^4)} } =
 {2 q } \int_{y_c}^{\Lam} dy \, {\sqrt{y^4-\cosh^2\eta}  \ov
(y^4-1)\sqrt{(y^4- y_c^4)} }\ ,
 \ee
with $y_c$ given by (\ref{bb.8}).
The action can be written as
 \be \label{ew2}
S(l) =\sqrt{\lambda}T{\cal T} \int_{y_c}^\Lam dy \;{\sqrt{y^4-\cosh^2\eta}  \ov
\sqrt{y^4- y_c^4} } \ ,
 \ee
from which the quark-antiquark potential $E(L)$ can be obtained as in (\ref{3.38})
using the same subtraction $S_0$ given in (\ref{3.39}).   In plotting Fig.~\ref{figappb},
we have used (\ref{peo2}) and (\ref{ew2}) to evaluate $l$ and $E$ for $q>\sinh\eta$.
We see that the stable, lower energy branch of solutions is similar to
those we have obtained previously. For this branch of solutions, the $\theta\rightarrow 0$
limit is smooth.  The inaccessible range of $q$, namely $q<\sinh \eta$,
where as described
above the string world sheet develops a cusp in the $\theta\rightarrow 0$ limit, corresponds
to the ``missing parts'' of the unstable high energy branch of solutions   in
Fig.~\ref{figappb}.

%%%%%%%%%%%%%%%%%%%%%%%%%%%%%%%%%%%%%%%%%%%%%%%%%%%%%%%%%%%%%%%%%%%%%%%%%%%%%%%%


\begin{thebibliography}{99}
%%%%%%%%%%%%%%%%%%%%%%%%%%%%%%%%%%%%%%%%%%%%%%%%%%%%%%%%%%%%%%%%%%%%%%%%%%%%%%%%


\bibitem{RHIC}
  K.~Adcox {\it et al.}  [PHENIX Collaboration],
  %``Formation of dense partonic matter in relativistic nucleus nucleus
  %collisions at RHIC: Experimental evaluation by the PHENIX  collaboration,''
  Nucl.\ Phys.\ A {\bf 757}, 184 (2005)
    [arXiv:nucl-ex/0410003];
  %%CITATION = NUCL-EX 0410003;%%
%
%\cite{Back:2004je}
%\bibitem{Back:2004je}
  B.~B.~Back {\it et al.} [PHOBOS Collaboration],
  %``The PHOBOS perspective on discoveries at RHIC,''
  Nucl.\ Phys.\ A {\bf 757}, 28  (2005)
    [arXiv:nucl-ex/0410022];
  %%CITATION = NUCL-EX 0410022;%%
%
%\cite{Arsene:2004fa}
%\bibitem{Arsene:2004fa}
  I.~Arsene {\it et al.}  [BRAHMS Collaboration];
  %``Quark gluon plasma and color glass condensate at RHIC? The perspective
  %from the BRAHMS experiment,''
  Nucl.\ Phys.\ A {\bf 757}, 1 (2005)
   [arXiv:nucl-ex/0410020];
  %%CITATION = NUCL-EX 0410020;%%
%
%\cite{Adams:2005dq}
%\bibitem{Adams:2005dq}
  J.~Adams {\it et al.}  [STAR Collaboration],
  %``Experimental and theoretical challenges in the search for the quark  gluon
  %plasma: The STAR collaboration's critical assessment of the  evidence from
  %RHIC collisions,''
  Nucl.\ Phys.\ A {\bf 757}, 102 (2005)
 [arXiv:nucl-ex/0501009].
  %%CITATION = NUCL-EX 0501009;%%
%

%\cite{AdS/CFT}
\bibitem{AdS/CFT}
%\cite{Maldacena:1997re}
%\bibitem{Maldacena:1997re}
  J.~M.~Maldacena,
  %``The large N limit of superconformal field theories and supergravity,''
  Adv.\ Theor.\ Math.\ Phys.\  {\bf 2}, 231 (1998)
   [Int.\ J.\ Theor.\ Phys.\  {\bf 38}, 1113 (1999)]   [arXiv:hep-th/9711200];
  %%CITATION = HEP-TH 9711200;%%
%\cite{Witten:1998qj}
  E.~Witten,
  %``Anti-de Sitter space and holography,''
  Adv.\ Theor.\ Math.\ Phys.\  {\bf 2}, 253 (1998)
  [arXiv:hep-th/9802150];
  %%CITATION = HEP-TH 9802150;%%
%\cite{Gubser:1998bc}
%\bibitem{Gubser:1998bc}
  S.~S.~Gubser, I.~R.~Klebanov and A.~M.~Polyakov,
  %``Gauge theory correlators from non-critical string theory,''
  Phys.\ Lett.\ B {\bf 428}, 105 (1998)  [arXiv:hep-th/9802109];
  %%CITATION = HEP-TH 9802109;%%
  %\cite{Aharony:1999ti}
%\bibitem{Aharony:1999ti}
 O.~Aharony, S.~S.~Gubser, J.~M.~Maldacena, H.~Ooguri and Y.~Oz,
  %``Large N field theories, string theory and gravity,''
Phys.\ Rept.\  {\bf 323}, 183 (2000) [arXiv:hep-th/9905111].
  %%CITATION = HEP-TH 9905111;%%



%\cite{Policastro:2001yc}
\bibitem{Policastro:2001yc}
  G.~Policastro, D.~T.~Son and A.~O.~Starinets,
  %``The shear viscosity of strongly coupled N = 4 supersymmetric Yang-Mills
  %plasma,''
  Phys.\ Rev.\ Lett.\  {\bf 87}, 081601 (2001) [arXiv:hep-th/0104066].
  %%CITATION = HEP-TH 0104066;%%


%\cite{Kovtun:2003wp}
\bibitem{Kovtun:2003wp}
  P.~Kovtun, D.~T.~Son and A.~O.~Starinets,
  %``Holography and hydrodynamics: Diffusion on stretched horizons,''
  JHEP {\bf 0310}, 064 (2003)   [arXiv:hep-th/0309213].
  %%CITATION = HEP-TH 0309213;%%

%\cite{Buchel:2003tz}
\bibitem{Buchel:2003tz}
  A.~Buchel and J.~T.~Liu,
  %``Universality of the shear viscosity in supergravity,''
  Phys.\ Rev.\ Lett.\  {\bf 93}, 090602 (2004)  [arXiv:hep-th/0311175].
  %%CITATION = HEP-TH 0311175;%%

%\cite{Buchel:2004hw}
\bibitem{Buchel:2004hw}
  A.~Buchel,
  %``N = 2* hydrodynamics,''
  Nucl.\ Phys.\ B {\bf 708}, 451 (2005)
  [arXiv:hep-th/0406200].
  %%CITATION = HEP-TH 0406200;%%

%\cite{Buchel:2004di}
\bibitem{Buchel:2004di}
A.~Buchel, J.~T.~Liu and A.~O.~Starinets,
%``Coupling constant dependence of the shear viscosity in N = 4
%supersymmetric Yang-Mills theory,''
Nucl.\ Phys.\ B {\bf 707}, 56 (2005) [arXiv:hep-th/0406264].
%%CITATION = HEP-TH 0406264;%%

%\cite{Buchel:2004qq}
\bibitem{Buchel:2004qq}
  A.~Buchel,
  %``On universality of stress-energy tensor correlation functions in
  %supergravity,''
  Phys.\ Lett.\ B {\bf 609}, 392 (2005)
  [arXiv:hep-th/0408095].
  %%CITATION = HEP-TH 0408095;%%


%\cite{Son:2006em}
\bibitem{Son:2006em}
%\cite{Mas:2006dy}
%\bibitem{Mas:2006dy}
  J.~Mas,
  %``Shear viscosity from R-charged AdS black holes,''
  JHEP {\bf 0603}, 016 (2006)  [arXiv:hep-th/0601144];
  %%CITATION = HEP-TH 0601144;%%
  D.~T.~Son and A.~O.~Starinets,
  %``Hydrodynamics of R-charged black holes,''
  JHEP {\bf 0603}, 052 (2006)   [arXiv:hep-th/0601157];
  %%CITATION = HEP-TH 0601157;%%
%\cite{Saremi:2006ep}
%\bibitem{Saremi:2006ep}
  O.~Saremi,
  %``The viscosity bound conjecture and hydrodynamics of M2-brane theory at
  %finite chemical potential,''
  arXiv:hep-th/0601159;
  %%CITATION = HEP-TH 0601159;%%
%\cite{Maeda:2006by}
%\bibitem{Maeda:2006by}
  K.~Maeda, M.~Natsuume and T.~Okamura,
  %``Viscosity of gauge theory plasma with a chemical potential from AdS/CFT,''
  Phys.\ Rev.\ D {\bf 73}, 066013 (2006)  [arXiv:hep-th/0602010].
  %%CITATION = HEP-TH 0602010;%%

%\cite{Benincasa:2006fu}
\bibitem{Benincasa:2006fu}
  P.~Benincasa, A.~Buchel and R.~Naryshkin,
  %``The shear viscosity of gauge theory plasma with chemical potentials,''
  arXiv:hep-th/0610145.
  %%CITATION = HEP-TH 0610145;%%

%\cite{Teaney:2003kp}
\bibitem{Teaney:2003kp}
  D.~Teaney,
  %``Effect of shear viscosity on spectra, elliptic flow, and Hanbury
  %Brown-Twiss radii,''
  Phys.\ Rev.\ C {\bf 68}, 034913 (2003)
   [arXiv:nucl-th/0301099].
  %%CITATION = NUCL-TH 0301099;%%

%\cite{Policastro:2002se}
\bibitem{Policastro:2002se}
  G.~Policastro, D.~T.~Son and A.~O.~Starinets,
  %``From AdS/CFT correspondence to hydrodynamics,''
  JHEP {\bf 0209}, 043 (2002)  [arXiv:hep-th/0205052].
  %%CITATION = HEP-TH 0205052;%%
%\cite{Policastro:2002tn}
%\bibitem{Policastro:2002tn}
G.~Policastro, D.~T.~Son and A.~O.~Starinets,
% ``From AdS/CFT correspondence to hydrodynamics. II: Sound waves,''
JHEP {\bf 0212}, 054 (2002)  [arXiv:hep-th/0210220].
%%CITATION = HEP-TH 0210220;%%


%\cite{Teaney:2006nc}
\bibitem{Teaney:2006nc}
  D.~Teaney,
  %``Finite temperature spectral densities of momentum and R-charge correlators
  %in N = 4 Yang Mills theory,''
  arXiv:hep-ph/0602044;
  %%CITATION = HEP-PH 0602044;%%
%\cite{Kovtun:2006pf}
%\bibitem{Kovtun:2006pf}
  P.~Kovtun and A.~Starinets,
  %``Thermal spectral functions of strongly coupled N = 4 supersymmetric
  %Yang-Mills theory,''
  Phys.\ Rev.\ Lett.\  {\bf 96}, 131601 (2006) [arXiv:hep-th/0602059].
  %%CITATION = HEP-TH 0602059;%%

%\cite{Sin:2004yx}
\bibitem{Sin:2004yx}
  S.~J.~Sin and I.~Zahed,
  %``Holography of radiation and jet quenching,''
  Phys.\ Lett.\ B {\bf 608}, 265 (2005)  [arXiv:hep-th/0407215];
  %\cite{Nastase:2005rp}
%\bibitem{Nastase:2005rp}
  H.~Nastase,
  %``The RHIC fireball as a dual black hole,''
  arXiv:hep-th/0501068;
  %%CITATION = HEP-TH 0501068;%%
  %%CITATION = HEP-TH 0407215;%%
%\cite{Shuryak:2005ia}
%\bibitem{Shuryak:2005ia}
  E.~Shuryak, S.~J.~Sin and I.~Zahed,
  %``A gravity dual of RHIC collisions,''
  arXiv:hep-th/0511199;
  %%CITATION = HEP-TH 0511199;%%
%\cite{Nastase:2006eb}
%\bibitem{Nastase:2006eb}
  H.~Nastase,
  %``More on the RHIC fireball and dual black holes,''
  arXiv:hep-th/0603176;
  %%CITATION = HEP-TH 0603176;%%
%\cite{Lin:2006rf}
%\bibitem{Lin:2006rf}
  S.~Lin and E.~Shuryak,
  %``Toward the AdS/CFT gravity dual for high energy heavy ion collisions,''
  arXiv:hep-ph/0610168;
  %%CITATION = HEP-PH 0610168;%%
%\cite{Friess:2006kw}
%\bibitem{Friess:2006kw}
  J.~J.~Friess, S.~S.~Gubser, G.~Michalogiorgakis and S.~S.~Pufu,
  %``Expanding plasmas and quasinormal modes of anti-de Sitter black holes,''
  arXiv:hep-th/0611005.
  %%CITATION = HEP-TH 0611005;%%

%\cite{Liu:2006ug}
\bibitem{Liu:2006ug}
  H.~Liu, K.~Rajagopal and U.~A.~Wiedemann,
  %``Calculating the jet quenching parameter from AdS/CFT,''
  Phys.\ Rev.\ Lett.\  {\bf 97}, 182301 (2006)
  [arXiv:hep-ph/0605178].
  %%CITATION = HEP-PH 0605178;%%


%\cite{Buchel:2006bv}
\bibitem{Buchel:2006bv}
A.~Buchel,
%``On jet quenching parameters in strongly coupled non-conformal gauge
%theories,''
arXiv:hep-th/0605178.
%%CITATION = HEP-TH 0605178;%%


%\cite{Vazquez-Poritz:2006ba}
\bibitem{Vazquez-Poritz:2006ba}
J.~F.~Vazquez-Poritz,
%``Enhancing the jet quenching parameter from marginal deformations,''
arXiv:hep-th/0605296.
%%CITATION = HEP-TH 0605296;%%

%\cite{Caceres:2006as}
\bibitem{Caceres:2006as}
E.~Caceres and A.~Guijosa,
%``On Drag Forces and Jet Quenching in Strongly Coupled Plasmas,''
arXiv:hep-th/0606134.
%%CITATION = HEP-TH 0606134;%%

%\cite{Lin:2006au}
\bibitem{Lin:2006au}
F.~L.~Lin and T.~Matsuo,
%``Jet Quenching Parameter in Medium with Chemical Potential from AdS/CFT,''\!\(\(65.37688123488302`\ T0\^3\ tau0\)\/z\)
arXiv:hep-th/0606136.
%%CITATION = HEP-TH 0606136;%%

%\cite{Avramis:2006ip}
\bibitem{Avramis:2006ip}
S.~D.~Avramis and K.~Sfetsos,
%``Supergravity and the jet quenching parameter in the presence of R-charge
%densities,''
arXiv:hep-th/0606190.
%%CITATION = HEP-TH 0606190;%%

%\cite{Armesto:2006zv}
\bibitem{Armesto:2006zv}
  N.~Armesto, J.~D.~Edelstein and J.~Mas,
%   ``Jet quenching at finite 't Hooft coupling and chemical potential from
%  AdS/CFT,''
  JHEP {\bf 0609}, 039 (2006)
  [arXiv:hep-ph/0606245].
  %%CITATION = HEP-PH 0606245;%%

%\cite{Nakano:2006js}
\bibitem{Nakano:2006js}
  E.~Nakano, S.~Teraguchi and W.~Y.~Wen,
  %``Drag force, jet quenching, and AdS/QCD,''
  arXiv:hep-ph/0608274.
  %%CITATION = HEP-PH 0608274;%%

%\cite{Herzog:2006gh}
\bibitem{Herzog:2006gh}
  C.~P.~Herzog, A.~Karch, P.~Kovtun, C.~Kozcaz and L.~G.~Yaffe,
  %``Energy loss of a heavy quark moving through N = 4 supersymmetric Yang-Mills
  %plasma,''
  JHEP {\bf 0607}, 013 (2006)
  [arXiv:hep-th/0605158].
  %%CITATION = HEP-TH 0605158;%%

%\cite{Gubser:2006bz}
\bibitem{Gubser:2006bz}
S.~S.~Gubser,
%``Drag force in AdS/CFT,''
arXiv:hep-th/0605182.
%%CITATION = HEP-TH 0605182;%%


%\cite{Casalderrey-Solana:2006rq}
\bibitem{Casalderrey-Solana:2006rq}
  J.~Casalderrey-Solana and D.~Teaney,
  %``Heavy quark diffusion in strongly coupled N = 4 Yang Mills,''
  Phys.\ Rev.\ D {\bf 74}, 085012 (2006)
  [arXiv:hep-ph/0605199].
  %%CITATION = HEP-PH 0605199;%%

\bibitem{Drag1}
%\cite{Herzog:2006se}
%\bibitem{Herzog:2006se}
  C.~P.~Herzog,
  %``Energy loss of heavy quarks from asymptotically AdS geometries,''
  JHEP {\bf 0609}, 032 (2006)
  [arXiv:hep-th/0605191];
  %%CITATION = HEP-TH 0605191;%%
%\cite{Caceres:2006dj}
%\bibitem{Caceres:2006dj}
E.~Caceres and A.~Guijosa,
%``Drag force in charged N = 4 SYM plasma,''
arXiv:hep-th/0605235;
%%CITATION = HEP-TH 0605235;%%
%\cite{Sin:2006yz}
%\bibitem{Sin:2006yz}
S.~J.~Sin and I.~Zahed,
%``Ampere's law and energy loss in AdS/CFT duality,''
arXiv:hep-ph/0606049.
%%CITATION = HEP-PH 0606049;%%
%\cite{Gao:2006se}
%\bibitem{Gao:2006se}
  Y.~h.~Gao, W.~s.~Xu and D.~f.~Zeng,
  %``Wake of color fileds in charged N = 4 SYM plasmas,''
  arXiv:hep-th/0606266;
  %%CITATION = HEP-TH 0606266;%%
%\cite{Matsuo:2006ws}
%\bibitem{Matsuo:2006ws}
  T.~Matsuo, D.~Tomino and W.~Y.~Wen,
  %``Drag force in SYM plasma with B field from AdS/CFT,''
  JHEP {\bf 0610}, 055 (2006)
  [arXiv:hep-th/0607178];
  %%CITATION = HEP-TH 0607178;%%


%\cite{Friess:2006aw}
\bibitem{Friess:2006aw}
  J.~J.~Friess, S.~S.~Gubser and G.~Michalogiorgakis,
  %``Dissipation from a heavy quark moving through N = 4 super-Yang-Mills
  %plasma,''
  JHEP {\bf 0609}, 072 (2006)
  [arXiv:hep-th/0605292];
  %%CITATION = HEP-TH 0605292;%%
%\cite{Friess:2006fk}
%\bibitem{Friess:2006fk}
  J.~J.~Friess, S.~S.~Gubser, G.~Michalogiorgakis and S.~S.~Pufu,
  %``The stress tensor of a quark moving through N = 4 thermal plasma,''
  arXiv:hep-th/0607022.
  %%CITATION = HEP-TH 0607022;%%





%\cite{Rey:1998bq}
 \bibitem{Rey:1998bq}
  S.~J.~Rey, S.~Theisen and J.~T.~Yee,
  %``Wilson-Polyakov loop at finite temperature in large N gauge theory and
  %anti-de Sitter supergravity,''
  Nucl.\ Phys.\ B {\bf 527}, 171 (1998) [arXiv:hep-th/9803135];
  %%CITATION = HEP-TH 9803135;%%
  %\cite{Brandhuber:1998bs}
% \bibitem{Brandhuber:1998bs}
  A.~Brandhuber,
  N.~Itzhaki, J.~Sonnenschein and S.~Yankielowicz,
  %``Wilson loops in the large N limit at finite temperature,''
  Phys.\ Lett.\ B {\bf 434}, 36 (1998) [arXiv:hep-th/9803137];
  %%CITATION = HEP-TH 9803137;%%
%\cite{Sonnenschein:1999if}
%\bibitem{Sonnenschein:1999if}
  J.~Sonnenschein,
  %``What does the string / gauge correspondence teach us about Wilson loops?,''
  arXiv:hep-th/0003032.
  %%CITATION = HEP-TH 0003032;%%

%\cite{Mateos:2006nu}
\bibitem{Mateos:2006nu}
  D.~Mateos, R.~C.~Myers and R.~M.~Thomson,
  %``Holographic phase transitions with fundamental matter,''
  Phys.\ Rev.\ Lett.\  {\bf 97}, 091601 (2006)
  [arXiv:hep-th/0605046].
  %%CITATION = HEP-TH 0605046;%%

%\cite{Peeters:2006iu}
\bibitem{Peeters:2006iu}
K.~Peeters, J.~Sonnenschein and M.~Zamaklar,
%``Holographic melting and related properties of mesons in a quark gluon
%plasma,''
arXiv:hep-th/0606195.
%%CITATION = HEP-TH 0606195;%%

% \cite{Liu:2006nn}
\bibitem{Liu:2006nn}
  H.~Liu, K.~Rajagopal and U.~A.~Wiedemann,
  %``An AdS/CFT calculation of screening in a hot wind,''
  arXiv:hep-ph/0607062.
  %%CITATION = HEP-PH 0607062;%%

%\cite{Chernicoff:2006hi}
\bibitem{Chernicoff:2006hi}
  M.~Chernicoff, J.~A.~Garcia and A.~Guijosa,
  %``The energy of a moving quark-antiquark pair in an N = 4 SYM plasma,''
  JHEP {\bf 0609}, 068 (2006)
  [arXiv:hep-th/0607089].
  %%CITATION = HEP-TH 0607089;%%

%\cite{Caceres:2006ta}
\bibitem{Caceres:2006ta}
  E.~Caceres, M.~Natsuume and T.~Okamura,
  %``Screening length in plasma winds,''
  JHEP {\bf 0610}, 011 (2006)
  [arXiv:hep-th/0607233].
  %%CITATION = HEP-TH 0607233;%%

%\cite{Argyres:2006vs}
\bibitem{Argyres:2006vs}
  P.~C.~Argyres, M.~Edalati and J.~F.~Vazquez-Poritz,
  %``No-drag string configurations for steadily moving quark-antiquark pairs in
  %a thermal bath,''
  arXiv:hep-th/0608118.
  %%CITATION = HEP-TH 0608118;%%

  %\cite{Avramis:2006em}
\bibitem{Avramis:2006em}
  S.~D.~Avramis, K.~Sfetsos and D.~Zoakos,
  %``On the velocity and chemical-potential dependence of the heavy-quark
  %interaction in N = 4 SYM plasmas,''
  arXiv:hep-th/0609079.
  %%CITATION = HEP-TH 0609079;%%

  %\cite{Friess:2006rk}
\bibitem{Friess:2006rk}
  J.~J.~Friess, S.~S.~Gubser, G.~Michalogiorgakis and S.~S.~Pufu,
  %``Stability of strings binding heavy-quark mesons,''
  arXiv:hep-th/0609137.
  %%CITATION = HEP-TH 0609137;%%

  %\cite{Talavera:2006tj}
\bibitem{Talavera:2006tj}
  P.~Talavera,
  %``Drag force in a string model dual to large-N QCD,''
  arXiv:hep-th/0610179.
  %%CITATION = HEP-TH 0610179;%%

  %\cite{Chernicoff:2006yp}
\bibitem{Chernicoff:2006yp}
  M.~Chernicoff and A.~Guijosa,
  %``Energy loss of gluons, baryons and k-quarks in an N = 4 SYM plasma,''
  arXiv:hep-th/0611155.
  %%CITATION = HEP-TH 0611155;%%

\bibitem{SeeAlso}
See also Appendix A of Ref.~\cite{Herzog:2006gh}.

%\cite{Wilson:1974sk}
\bibitem{Wilson:1974sk}
  K.~G.~Wilson,
  %``CONFINEMENT OF QUARKS,''
  Phys.\ Rev.\ D {\bf 10}, 2445 (1974).
  %%CITATION = PHRVA,D10,2445;%%

%\cite{McLerran:1980pk}
\bibitem{McLerran:1980pk}
  L.~D.~McLerran and B.~Svetitsky,
  %``A Monte Carlo Study Of SU(2) Yang-Mills Theory At Finite Temperature,''
  Phys.\ Lett.\ B {\bf 98}, 195 (1981);
 %%CITATION = PHLTA,B98,195;%%
 %\cite{Kuti:1980gh}
%\bibitem{Kuti:1980gh}
  J.~Kuti, J.~Polonyi and K.~Szlachanyi,
  %``Monte Carlo Study Of SU(2) Gauge Theory At Finite Temperature,''
  Phys.\ Lett.\ B {\bf 98}, 199 (1981);
  %%CITATION = PHLTA,B98,199;%%
   % \cite{McLerran:1981pb}
%\bibitem{McLerran:1981pb}
  L.~D.~McLerran and B.~Svetitsky,
  %``Quark Liberation At High Temperature: A Monte Carlo Study Of SU(2) Gauge
  %Theory,''
  Phys.\ Rev.\ D {\bf 24}, 450 (1981).
  %%CITATION = PHRVA,D24,450;%%

%\cite{Philipsen:2002az}
\bibitem{Philipsen:2002az}
  O.~Philipsen,
  %``Non-perturbative formulation of the static color octet potential,''
  Phys.\ Lett.\ B {\bf 535} (2002) 138
  [arXiv:hep-lat/0203018];
  %%CITATION = HEP-LAT 0203018;%%
%\cite{Jahn:2004qr}
%\bibitem{Jahn:2004qr}
  O.~Jahn and O.~Philipsen,
  %``The Polyakov loop and its relation to static quark potentials and free
  %energies,''
  Phys.\ Rev.\ D {\bf 70}, 074504 (2004)
  [arXiv:hep-lat/0407042].
  %%CITATION = HEP-LAT 0407042;%%

\bibitem{Kaczmarek:2002mc}
  O.~Kaczmarek, F.~Karsch, P.~Petreczky and F.~Zantow,
  %``Heavy quark anti-quark free energy and the renormalized Polyakov loop,''
  Phys.\ Lett.\ B {\bf 543} (2002) 41
  [arXiv:hep-lat/0207002].
  %%CITATION = HEP-LAT 0207002;%%


%\cite{Laine:2006ns}
\bibitem{Laine:2006ns}
   M.~Laine, O.~Philipsen, P.~Romatschke and M.~Tassler,
   %``Real-time static potential in hot QCD,''
   arXiv:hep-ph/0611300.
   %%CITATION = HEP-PH 0611300;%%

%\cite{Rey:1998ik}
\bibitem{Rey:1998ik}
%\cite{Maldacena:1998im}
%\bibitem{Maldacena:1998im}
  J.~M.~Maldacena,
  %``Wilson loops in large N field theories,''
  Phys.\ Rev.\ Lett.\  {\bf 80}, 4859 (1998)  [arXiv:hep-th/9803002];
  %%CITATION = HEP-TH 9803002;%%
  S.~J.~Rey and J.~T.~Yee,
  %``Macroscopic strings as heavy quarks in large N gauge theory and  anti-de
  %Sitter supergravity,''
  Eur.\ Phys.\ J.\ C {\bf 22}, 379 (2001)  [arXiv:hep-th/9803001].
  %%CITATION = HEP-TH 9803001;%%

%\cite{Matsui:1986dk}
\bibitem{Matsui:1986dk}
  T.~Matsui and H.~Satz,
  %``J / Psi Suppression By Quark - Gluon Plasma Formation,''
  Phys.\ Lett.\ B {\bf 178}, 416 (1986).
  %%CITATION = PHLTA,B178,416;%%

%\cite{Satz:2005hx}
\bibitem{Satz:2005hx}
  H.~Satz,
  %``Colour deconfinement and quarkonium binding,''
  J.\ Phys.\ G {\bf 32} (2006) R25
  [arXiv:hep-ph/0512217].
  %%CITATION = HEP-PH 0512217;%%

\bibitem{Kovner:2001vi}
  A.~Kovner and U.~A.~Wiedemann,
  %``Eikonal evolution and gluon radiation,''
  Phys.\ Rev.\ D {\bf 64}, 114002 (2001) [arXiv:hep-ph/0106240].
  %%CITATION = HEP-PH 0106240;%%

%\cite{Kovner:2003zj}
\bibitem{Kovner:2003zj}
  A.~Kovner and U.~A.~Wiedemann,
  %``Gluon radiation and parton energy loss,''
  arXiv:hep-ph/0304151.
  %%CITATION = HEP-PH 0304151;%%

% virtual photon absorption start

%\cite{Mueller:1989st}
\bibitem{Mueller:1989st}
  A.~H.~Mueller,
  %``SMALL x BEHAVIOR AND PARTON SATURATION: A QCD MODEL,''
  Nucl.\ Phys.\ B {\bf 335} (1990) 115.
  %%CITATION = NUPHA,B335,115;%%

 %\cite{Nikolaev:1990ja}
\bibitem{Nikolaev:1990ja}
  N.~N.~Nikolaev and B.~G.~Zakharov,
%   ``Colour transparency and scaling properties of nuclear shadowing in deep
  %inelastic scattering,''
  Z.\ Phys.\ C {\bf 49}, 607 (1991).
  %%CITATION = ZEPYA,C49,607;%%

  %\cite{Piller:1999wx}
\bibitem{Piller:1999wx}
  G.~Piller and W.~Weise,
  %``Nuclear deep-inelastic lepton scattering and coherence phenomena,''
  Phys.\ Rept.\  {\bf 330}, 1 (2000)
  [arXiv:hep-ph/9908230].
  %%CITATION = HEP-PH 9908230;%%

  %\cite{Kovchegov:1999kx}
\bibitem{Kovchegov:1999kx}
  Y.~V.~Kovchegov and L.~D.~McLerran,
  %``Diffractive structure function in a quasi-classical approximation,''
  Phys.\ Rev.\ D {\bf 60} (1999) 054025
  [Erratum-ibid.\ D {\bf 62} (2000) 019901]
  [arXiv:hep-ph/9903246].
  %%CITATION = HEP-PH 9903246;%%

 %\cite{Wiedemann:2000ez}
\bibitem{Wiedemann:2000ez}
  U.~A.~Wiedemann,
  %``Transverse dynamics of hard partons in nuclear media and the QCD  dipole,''
  Nucl.\ Phys.\ B {\bf 582}, 409 (2000)
  [arXiv:hep-ph/0003021].
  %%CITATION = HEP-PH 0003021;%%

%%% virtual photon absorption end

%%% REFS on CGC

%\cite{Kovner:2005pe}
\bibitem{Kovner:2005pe}
  A.~Kovner,
  %``High energy evolution: The wave function point of view,''
  Acta Phys.\ Polon.\ B {\bf 36} (2005) 3551
  [arXiv:hep-ph/0508232].
  %%CITATION = HEP-PH 0508232;%%

%\cite{Weigert:2005us}
\bibitem{Weigert:2005us}
  H.~Weigert,
  %``Evolution at small x(bj): The color glass condensate,''
  Prog.\ Part.\ Nucl.\ Phys.\  {\bf 55} (2005) 461
  [arXiv:hep-ph/0501087].
  %%CITATION = HEP-PH 0501087;%%

%\cite{Iancu:2003xm}
\bibitem{Iancu:2003xm}
  E.~Iancu and R.~Venugopalan,
  %``The color glass condensate and high energy scattering in QCD,''
  arXiv:hep-ph/0303204.
  %%CITATION = HEP-PH 0303204;%%
%%% end REFS on CGC

%%%% Cronin

%\bibitem{Cronin:zm}
\bibitem{CRONIN}
J.~W.~Cronin {\it et al.},
%``Production Of Hadrons With Large Transverse Momentum At 200-Gev,
% 300-Gev, And 400-Gev,''
Phys.\ Rev.\ D {\bf 11}, 3105 (1975),
%%CITATION = PHRVA,D11,3105;%%
%
%
%\cite{Adler:2003ii}
%\bibitem{Adler:2003ii}
  S.~S.~Adler {\it et al.}  [PHENIX Collaboration],
%   ``Absence of suppression in particle production at large transverse  momentum
  %in s(NN)**(1/2) = 200-GeV d + Au collisions,''
  Phys.\ Rev.\ Lett.\  {\bf 91} (2003) 072303
  [arXiv:nucl-ex/0306021],
  %%CITATION = NUCL-EX 0306021;%%
 %
  %\cite{Adams:2003im}
%\bibitem{Adams:2003im}
  J.~Adams {\it et al.}  [STAR Collaboration],
%   ``Evidence from d + Au measurements for final-state suppression of high  p(T)
  %hadrons in Au + Au collisions at RHIC,''
  Phys.\ Rev.\ Lett.\  {\bf 91} (2003) 072304
  [arXiv:nucl-ex/0306024],
  %%CITATION = NUCL-EX 0306024;%%
%
 %\cite{Back:2003ns}
%\bibitem{Back:2003ns}
  B.~B.~Back {\it et al.}  [PHOBOS Collaboration],
 %  ``Centrality dependence of charged hadron transverse momentum spectra in  d +
  %Au collisions at s(NN)**(1/2) = 200-GeV,''
  Phys.\ Rev.\ Lett.\  {\bf 91} (2003) 072302
  [arXiv:nucl-ex/0306025],
  %%CITATION = NUCL-EX 0306025;%%
%
%\cite{Arsene:2003yk}
%\bibitem{Arsene:2003yk}
  I.~Arsene {\it et al.}  [BRAHMS Collaboration],
 %  ``Transverse momentum spectra in Au + Au and d + Au collisions at
 %  s(NN)**(1/2) = 200-GeV and the pseudorapidity dependence of high p(T)
  %suppression,''
  Phys.\ Rev.\ Lett.\  {\bf 91} (2003) 072305
  [arXiv:nucl-ex/0307003].
  %%CITATION = NUCL-EX 0307003;%%

%%%%% REFS on pA

%\cite{Kharzeev:2002pc}
\bibitem{Kharzeev:2002pc}
  D.~Kharzeev, E.~Levin and L.~McLerran,
  %``Parton saturation and N(part) scaling of semi-hard processes in QCD,''
  Phys.\ Lett.\ B {\bf 561}, 93 (2003)
  [arXiv:hep-ph/0210332].
  %%CITATION = HEP-PH 0210332;%%

%\cite{Baier:2003hr}
\bibitem{Baier:2003hr}
  R.~Baier, A.~Kovner and U.~A.~Wiedemann,
%   ``Saturation and parton level Cronin effect: Enhancement vs suppression  of
  %gluon production in p A and A A collisions,''
  Phys.\ Rev.\ D {\bf 68} (2003) 054009
  [arXiv:hep-ph/0305265].
  %%CITATION = HEP-PH 0305265;%%

%\cite{Kharzeev:2003wz}
\bibitem{Kharzeev:2003wz}
  D.~Kharzeev, Y.~V.~Kovchegov and K.~Tuchin,
  %``Cronin effect and high-p(T) suppression in p A collisions,''
  Phys.\ Rev.\ D {\bf 68} (2003) 094013
  [arXiv:hep-ph/0307037].
  %%CITATION = HEP-PH 0307037;%%

%\cite{Albacete:2003iq}
\bibitem{Albacete:2003iq}
  J.~L.~Albacete, N.~Armesto, A.~Kovner, C.~A.~Salgado and U.~A.~Wiedemann,
  %``Energy dependence of the Cronin effect from non-linear QCD evolution,''
  Phys.\ Rev.\ Lett.\  {\bf 92} (2004) 082001
  [arXiv:hep-ph/0307179].
  %%CITATION = HEP-PH 0307179;%%

%\cite{Jalilian-Marian:2003mf}
\bibitem{Jalilian-Marian:2003mf}
  J.~Jalilian-Marian, Y.~Nara and R.~Venugopalan,
  %``The Cronin effect, quantum evolution and the color glass condensate,''
  Phys.\ Lett.\ B {\bf 577}, 54 (2003)
  [arXiv:nucl-th/0307022].
  %%CITATION = NUCL-TH 0307022;%%

%\cite{Blaizot:2004wu}
\bibitem{Blaizot:2004wu}
  J.~P.~Blaizot, F.~Gelis and R.~Venugopalan,
%   ``High energy p A collisions in the color glass condensate approach. I:
  %Gluon production and the Cronin effect,''
  Nucl.\ Phys.\ A {\bf 743} (2004) 13
  [arXiv:hep-ph/0402256].
  %%CITATION = HEP-PH 0402256;%%

%% end REFS on pA

%\cite{Kovchegov:1998bi}
\bibitem{Kovchegov:1998bi}
Y.~V.~Kovchegov and A.~H.~Mueller,
%``Gluon production in current nucleus and nucleon nucleus collisions
% in  a quasi-classical approximation,''
Nucl.\ Phys.\ B {\bf 529} (1998) 451
[arXiv:hep-ph/9802440].
%%CITATION = HEP-PH 9802440;%%

%\cite{Baier:2002tc}
\bibitem{Baier:2002tc}
R.~Baier,
%``Jet quenching,''
Nucl.\ Phys.\ A {\bf 715}, 209 (2003) [arXiv:hep-ph/0209038].
%%CITATION = HEP-PH 0209038;%%

%\cite{Baier:1996sk}
\bibitem{Baier:1996sk}
  R.~Baier, Y.~L.~Dokshitzer, A.~H.~Mueller, S.~Peigne and D.~Schiff,
%   ``Radiative energy loss and p(T)-broadening of high energy partons in
  %nuclei,''
  Nucl.\ Phys.\ B {\bf 484} (1997) 265
  [arXiv:hep-ph/9608322].
  %%CITATION = HEP-PH 9608322;%%

%\cite{Zakharov:1997uu}
\bibitem{Zakharov:1997uu}
B.~G.~Zakharov,
%\cite{Zakharov:1996fv}
% \bibitem{Zakharov:1996fv}
%  B.~G.~Zakharov,
  %``Fully quantum treatment of the Landau-Pomeranchuk-Migdal effect in QED  and
  %QCD,''
  JETP Lett.\  {\bf 63} (1996) 952
  [arXiv:hep-ph/9607440];
  %%CITATION = HEP-PH 9607440;%%
%
%``Radiative energy loss of high energy quarks in finite-size nuclear
% matter and quark-gluon plasma,''
JETP Lett.\  {\bf 65}, 615 (1997) [arXiv:hep-ph/9704255].
%%CITATION = HEP-PH 9704255;%%
%

%\cite{Wiedemann:2000za}
\bibitem{Wiedemann:2000za}
U.~A.~Wiedemann,
%``Gluon radiation off hard quarks in a nuclear environment:
% Opacity expansion,''
Nucl.\ Phys.\ B {\bf 588}, 303 (2000) [arXiv:hep-ph/0005129].
%%CITATION = HEP-PH 0005129;%%

%\cite{Kopeliovich:1998nw}
\bibitem{Kopeliovich:1998nw}
  B.~Z.~Kopeliovich, A.~V.~Tarasov and A.~Schafer,
  %``Bremsstrahlung of a quark propagating through a nucleus,''
  Phys.\ Rev.\ C {\bf 59} (1999) 1609
  [arXiv:hep-ph/9808378].
  %%CITATION = HEP-PH 9808378;%%

%
  %\cite{Buchmuller:1995mr}
\bibitem{Buchmuller:1995mr}
  W.~Buchmuller and A.~Hebecker,
  %``Semiclassical Approach to Structure Functions at Small x,''
  Nucl.\ Phys.\ B {\bf 476} (1996) 203
  [arXiv:hep-ph/9512329].
  %%CITATION = HEP-PH 9512329;%%
%

%\cite{Salgado:2003rv}
\bibitem{Salgado:2003rv}
  C.~A.~Salgado and U.~A.~Wiedemann,
  %``Medium modification of jet shapes and jet multiplicities,''
  Phys.\ Rev.\ Lett.\  {\bf 93}, 042301 (2004)
  [arXiv:hep-ph/0310079].
  %%CITATION = HEP-PH 0310079;%%

%\cite{Salgado:2003gb}
\bibitem{Salgado:2003gb}
  C.~A.~Salgado and U.~A.~Wiedemann,
  %``Calculating quenching weights,''
  Phys.\ Rev.\ D {\bf 68} (2003) 014008
  [arXiv:hep-ph/0302184].
  %%CITATION = HEP-PH 0302184;%%

%\cite{Zakharov:2000iz}
\bibitem{Zakharov:2000iz}
  B.~G.~Zakharov,
  %``On the energy loss of high energy quarks in a finite-size quark gluon
  %plasma,''
  JETP Lett.\  {\bf 73}, 49 (2001)
  [Pisma Zh.\ Eksp.\ Teor.\ Fiz.\  {\bf 73}, 55 (2001)]
  [arXiv:hep-ph/0012360].
  %%CITATION = HEP-PH 0012360;%%


%\cite{Gyulassy:2000er}
\bibitem{Gyulassy:2000er}
M.~Gyulassy, P.~Levai and I.~Vitev,
%``Reaction operator approach to non-Abelian energy loss,''
Nucl.\ Phys.\ B {\bf 594}, 371 (2001)  [arXiv:nucl-th/0006010];
%%CITATION = NUCL-TH 0006010;%%
%\cite{Wang:2001if}
%\bibitem{Wang:2001if}
X.~N.~Wang and X.~f.~Guo,
%``Multiple parton scattering in nuclei: Parton energy loss,''
Nucl.\ Phys.\ A {\bf 696}, 788 (2001)  [arXiv:hep-ph/0102230];
%%CITATION = HEP-PH 0102230;%%
%\cite{Jeon:2003gi}
%\bibitem{Jeon:2003gi}
S.~Jeon and G.~D.~Moore,
%``Energy loss of leading partons in a thermal QCD medium,''
Phys.\ Rev.\ C {\bf 71}, 034901 (2005) [arXiv:hep-ph/0309332].
%%CITATION = HEP-PH 0309332;%%






\bibitem{jetquenchrev}
%\cite{Baier:2000mf}
%\bibitem{Baier:2000mf}
 For reviews, see
  R.~Baier, D.~Schiff and B.~G.~Zakharov,
  %``Energy loss in perturbative QCD,''
  Ann.\ Rev.\ Nucl.\ Part.\ Sci.\  {\bf 50}, 37 (2000)  [arXiv:hep-ph/0002198];
  %%CITATION = HEP-PH 0002198;%%
  %%
%\cite{Gyulassy:2003mc}
% \bibitem{Gyulassy:2003mc}
  M.~Gyulassy, I.~Vitev, X.~N.~Wang and B.~W.~Zhang,
  %``Jet quenching and radiative energy loss in dense nuclear matter,''
  arXiv:nucl-th/0302077;
  %%CITATION = NUCL-TH 0302077;%%
%
%\cite{Jacobs:2004qv}
%\bibitem{Jacobs:2004qv}
  P.~Jacobs and X.~N.~Wang,
  %``Matter in extremis: Ultrarelativistic nuclear collisions at RHIC,''
  Prog.\ Part.\ Nucl.\ Phys.\  {\bf 54}, 443 (2005) [arXiv:hep-ph/0405125].
  %%CITATION = HEP-PH 0405125;%%


\bibitem{LHC}
%
%\cite{Carminati:2004fp}
% \bibitem{Carminati:2004fp}
  F.~Carminati {\it et al.}  [ALICE Collaboration],
  %``ALICE: Physics performance report, volume I,''
  J.\ Phys.\ G {\bf 30}, 1517 (2004);
  %%CITATION = JPHGB,G30,1517;%%
%
%\cite{CMS-Loi}
%\bibitem{CMS-Loi}
G.~Baur {\it et al.} [CMS Collaboration], % "Heavy-Ion Physics at the LHC"
CMS-Note-2000-060;
%
%\cite{Takai:2004nm}
% \bibitem{Takai:2004nm}
  H.~Takai, [for the ATLAS Collaboration]
  %``Heavy ion physics with the ATLAS detector,''
  Eur.\ Phys.\ J.\ C {\bf 34}, S307 (2004).
  %%CITATION = EPHJA,C34,S307;%%



%\cite{Baier:2006fr}
\bibitem{Baier:2006fr}
  R.~Baier and D.~Schiff,
  %``Deciphering the properties of the medium produced in heavy ion collisions
  %at RHIC by a pQCD analysis of quenched large p(T) pi0 spectra,''
  JHEP {\bf 0609}, 059 (2006)
  [arXiv:hep-ph/0605183].
  %%CITATION = HEP-PH 0605183;%%

 \bibitem{IRUV}
 %\cite{Susskind:1998dq}
%\bibitem{Susskind:1998dq}
  L.~Susskind and E.~Witten,
  %``The holographic bound in anti-de Sitter space,''
  arXiv:hep-th/9805114;
  %%CITATION = HEP-TH 9805114;%%
%\cite{Peet:1998wn}
%\bibitem{Peet:1998wn}
  A.~W.~Peet and J.~Polchinski,
  %``UV/IR relations in AdS dynamics,''
  Phys.\ Rev.\ D {\bf 59}, 065011 (1999)
  [arXiv:hep-th/9809022].
  %%CITATION = HEP-TH 9809022;%%





%\cite{Witten:1998zw}
\bibitem{Witten:1998zw}
E.~Witten, in Ref.~\cite{AdS/CFT}
%``Anti-de Sitter space, thermal phase transition, and confinement in  gauge
%theories,''
%Adv.\ Theor.\ Math.\ Phys.\  {\bf 2}, 505 (1998).
%[arXiv:hep-th/9803131].
%%CITATION = HEP-TH 9803131;%%

%\cite{Karsch:2005nk}
\bibitem{Karsch:2005nk}
  F.~Karsch, D.~Kharzeev and H.~Satz,
  %``Sequential charmonium dissociation,''
  Phys.\ Lett.\ B {\bf 637}, 75 (2006) [arXiv:hep-ph/0512239].
  %%CITATION = HEP-PH 0512239;%%


\bibitem{Satz}
%\cite{Digal:2001iu}
%\bibitem{Digal:2001iu}
 S.~Digal, P.~Petreczky and H.~Satz,
  %``String breaking and quarkonium dissociation at finite temperatures,''
  Phys.\ Lett.\ B {\bf 514}, 57 (2001);
%  [arXiv:hep-ph/0105234].
  %%CITATION = HEP-PH 0105234;%%
%
%\cite{Satz:2006kb}
% \bibitem{Satz:2006kb}
  H.~Satz,
  %``Quarkonium binding and dissociation: The spectral analysis of the QGP,''
  arXiv:hep-ph/0609197.
  %%CITATION = HEP-PH 0609197;%%

%\cite{Karsch:1987zw}
\bibitem{Karsch:1987zw}
  F.~Karsch and R.~Petronzio,
  %``chi AND J / psi SUPPRESSION IN HEAVY ION COLLISIONS AND A MODEL FOR ITS
  %MOMENTUM DEPENDENCE,''
  Z.\ Phys.\ C {\bf 37}, 627 (1988).
  %%CITATION = ZEPYA,C37,627;%%


%\cite{Armesto:2004vz}
\bibitem{Armesto:2004vz}
  N.~Armesto, C.~A.~Salgado and U.~A.~Wiedemann,
  %``Low-p(T) collective flow induces high-p(T) jet quenching,''
  Phys.\ Rev.\ C {\bf 72}, 064910 (2005)
  [arXiv:hep-ph/0411341].
  %%CITATION = HEP-PH 0411341;%%

  %\cite{Renk:2005ta}
\bibitem{Renk:2005ta}
  T.~Renk and J.~Ruppert,
  %``Flow dependence of high p(T) parton energy loss in heavy-ion  collisions,''
  Phys.\ Rev.\ C {\bf 72}, 044901 (2005)
  [arXiv:hep-ph/0507075].
  %%CITATION = HEP-PH 0507075;%%

\bibitem{BaierPrivate}
%\cite{Baier:2006pt}
%\bibitem{Baier:2006pt}
R. Baier, private communication,
  R.~Baier, A.~H.~Mueller and D.~Schiff,
  %``How does transverse (hydrodynamic) flow affect jet-broadening and
  %jet-quenching ?,''
  arXiv:nucl-th/0612068.
  %%CITATION = NUCL-TH 0612068;%%


%\cite{Renk:2006sx}
\bibitem{Renk:2006sx}
  T.~Renk, J.~Ruppert, C.~Nonaka and S.~A.~Bass,
  %``Jet-quenching in a 3D hydrodynamic medium,''
  arXiv:nucl-th/0611027.
  %%CITATION = NUCL-TH 0611027;%%

%\cite{Klebanov:2000hb}
\bibitem{Klebanov:2000hb}
  I.~R.~Klebanov and M.~J.~Strassler,
   %``Supergravity and a confining gauge theory: Duality cascades and
  %chiSB-resolution of naked singularities,''
  JHEP {\bf 0008}, 052 (2000)
  [arXiv:hep-th/0007191].
  %%CITATION = HEP-TH 0007191;%%

%\cite{Karsch:2006xs}
\bibitem{Karsch:2006xs}
For recent reviews, see
  %``Lattice simulations of the thermodynamics of strongly interacting
  %elementary particles and the exploration of new phases of matter in
  %relativistic heavy ion collisions,''
  J.\ Phys.\ Conf.\ Ser.\  {\bf 46}, 122 (2006)
  [arXiv:hep-lat/0608003];
  %%CITATION = HEP-LAT 0608003;%%
%\cite{Karsch:2006sf}
%\bibitem{Karsch:2006sf}
  F.~Karsch,
  %``Properties of the quark gluon plasma: A lattice perspective,''
  arXiv:hep-ph/0610024.
  %%CITATION = HEP-PH 0610024;%%

%\cite{Caron-Huot:2006te}
\bibitem{Caron-Huot:2006te}
  S.~Caron-Huot, P.~Kovtun, G.~D.~Moore, A.~Starinets and L.~G.~Yaffe,
  %``Photon and dilepton production in supersymmetric Yang-Mills plasma,''
  arXiv:hep-th/0607237;
  %%CITATION = HEP-TH 0607237;%%
%\cite{Huot:2006ys}
%\bibitem{Huot:2006ys}
  S.~C.~Huot, S.~Jeon and G.~D.~Moore,
  %``Shear viscosity in weakly coupled N = 4 super Yang-Mills theory compared to
  %QCD,''
  arXiv:hep-ph/0608062.
  %%CITATION = HEP-PH 0608062;%%

%\cite{Gubser:2006qh}
\bibitem{Gubser:2006qh}
  S.~S.~Gubser,
  %``Comparing the drag force on heavy quarks in N=4 super-Yang-Mills theory and
  %QCD,''
  arXiv:hep-th/0611272.
  %%CITATION = HEP-TH 0611272;%%


%\cite{Karsch:2001vs}
\bibitem{Karsch:2001vs}
  F.~Karsch,
  %``Lattice results on QCD thermodynamics,''
  Nucl.\ Phys.\ A {\bf 698}, 199 (2002)
  [arXiv:hep-ph/0103314].
  %%CITATION = HEP-PH 0103314;%%

%\cite{Gubser:1996de}
\bibitem{Gubser:1996de}
  S.~S.~Gubser, I.~R.~Klebanov and A.~W.~Peet,
  %``Entropy and Temperature of Black 3-Branes,''
  Phys.\ Rev.\ D {\bf 54}, 3915 (1996)
  [arXiv:hep-th/9602135].
  %%CITATION = HEP-TH 9602135;%%

%\cite{Gubser:1998nz}
\bibitem{Gubser:1998nz}
  S.~S.~Gubser, I.~R.~Klebanov and A.~A.~Tseytlin,
  %``Coupling constant dependence in the thermodynamics of N = 4  supersymmetric
  %Yang-Mills theory,''
  Nucl.\ Phys.\ B {\bf 534}, 202 (1998)
  [arXiv:hep-th/9805156].
  %%CITATION = HEP-TH 9805156;%%

  %\cite{Bringoltz:2005rr}
\bibitem{Bringoltz:2005rr}
  B.~Bringoltz and M.~Teper,
  %``The pressure of the SU(N) lattice gauge theory at large-N,''
  Phys.\ Lett.\ B {\bf 628}, 113 (2005)
  [arXiv:hep-lat/0506034].
  %%CITATION = HEP-LAT 0506034;%%

  %
%\cite{Eskola:2004cr}
\bibitem{Eskola:2004cr}
K.~J.~Eskola {\it et al.},
%H.~Honkanen, C.~A.~Salgado and U.~A.~Wiedemann,
%``The fragility of high-p(T) hadron spectra as a hard probe,''
Nucl.\ Phys.\ A {\bf 747}, 511 (2005) [arXiv:hep-ph/0406319].
%%CITATION = HEP-PH 0406319;%%

%\cite{Dainese:2004te}
\bibitem{Dainese:2004te}
  A.~Dainese, C.~Loizides and G.~Paic,
  %``Leading-particle suppression in high energy nucleus nucleus collisions,''
  Eur.\ Phys.\ J.\ C {\bf 38}, 461 (2005)  [arXiv:hep-ph/0406201].
  %%CITATION = HEP-PH 0406201;%%

%\cite{Kolb:2003dz}
\bibitem{Kolb:2003dz}
  P.~F.~Kolb and U.~W.~Heinz,
  %``Hydrodynamic description of ultrarelativistic heavy-ion collisions,''
  arXiv:nucl-th/0305084.
  %%CITATION = NUCL-TH 0305084;%%
%


%\cite{Moore:2004tg}
\bibitem{Moore:2004tg}
  G.~D.~Moore and D.~Teaney,
  %``How much do heavy quarks thermalize in a heavy ion collision?,''
  Phys.\ Rev.\ C {\bf 71}, 064904 (2005)
  [arXiv:hep-ph/0412346].
  %%CITATION = HEP-PH 0412346;%%

\bibitem{Casalderrey-Solana-Private}
J. Casalderrey-Solana, private communication.


%\cite{Karsch:2000ps}
\bibitem{Karsch:2000ps}
  F.~Karsch, E.~Laermann and A.~Peikert,
  %``The pressure in 2, 2+1 and 3 flavour QCD,''
  Phys.\ Lett.\ B {\bf 478}, 447 (2000)
  [arXiv:hep-lat/0002003].
  %%CITATION = HEP-LAT 0002003;%%

%\cite{Boyd:1996bx}
\bibitem{Boyd:1996bx}
  G.~Boyd, J.~Engels, F.~Karsch, E.~Laermann, C.~Legeland, M.~Lutgemeier and B.~Petersson,
  %``Thermodynamics of SU(3) Lattice Gauge Theory,''
  Nucl.\ Phys.\ B {\bf 469}, 419 (1996)
  [arXiv:hep-lat/9602007].
  %%CITATION = HEP-LAT 9602007;%%

%\cite{AliKhan:2001ek}
\bibitem{AliKhan:2001ek}
  A.~Ali Khan {\it et al.}  [CP-PACS collaboration],
  %``Equation of state in finite-temperature QCD with two flavors of  improved
  %Wilson quarks,''
  Phys.\ Rev.\ D {\bf 64}, 074510 (2001)
  [arXiv:hep-lat/0103028].
  %%CITATION = HEP-LAT 0103028;%%

%\cite{Aoki:2005vt}
\bibitem{Aoki:2005vt}
  Y.~Aoki, Z.~Fodor, S.~D.~Katz and K.~K.~Szabo,
  %``The equation of state in lattice QCD: With physical quark masses  towards
  %the continuum limit,''
  JHEP {\bf 0601}, 089 (2006)
  [arXiv:hep-lat/0510084].
  %%CITATION = HEP-LAT 0510084;%%

  %\cite{Ejiri:2005uv}
\bibitem{Ejiri:2005uv}
  S.~Ejiri, F.~Karsch, E.~Laermann and C.~Schmidt,
  %``The isentropic equation of state of 2-flavor QCD,''
  Phys.\ Rev.\ D {\bf 73}, 054506 (2006)
  [arXiv:hep-lat/0512040].
  %%CITATION = HEP-LAT 0512040;%%

%\cite{Kaczmarek:2004gv}
\bibitem{Kaczmarek:2004gv}
  O.~Kaczmarek, F.~Karsch, F.~Zantow and P.~Petreczky,
  %``Static quark anti-quark free energy and the running coupling at finite
  %temperature,''
  Phys.\ Rev.\ D {\bf 70}, 074505 (2004)
  [Erratum-ibid.\ D {\bf 72}, 059903 (2005)] [arXiv:hep-lat/0406036].
  %%CITATION = HEP-LAT 0406036;%%

%\cite{Kaczmarek:2005ui}
\bibitem{Kaczmarek:2005ui}
  O.~Kaczmarek and F.~Zantow,
  %``Static quark anti-quark interactions in zero and finite temperature  QCD.
  %I: Heavy quark free energies, running coupling and quarkonium  binding,''
  Phys.\ Rev.\ D {\bf 71}, 114510 (2005)
  [arXiv:hep-lat/0503017].
  %%CITATION = HEP-LAT 0503017;%%

%\cite{Muronga:2001zk}
\bibitem{Muronga:2001zk}
   A.~Muronga,
   %``Second order dissipative fluid dynamics for ultra-relativistic nuclear
   %collisions,''
   Phys.\ Rev.\ Lett.\  {\bf 88} (2002) 062302
   [Erratum-ibid.\  {\bf 89} (2002) 159901]
   [arXiv:nucl-th/0104064];
   %%CITATION = NUCL-TH 0104064;%%
%\cite{Muronga:2003ta}
%\bibitem{Muronga:2003ta}
   A.~Muronga,
   %``Causal Theories of Dissipative Relativistic Fluid Dynamics for Nuclear
   %Collisions,''
   Phys.\ Rev.\ C {\bf 69} (2004) 034903
   [arXiv:nucl-th/0309055];
   %%CITATION = NUCL-TH 0309055;%%
   %\cite{Heinz:2005bw}
%\bibitem{Heinz:2005bw}
   U.~W.~Heinz, H.~Song and A.~K.~Chaudhuri,
   %``Dissipative hydrodynamics for viscous relativistic fluids,''
   Phys.\ Rev.\ C {\bf 73} (2006) 034904
   [arXiv:nucl-th/0510014];
   %%CITATION = NUCL-TH 0510014;%%
   %\cite{Baier:2006um}
%\bibitem{Baier:2006um}
   R.~Baier, P.~Romatschke and U.~A.~Wiedemann,
   %``Dissipative hydrodynamics and heavy ion collisions,''
   Phys.\ Rev.\ C {\bf 73} (2006) 064903
   [arXiv:hep-ph/0602249];
   %%CITATION = HEP-PH 0602249;%%
%\bibitem{Baier:2006gy}
   R.~Baier and P.~Romatschke,
   %``Causal viscous hydrodynamics for central heavy-ion collisions,''
   arXiv:nucl-th/0610108.
   %%CITATION = NUCL-TH 0610108;%%



%\cite{Gauntlett:2004yd}
\bibitem{gauntlett}
  J.~P.~Gauntlett, D.~Martelli, J.~Sparks and D.~Waldram,
  %``Sasaki-Einstein metrics on S(2) x S(3),''
  Adv.\ Theor.\ Math.\ Phys.\  {\bf 8}, 711 (2004)
  [arXiv:hep-th/0403002].
  %%CITATION = HEP-TH 0403002;%%

%\cite{Benvenuti:2004dy}
\bibitem{ami}
  S.~Benvenuti, S.~Franco, A.~Hanany, D.~Martelli and J.~Sparks,
   %``An infinite family of superconformal quiver gauge theories with
  %Sasaki-Einstein duals,''
  JHEP {\bf 0506}, 064 (2005)
  [arXiv:hep-th/0411264].
  %%CITATION = HEP-TH 0411264;%%

%\cite{Klebanov:1998hh}
\bibitem{klebanov}
  I.~R.~Klebanov and E.~Witten,
  %``Superconformal field theory on threebranes at a Calabi-Yau  singularity,''
  Nucl.\ Phys.\ B {\bf 536}, 199 (1998)
  [arXiv:hep-th/9807080].
  %%CITATION = HEP-TH 9807080;%%

%\cite{Henningson:1998gx}
\bibitem{skenderis}
  M.~Henningson and K.~Skenderis,
  %``The holographic Weyl anomaly,''
  JHEP {\bf 9807}, 023 (1998)
  [arXiv:hep-th/9806087].
  %%CITATION = HEP-TH 9806087;%%

%\cite{Gubser:1998vd}
\bibitem{gubser}
  S.~S.~Gubser,
  %``Einstein manifolds and conformal field theories,''
  Phys.\ Rev.\ D {\bf 59}, 025006 (1999)
  [arXiv:hep-th/9807164].
  %%CITATION = HEP-TH 9807164;%%


%\cite{Pilch:2000ej}
\bibitem{pilch}
  K.~Pilch and N.~P.~Warner,
  %``A new supersymmetric compactification of chiral IIB supergravity,''
  Phys.\ Lett.\ B {\bf 487}, 22 (2000)
  [arXiv:hep-th/0002192].
  %%CITATION = HEP-TH 0002192;%%


%\cite{Leigh:1995ep}
\bibitem{leigh}
  R.~G.~Leigh and M.~J.~Strassler,
  % ``Exactly Marginal Operators And Duality In Four-Dimensional N=1
  %Supersymmetric Gauge Theory,''
  Nucl.\ Phys.\ B {\bf 447}, 95 (1995)
  [arXiv:hep-th/9503121].
  %%CITATION = HEP-TH 9503121;%%



%\cite{Gauntlett:2005ww}
\bibitem{mgauntlett}
  J.~P.~Gauntlett, D.~Martelli, J.~Sparks and D.~Waldram,
  %``Supersymmetric AdS(5) solutions of type IIB supergravity,''
  Class.\ Quant.\ Grav.\  {\bf 23}, 4693 (2006)
  [arXiv:hep-th/0510125].
  %%CITATION = HEP-TH 0510125;%%



%\cite{Freedman:1999gp}
\bibitem{FGPW}
  D.~Z.~Freedman, S.~S.~Gubser, K.~Pilch and N.~P.~Warner,
   %``Renormalization group flows from holography supersymmetry and a
  %c-theorem,''
  Adv.\ Theor.\ Math.\ Phys.\  {\bf 3}, 363 (1999)
  [arXiv:hep-th/9904017].
  %%CITATION = HEP-TH 9904017;%%

%\cite{Buchel:2003ah}
\bibitem{Buchel:2003ah}
  A.~Buchel and J.~T.~Liu,
  %``Thermodynamics of the N = 2* flow,''
  JHEP {\bf 0311}, 031 (2003)   [arXiv:hep-th/0305064].
  %%CITATION = HEP-TH 0305064;%%

%\cite{Itzhaki:1998dd}
\bibitem{Itzhaki:1998dd}
N.~Itzhaki, J.~M.~Maldacena, J.~Sonnenschein and S.~Yankielowicz,
%``Supergravity and the large N limit of theories with sixteen
%supercharges,''
Phys.\ Rev.\ D {\bf 58}, 046004 (1998)
[arXiv:hep-th/9802042].
%%CITATION = HEP-TH 9802042;%%

%\cite{Gubser:2006nz}
\bibitem{Gubser:2006nz}
  S.~S.~Gubser,
  %``Jet-quenching and momentum correlators from the gauge-string duality,''
  arXiv:hep-th/0612143.
  %%CITATION = HEP-TH 0612143;%%

%\cite{Casalderrey-Solana:2007qw}
\bibitem{calsa}
  J.~Casalderrey-Solana and D.~Teaney,
  %``Transverse momentum broadening of a fast quark in a N = 4 Yang Mills
  %plasma,''
  arXiv:hep-th/0701123.
  %%CITATION = HEP-TH 0701123;%%



%\cite{Argyres:2006yz}
\bibitem{Argyres:2006yz}
  P.~C.~Argyres, M.~Edalati and J.~F.~Vazquez-Poritz,
  %``Spacelike strings and jet quenching from a Wilson loop,''
  arXiv:hep-th/0612157.
  %%CITATION = HEP-TH 0612157;%%


%\cite{Dokshitzer:2001zm}
\bibitem{Dokshitzer:2001zm}
 Y.~L.~Dokshitzer and D.~E.~Kharzeev,
 %``Heavy quark colorimetry of QCD matter,''
 Phys.\ Lett.\ B {\bf 519} (2001) 199
 [arXiv:hep-ph/0106202];
 %%CITATION = HEP-PH 0106202;%%
 %\cite{Armesto:2003jh}
%\bibitem{Armesto:2003jh}
N.~Armesto, C.~A.~Salgado and U.~A.~Wiedemann,
%``Medium-induced gluon radiation off massive quarks fills the dead cone,''
Phys.\ Rev.\ D {\bf 69} (2004) 114003
 [arXiv:hep-ph/0312106].
 %%CITATION = HEP-PH 0312106;%%


\end{thebibliography}
\end{document}